# Theoretische Modellierung granularer Ströme in dünnen Röhren mit Langevin-Gleichungen.

Kinetischen Gleichungen, Simulation und Experimente von Verklumpungserscheinungen.

Diplomarbeit
vorgelegt von
Tino Riethmüller

Humboldt-Universität zu Berlin
Mathematisch-Naturwissenschaftliche Fakultät I
Institut für Physik
Lehrstuhl Stochastische Prozesse

Aufgabensteller und Betreuer: Prof. Dr. Lutz Schimansky-Geier

Berlin, am 13. September 1995

# Inhaltsverzeichnis

# 0 Einleitung

Die vorliegende Arbeit behandelt die stochastische Beschreibung granularer Ströme in dünnen Röhren. Granulare Medien zeigen eine Vielzahl interessanter Phänomene auf und werden seit mehr als 200 Jahren wissenschaftlich untersucht. Als Beispiele für solche Phänomene seien hier das Entstehen eines bestimmten Schüttwinkels bei der Haufenbildung [28], [24], [15], die Ausbildung von Konvektionszellen bei periodischer Erregung [29], [14], die Größenseparation eines Granulatgemisches in einem rotierenden Zylinder [57], [60], das Verhalten von Granulaten in Kugelmühlen [24], [58], [59], das Entstehen von Dichtewellen beim Ausfluß aus einem Trichter [61], [62], [63] und beim Durchfluß durch ein dünnes Rohr [8], [13], [17], genannt.

Unter einem granularen Medium oder Granulat versteht man dabei einen Stoff, der aus vielen einzelnen Teilchen (Körnern) besteht. Die einzelnen Teilchen sind dabei frei beweglich wie in einem Gas und sind aber selbst als ein Stück Festkörper anzusehen, d.h. ein Teilchen besteht selbst aus sehr, sehr vielen gebundenen Atomen oder Molekülen. Beispiele für Granulate sind Sand, Kies, Pulver, Mohnkörner, Glaskugeln oder ähnliches. Wenn man sich nun überlegt, wieviel Energie $W_d$ notwendig ist, um ein Teilchen der Masse m um einen Teilchendurchmesser d entgegen der Schwerkraft anzuheben ($W_d = mgd$), so kommt man für ein Gas aus $0_2$-Molekülen auf Größenordnungen um die $10^{-13}$ meV und für einen typischen Sand auf Größenordnungen um die $10^{+13}$ meV. Ein Vergleich mit der thermischen Energie $k_BT$ bei Raumtemperatur ($k_BT = 25$meV) ergibt folgendes:

$$W_d \ll k_BT \qquad \text{für ein Gas} \qquad (1)$$

$$W_d \gg k_BT \qquad \text{für ein Granulat} \qquad (2)$$

Das heißt, daß sich die Teilchen eines Gases aufgrund der Raumtemperatur ständig bewegen, während die Teilchen eines Granulats in Ruhe sind. Als Folge davon kann sich ein Granulat in bestimmten Situationen ähnlich wie eine Flüssigkeit verhalten (z.B. beim Ausfließen aus einem Trichter) und in anderen Situationen ähnlich wie ein Festkörper (z.B. hat man relativ festen Boden unter den Füßen, wenn man einen Sandstrand entlang läuft).

Kenntnisse über das Verhalten granularer Medien sind nicht nur aus theoretischer Sicht von Interesse, sondern auch für industrielle Anwendungen. Man denke z.B. an Kugelmühlen, wo man mit möglichst wenig Energie Stoffe zerkleinern will oder an die gleichmäßige Dosierung von pulvrigen Zusatzstoffen mittels Trichter.

Wie eben dargelegt, existiert eine Vielzahl interessanter Effekte mit granularen Medien, jedoch steht die Theorie zu deren Beschreibung erst ganz am Anfang der Entwicklung. In dieser Arbeit soll nun eines der einfachsten Phänomene, das Entstehen von Verklumpungserscheinungen (Dichtewellen) in einem dünnen Rohr sowohl experimentell als auch theoretisch untersucht werden.

Die einzelnen Teilchen eines Granulats können durch inelastische Stöße miteinander wechselwirken. Dabei wird allerdings Energie dissipiert, so daß ein sich selbst überlassenes granulares System unweigerlich in den Zustand der Ruhe kehrt. Um bestimmte Effekte beobachten zu können, muß dem System ständig Energie zugeführt werden. In manchen Experimenten geschieht das durch Schütteln (Vibrationen) oder Drehen des Behälters (z.B. bei Kugelmühlen), in dem hier betrachteten Experiment pumpt die Gravitationskraft ständig Energie in das granulare System im Rohr.

# 1 Theorie

## 1.1 Ableitung einer Boltzmann-Gleichung

Es soll die Bewegung von N zunächst nichtwechselwirkenden granularen Teilchen in einem senkrecht angebrachten dünnen Rohr der Länge L beschrieben werden. Der Querschnitt A des Rohres sei an allen Stellen konstant, so daß der Teilchenfluß eindimensional angesehen werden kann. Die physikalische Beschreibung soll im Modell der Brownschen Teilchen, die sich in einem Bad der Temperatur $T_B$ befinden, erfolgen. Man betrachte die folgenden Langevin-Gleichungen:

$$\dot{x}_i = v_i \tag{3}$$

$$m\dot{v}_i = mg - \gamma v_i + \sqrt{2\varepsilon\gamma}\,\xi_i(t) \tag{4}$$

wobei:

| | | |
|---|---|---|
| $i$ | – Teilchennummer | $i = 1,2,...,N$ |
| $N$ | – Anzahl Teilchen | |
| $x_i$ | – Ort des Teilchens i | $x_i \in [0, L]$ |
| $v_i$ | – Geschwindigkeit des Teilchens i | $v_i \in (-\infty, \infty)$ |
| $m$ | – mittlere Masse eines Teilchens | |
| $g$ | – Erdbeschleunigung | |
| $\gamma$ | – mittlerer Reibungskoeffizient, aufgrund von Reibung und Stößen zwischen Teilchen und Rohrwand | |
| $\varepsilon = k_B T_B$ | – Rauschenergie | |
| $T_B$ | – Temperatur des Bades | |
| $k_B$ | – Boltzmann-Konstante | |
| $\xi_i(t)$ | – Langevin-Kraft zur Zeit t = weißes Gauß-Rauschen, d.h. $\langle\xi_i(t)\rangle = 0$ und $\langle\xi_i(t)\xi_j(t')\rangle = \delta_{ij}\delta(t - t')$ | |

Die zugehörige Fokker-Planck-Gleichung für die Wahrscheinlichkeitsdichte P(x,v,t) lautet:

$$\frac{\partial P}{\partial t} + \frac{\partial}{\partial x}[vP] + \frac{\partial}{\partial v}\left[\left(g - \frac{\gamma}{m}v\right)P\right] = \frac{\varepsilon\gamma}{m^2}\frac{\partial^2 P}{\partial v^2} \tag{5}$$

Nun sollen zusätzlich unelastische Stöße zwischen den Teilchen zugelassen werden. Dies läßt sich nach Prigogine und Herman [39] in einem Stoßintegral formulieren, wobei die Teilchen als punktförmig angenommen werden. Die Wahrscheinlichkeitsdichte P(x,v,t) wird von zwei Stoßtypen verändert. Einmal betrachtet man ein bestimmtes Teilchen am Ort x mit der Geschwindigkeit v, welches von P(x,v',t)(v'–v)dv schnelleren Teilchen der Geschwindikkeit v'>v erreicht wird. All diese Teilchen werden auf die Geschwindigkeit v abgebremst.

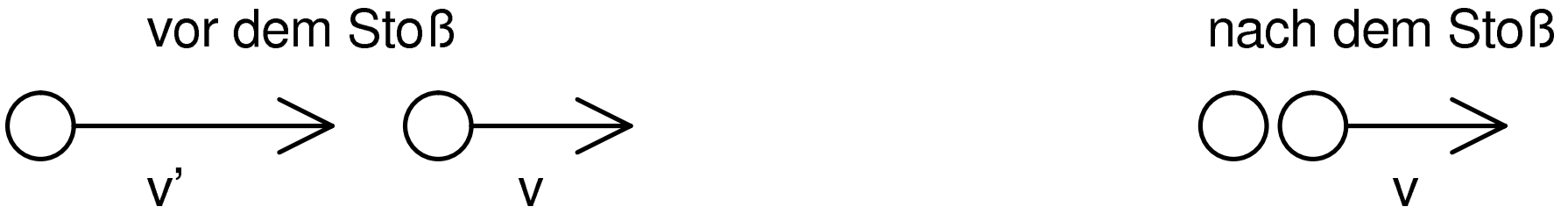

Bild 1 unelastischer Stoß, der zur Erhöhung von P(x,v,t) führt

Dadurch erhöht sich die Anzahl der Teilchen mit der Geschwindigkeit v in x (Gewinnterm):

$$\left(\frac{\partial P}{\partial t}\right)_S^{I} \sim \int_{v}^{\infty} P(x,v,t)P(x,v',t)(v' - v)dv' \tag{6}$$

Andererseits kann das Teilchen mit der Geschwindigkeit v auch mit einem langsameren Teilchen vor sich zusammenstoßen und dadurch selbst abgebremst werden.

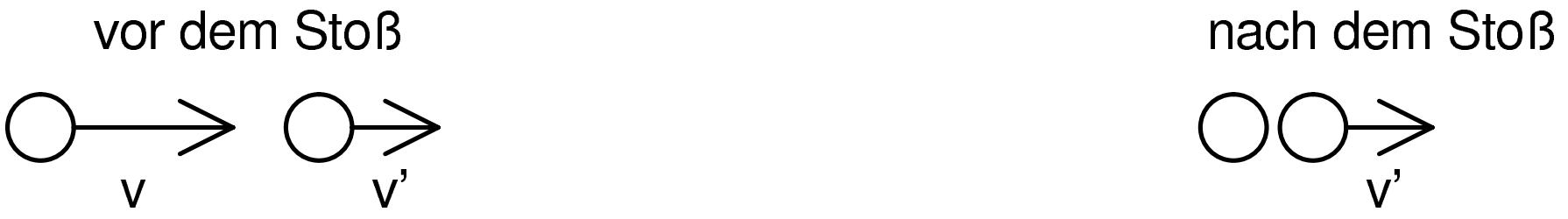

Bild 2 unelastischer Stoß, der zur Verringerung von P(x,v,t) führt

Dieser Stoßtyp bewirkt eine Verringerung der Wahrscheinlichkeitsdichte P(x,v,t) (Verlustterm):

$$\left(\frac{\partial P}{\partial t}\right)_S^{II} \sim -\int_{-\infty}^{v} P(x,v,t)P(x,v',t)(v - v')dv' \tag{7}$$

Werden beide Größen zusammengenommen, erhält man folgendes Stoßintegral:

$$\left(\frac{\partial P}{\partial t}\right)_S = C\int_{-\infty}^{\infty} P(x,v,t)P(x,v',t)(v' - v)dv' \tag{8}$$

Es sei bemerkt, daß die hier angenommenen Stöße den Impulserhaltungssatz verletzen, aber dafür den Vorteil besitzen, ein einfaches Stoßintegral zu liefern. Dieser Ansatz stammt aus der Verkehrstheorie und soll auch für granulare Teilchen angewendet werden. Man kann sich vorstellen, daß in einem realen Rohr ein gestoßenes Teilchen häufig zur Seite gestreut wird. Da sehr dünne Rohre vorausgesetzt werden, kommt es zwangsläufig nach kurzer Zeit zu einer Wechselwirkung vieler Teilchen mit der Rohrwand. Für das Gesamtsystem Teilchen + Rohrwand gilt selbstverständlich der Impulserhaltungssatz. Ohne Einbeziehung der Rohrwand in die eindimensionale Beschreibung soll das oben angeführte Stoßintegral gelten.

C ist eine Art effektiver Wirkungsquerschnitt für die Stöße zwischen den Teilchen und hängt von der Geometrie des Rohres und der Teilchen ab. Es werden nun die Teilchenzahldichte n, die mittlere Geschwindigkeit u und die Temperatur T der Brownschen Teilchen am Ort x zur Zeit t eingeführt:

$$n(x,t) = \int_{-\infty}^{\infty} P(x,v,t)dv \tag{9}$$

$$u(x,t) = \frac{1}{n(x,t)} \int_{-\infty}^{\infty} vP(x,v,t)dv \tag{10}$$

$$T(x,t) = \frac{m}{k_B n(x,t)} \int_{-\infty}^{\infty} (v - u(x,t))^2 P(x,v,t)dv \tag{11}$$

Es sei an dieser Stelle ausdrücklich darauf hingewiesen, daß T(x,t) nicht die Temperatur meint, die durch Hineinhalten eines Thermometers in das granulare Medium gemessen werden kann und daß $T_B$ nicht die mit einem Thermometer meßbare Temperatur der Rohrwand ist. Vielmehr meint T(x,t) im wesentlichen die mittlere quadratische Geschwindigkeitsabweichung der Brownschen Teilchen und nicht die mittlere quadratische Geschwindigkeitsabweichung der Atome bzw. Moleküle, aus denen ein einzelnes Brownsches Teilchen besteht (das mißt man nämlich mit einem Thermometer). Ebenso ist $T_B$ lediglich als Quotient aus Rauschenergie $\varepsilon$ und Boltzmann-Konstante $k_B$ aufzufassen.

Mit diesen Festlegungen läßt sich (8) wie folgt schreiben:

$$\left(\frac{\partial P}{\partial t}\right)_S = CP(x,v,t)n(x,t)(u(x,t) - v) \tag{12}$$

Durch Addition von (5) und (12) erhält man die folgende Boltzmann-Gleichung:

$$\frac{\partial P}{\partial t} + \frac{\partial}{\partial x}[vP] + \frac{\partial}{\partial v}\left[\left(g - \frac{\gamma}{m}v\right)P\right] = \frac{\varepsilon\gamma}{m^2}\frac{\partial^2 P}{\partial v^2} + CP(x,v,t)n(x,t)(u(x,t) - v) \tag{13}$$

Diese Gleichung beschreibt die Bewegung granularer Teilchen einschließlich unelastischer Stöße und Reibung. Eine homogene Anfangsbedingung für die Teilchenzahldichte (n(x,0) = $n_0$) führt nach einer Relaxationszeit der Größenordnung $\tau = m/\gamma$ auf die folgende stationäre homogene Wahrscheinlichkeitsdichte P:

$$P_0(v) = \sqrt{\frac{m}{2\pi k_B T_0}}\, n_0 \exp\left[-\frac{m}{2k_B T_0}(v - u_0)^2\right] \tag{14}$$

wobei:

$$u_0 = \frac{mg}{\gamma} - \frac{Ck_BT_0n_0}{\gamma} \qquad (15)$$

die stationäre homogene mittlere Geschwindigkeit und

$$T_0 = T_B = \frac{\varepsilon}{k_B} \qquad (16)$$

die Temperatur der Brownschen Teilchen ist. Im homogenen Falle herrscht thermodynamisches Gleichgewicht, d.h. Bad und Teilchen besitzen dieselbe Temperatur.

(14) ist eine Gaußglocke mit dem Zentrum bei $u_0$ und der Varianz $k_BT_0/m$. Definiert man die Geschwindigkeitsverteilung als:

$$w(v,t) = \int_0^L P(x,v,t)dx \qquad (17)$$

wobei auf die Teilchenzahl N normiert wurde:

$$\int_{-\infty}^{\infty} w(v,t)dv = N \qquad (18)$$

so ergibt sich aus (14) folgende stationäre Geschwindigkeitsverteilung:

$$w_0(v) = \sqrt{\frac{m}{2\pi\varepsilon}} N \exp\left[-\frac{m}{2\varepsilon}(v-u_0)^2\right] \qquad (19)$$

Die stationäre Wahrscheinlichkeitsdichte (14) ist die Lösung der folgenden Langevin-Gleichungen:

$$\dot{x}_i = v_i \qquad (20)$$

$$m\dot{v}_i = mg - \gamma v_i + \sqrt{2\varepsilon\gamma}\,\xi_i(t) - Ck_BT_0n_0 \qquad (21)$$

Der letzte Term stellt eine effektive mittlere Kraft dar, die ihre Ursache in den Wechselwirkungen der Teilchen hat.

## 1.2 Hydrodynamische Beschreibung

Um Aussagen über die Stabilität des homogenen Zustands machen zu können, soll nun zu einer hydrodynamischen Approximation der Boltzmann-Gleichung (13) übergegangen werden. Diese Methode ist z.B. in [55] und ansatzweise auch schon in [56] beschrieben.

Man integriere die Boltzmann-Gleichung (13) über alle v und erhält die Kontinuitätsgleichung:

$$\frac{\partial n}{\partial t} + \frac{\partial}{\partial x}[nu] = 0 \tag{22}$$

Anschließend multipliziert man (13) erst mit v und integriert dann. Mit der Abkürzung:

$$F(n, T) = mg - Ck_B T(x, t) n(x, t) \tag{23}$$

ergibt sich:

$$\frac{\partial u}{\partial t} + u\frac{\partial u}{\partial x} = \frac{F(n, T)}{m} - \frac{\gamma}{m}u - \frac{k_B}{m}\frac{1}{n}\frac{\partial}{\partial x}[nT] \tag{24}$$

Als letztes wird (13) erst mit $v^2$ multipliziert und dann integriert. Unter der Annahme, daß für das granulare System lokales Gleichgewicht gilt, d.h.

$$\left\langle (v - u(x, t))^n \right\rangle = 0 \qquad \forall\ n \geq 3 \tag{25}$$

ergibt sich mit einigen Umformungen:

$$\frac{\partial T}{\partial t} + u\frac{\partial T}{\partial x} = -\frac{2\gamma}{m}T + \frac{2\varepsilon\gamma}{mk_B} - 2T\frac{\partial u}{\partial x} \tag{26}$$

Auf diese Weise erhält man drei partielle Differentialgleichungen: (22) ist die Teilchenzahlbilanz, (24) ist die Impulsbilanz und (26) ist die Temperaturbilanz.

Zum besseren Verständnis sei hier auf den letzten Term in (26) eingegangen. Ausgangspunkt ist die Kontinuitätsgleichung:

$$\frac{\partial n}{\partial t} + \frac{\partial}{\partial x}[nu] = 0 \tag{27}$$

$$\frac{\partial n}{\partial t} + u\frac{\partial n}{\partial x} + n\frac{\partial u}{\partial x} = 0 \tag{28}$$

$$\frac{dn}{dt} + n\frac{\partial u}{\partial x} = 0 \tag{29}$$

Die Dichte wird nun wie folgt umgeschrieben:

$$n = \frac{1}{v} \tag{30}$$

wobei v ein spezifisches Volumen ist. Dann ergibt sich:

$$-\frac{1}{v^2}\frac{dv}{dt} + \frac{1}{v}\frac{\partial u}{\partial x} = 0 \tag{31}$$

$$\frac{1}{v}\frac{dv}{dt} = \frac{\partial u}{\partial x} \tag{32}$$

Umschreibung auf die ursprüngliche Dichte liefert:

$$n\frac{dv}{dt} = \frac{\partial u}{\partial x} \tag{33}$$

Das ideale Gasgesetz lautet:

$$n = \frac{p}{k_B T} \tag{34}$$

was in (26) eingesetzt folgendes ergibt:

$$\frac{p}{k_B}\frac{dv}{dt} = T\frac{\partial u}{\partial x} \tag{35}$$

$T\frac{\partial u}{\partial x}$ ist also im wesentlichen eine Volumenarbeit pdv.

## 1.3 Aufstellen von Langevin-Gleichungen

Es sollen nun Langevin-Gleichungen aufgestellt werden, die auch zur Beschreibung inhomogener Flüsse geeignet sind. Man betrachte den folgenden Ansatz:

$$\dot{x}_i = v_i \tag{36}$$

$$m\dot{v}_i = mg - \gamma v_i + \sqrt{2\varepsilon\gamma}\,\xi_i(t) - Ck_B T(x_i, t)n(x_i, t) = -\gamma v_i + \sqrt{2\varepsilon\gamma}\,\xi_i(t) + F(n, T) \tag{37}$$

Behandelt man die Dichte n und die Temperatur T als Parameter, kann man die folgende Quasi-Fokker-Planck-Gleichung aufstellen:

$$\frac{\partial P}{\partial t} + \frac{\partial}{\partial x}[vP] + \frac{\partial}{\partial v}\left[\left(\frac{F(n, T)}{m} - \frac{\gamma}{m}v\right)P\right] = \frac{\varepsilon\gamma}{m^2}\frac{\partial^2 P}{\partial v^2} \tag{38}$$

Zwei Gründe motivieren den Ansatz (36), (37):

- Ansatz (36), (37) besitzt dieselbe homogene Lösung (14), wie die Boltzmann-Gleichung (13).
- Man hätte dieselben drei hydrodynamischen Gleichungen (22), (24), (26) erhalten, wenn man die Rechnung nicht mit der Boltzmann-Gleichung (13), sondern mit (38) begonnen hätte.

## 1.4 Der Einstein-Smoluchowski-Grenzübergang

Es soll nun das Verhalten granularer Medien bei sehr großen Reibungskoeffizienten von Interesse sein, d.h. man betrachte den Grenzfall $\gamma \to \infty$. In diesem Falle wird der Term $m\dot{v}$ klein gegenüber den anderen Termen, so daß man ihn Null setzen kann. Aus (37) folgt dann:

$$v_i = \frac{mg}{\gamma} + \sqrt{\frac{2\varepsilon}{\gamma}}\,\xi_i(t) - \frac{Ck_B T(x_i, t)n(x_i, t)}{\gamma} \tag{39}$$

Durch Einsetzen von (39) in (36) erhält man die folgende neue Langevin-Gleichung:

$$\dot{x}_i = \frac{mg}{\gamma} + \sqrt{\frac{2\varepsilon}{\gamma}}\xi_i(t) - \frac{Ck_BT(x_i,t)n(x_i,t)}{\gamma} \tag{40}$$

Das Problem hat sich durch den Grenzübergang vereinfacht, da man nun nur noch eine, von der Teilchengeschwindigkeit unabhängige Gleichung hat. Es wird nun die zu (40) gehörige Quasi-Fokker-Planck-Gleichung aufgestellt und man weiß dabei, daß P(x,t) = n(x,t) ist.

$$\frac{\partial n}{\partial t} + \frac{1}{\gamma}\frac{\partial}{\partial x}\Big[\big(mg - Ck_BT(x,t)n\big)n\Big] = \frac{\varepsilon}{\gamma}\frac{\partial^2 n}{\partial x^2} \tag{41}$$

Für den Fall des thermodynamischen Gleichgewichtes, d.h. $\varepsilon = k_BT(x,t)$, ist das die Burgers-Gleichung, die eine Kontinuitätsgleichung für die spezielle Teilchenstromdichte

$$j(x,t) = \frac{1}{\gamma}\left[(mg - C\varepsilon n)n - \varepsilon\frac{\partial n}{\partial x}\right] \tag{42}$$

ist.

Durch die Transformation $x' = x - \frac{mg}{\gamma}t$ gelangt man zu folgender Gleichung:

$$\frac{\partial n}{\partial t} - \frac{2C\varepsilon}{\gamma}n\frac{\partial n}{\partial x'} = \frac{\varepsilon}{\gamma}\frac{\partial^2 n}{\partial x'^2} \tag{43}$$

Man dividiert durch $2C\varepsilon/\gamma$ und transformiert erneut $t' = \frac{2C\varepsilon}{\gamma}t$:

$$\frac{\partial n}{\partial t'} = n\frac{\partial n}{\partial x'} + \nu\frac{\partial^2 n}{\partial x'^2} \tag{44}$$

Das ist die Burgers-Gleichung in ihrer Normalform, wobei

$$\nu = \frac{1}{2C} \tag{45}$$

die Viskosität ist. Die Burgers-Gleichung ist analytisch lösbar [20], [21], [22], [33] und war schon oft Gegenstand wissenschaftlicher Untersuchungen. Eine ausführliche Darstellung der statistischen Mechanik der Burgers-Gleichung findet man in [27], diverse andere Fragestellungen werden in [1], [7], [25], [26], [31], [32] behandelt.

## 1.5 Zusammenhang zwischen Verkehrsflüssen und granularen Flüssen

Es sollen nun einige Gemeinsamkeiten von granularen Teilchen, z.B. Sand in einem dünnen Rohr und den Teilchen eines Verkehrsflusses, also den Autos auf einer Autobahn aufgezählt werden. In beiden Fällen handelt es sich um wechselwirkende Teilchen. Die Wechselwir-kung besteht bei Sandkörnern in der Geschwindigkeitsänderung durch Stöße oder Reibung untereinander. Ähnlich ist es auch in einem Verkehrsfluß. Trifft ein Auto mit großer Geschwindigkeit auf eines mit kleinerer, so muß besonders bei hoher Verkehrsdichte abgebremst werden, weil ein Überholvorgang nicht immer möglich ist. Weiterhin handelt es sich in beiden Fällen um gepumpte Systeme. Bei einem Sandfluß stellt die Gravitation die Triebkraft dar, bei einem Verkehrsfluß ist es der Wille der Autofahrer, so schnell wie möglich vorwärtszukommen. In beiden Systemen ruft die Wechselwirkung der Teilchen ähnliche Phänomene hervor. Die Sandkörner sind in der Lage Klumpen zu bilden, was auf einer Autobahn der Entstehung eines Staus ohne ersichtlichen Grund entspricht. "Ohne ersichtlichen Grund" meint dabei, daß die Ursache der Staubildung weder ein Unfall, noch ein liegengebliebener LKW war, sondern lediglich ein hohes Verkehrsaufkommen.

Die spontane Entstehung von Staus auf einer Autobahn in einem anfangs homogenen Verkehrsfluß wurde sehr ausführlich von Kerner und Konhäuser numerisch untersucht, siehe [34] und [35]. In diesen Arbeiten wird eine hydrodynamische Beschreibung der Dichte und der mittleren Geschwindigkeit benutzt, wobei die Temperatur der Brownschen Teilchen (Autos) als konstant angenommen wird. Der als eindimensional und kompressibel aufgefaßte Verkehrsfluß wird durch zwei Gleichungen beschrieben, zum einen durch die Kontinuitätsgleichung und zum anderen durch eine Bewegungsgleichung, die durch die folgende Navier-Stokes-Gleichung gegeben ist:

$$\frac{\partial u}{\partial t} + u\frac{\partial u}{\partial x} = \frac{V(n)}{\tau} - \frac{u}{\tau} - c_0^2\frac{1}{n}\frac{\partial n}{\partial x} + \frac{\mu}{n}\frac{\partial^2 u}{\partial x^2} \tag{46}$$

wobei:

| | |
|---|---|
| $V(n)$ | – maximale, gerade noch sichere Geschwindigkeit in einem homogenen Verkehrsfluß der Dichte n |
| $\tau$ | – mittlere Relaxationszeit bei der Anpassung der Geschwindigkeiten an die aktuellen Verkehrsverhältnisse |
| $c_0^2$ | – Varianz der Geschwindigkeitsverteilung |
| $\mu$ | – dynamische Viskosität |

Es soll nun gezeigt werden, daß sich der Ansatz (46) aus den Gleichungen (22), (24) und (26) zur Beschreibung granularer Medien ableiten läßt. In (46) wird die Varianz der Geschwindigkeitsverteilung der Teilchen als konstant angesetzt. Diesem Sachverhalt soll Rechnung getragen werden, indem die Kraft F als temperaturunabängig angenommen wird:

$$F(n, T) \equiv F(n, T_B) = mg - C\varepsilon n(x, t) \tag{47}$$

Nun muß die Temperatur aus Gleichung (26) eliminiert und in Gleichung (24) eingesetzt werden.

**1. Fall:**

Man dividiere (26) durch $\gamma$ und erhält:

$$\frac{1}{\gamma}\frac{\partial T}{\partial t} + \frac{1}{\gamma}u\frac{\partial T}{\partial x} = -\frac{2}{m}T + \frac{2\varepsilon}{mk_B} - \frac{1}{\gamma}2T\frac{\partial u}{\partial x} \tag{48}$$

Im Grenzfall sehr großer $\gamma$, d.h.:

$$\gamma \gg \frac{\partial T}{\partial t},\ u\frac{\partial T}{\partial x},\ T\frac{\partial u}{\partial x} \tag{49}$$

ergibt sich:

$$\varepsilon = k_B T\ ,\ \text{d.h. } T(x,t) \equiv T_B \tag{50}$$

(50) heißt Einstein-Relation und liefert in (24) eingesetzt:

$$\frac{\partial u}{\partial t} + u\frac{\partial u}{\partial x} = \frac{1}{m}\left[F(n,T_B) - \gamma u - \frac{\varepsilon}{n}\frac{\partial n}{\partial x}\right] \tag{51}$$

Ein Vergleich mit (46) zeigt, daß der Viskositätsterm fehlt. Die Näherung für $\gamma$ war also zu stark.

**2. Fall:**

Es sei angenommen, daß:

$$\gamma \gg \frac{\partial T}{\partial t},\ u\frac{\partial T}{\partial x} \quad \text{, d.h} \quad \gamma \gg \frac{dT}{dt} \tag{52}$$

gilt, aber es muß <u>nicht</u> notwendigerweise gelten:

$$\gamma \gg T\frac{\partial u}{\partial x} \tag{53}$$

Es folgt dann aus (26)

$$k_B T = \frac{\varepsilon}{1 + \frac{m}{\gamma}\frac{\partial u}{\partial x}} \approx \varepsilon\left(1 - \frac{m}{\gamma}\frac{\partial u}{\partial x}\right) \tag{54}$$

(54) in (24) eingesetzt, liefert:

$$\frac{\partial u}{\partial t} + u\frac{\partial u}{\partial x} = \frac{1}{m}\left[F(n,T_B) - \gamma u - \varepsilon\left(1 - \frac{m}{\gamma}\frac{\partial u}{\partial x}\right)\frac{1}{n}\frac{\partial n}{\partial x} + \frac{\varepsilon m}{\gamma}\frac{\partial^2 u}{\partial x^2}\right] \tag{55}$$

Diese Gleichung hat nun die gesuchte Struktur. Nun sollen durch direkten Vergleich mit (46) die verschiedenen Parameter identifiziert werden:

| Riethmüller | Kerner, Konhäuser | |
|---|---|---|
| $\frac{m}{\gamma}$ | $\tau$ | (56) |
| $\frac{F(n, T_B)}{m}$ | $\frac{V(n)}{\tau}$ | (57) |
| $\frac{\varepsilon}{m}\left(1 - \frac{m}{\gamma}\frac{\partial u}{\partial x}\right) \approx \frac{\varepsilon}{m}$ | $c_0^2$ | (58) |
| $\frac{\varepsilon}{\gamma}$ | $\frac{\mu}{n}$ | (59) |

Tabelle 1 Zuordnung der Parameter des granularen Flusses zu denen des Verkehrsflusses in [34]

Die Beziehungen (56) und (58) sind nicht weiter verwunderlich. Es fällt auf, daß Kerner und Konhäuser einen Parameter ($\mu$) mehr verwenden. Würde man die Kraft F in der Langevin-Gleichung (37) noch um Scherkräfte erweitern, bekäme man diesen zusätzlichen Parameter ebenfalls. Die Einführung von Scherkräften erfordert jedoch mehrere Dimensionen der Ortsvariablen x. Da sich aber auf eine eindimensionale Beschreibung beschränkt werden soll und für die Entstehung von Instabilitäten kein zusätzlicher Parameter benötigt wird, soll darauf verzichtet werden.

Interessant ist der Vergleich (57). Die maximale, gerade noch sichere Geschwindigkeit der Autos V(n) ist das Analogon zur dichteabhängigen Kraft $F(n,T_B)$, bestehend aus Gravitationskraft mg und Effektivkraft der Teilchenwechselwirkung $C\varepsilon n$. In den bisherigen Annahmen war der Querschnitt C immer dichtunabhängig. Rechnet man dies in V(n) um, ergibt sich:

$$V(n) = \frac{1}{\gamma}(mg - C\varepsilon n) \qquad (60)$$

was einem linearen Abfall von V mit der Dichte n entspricht:

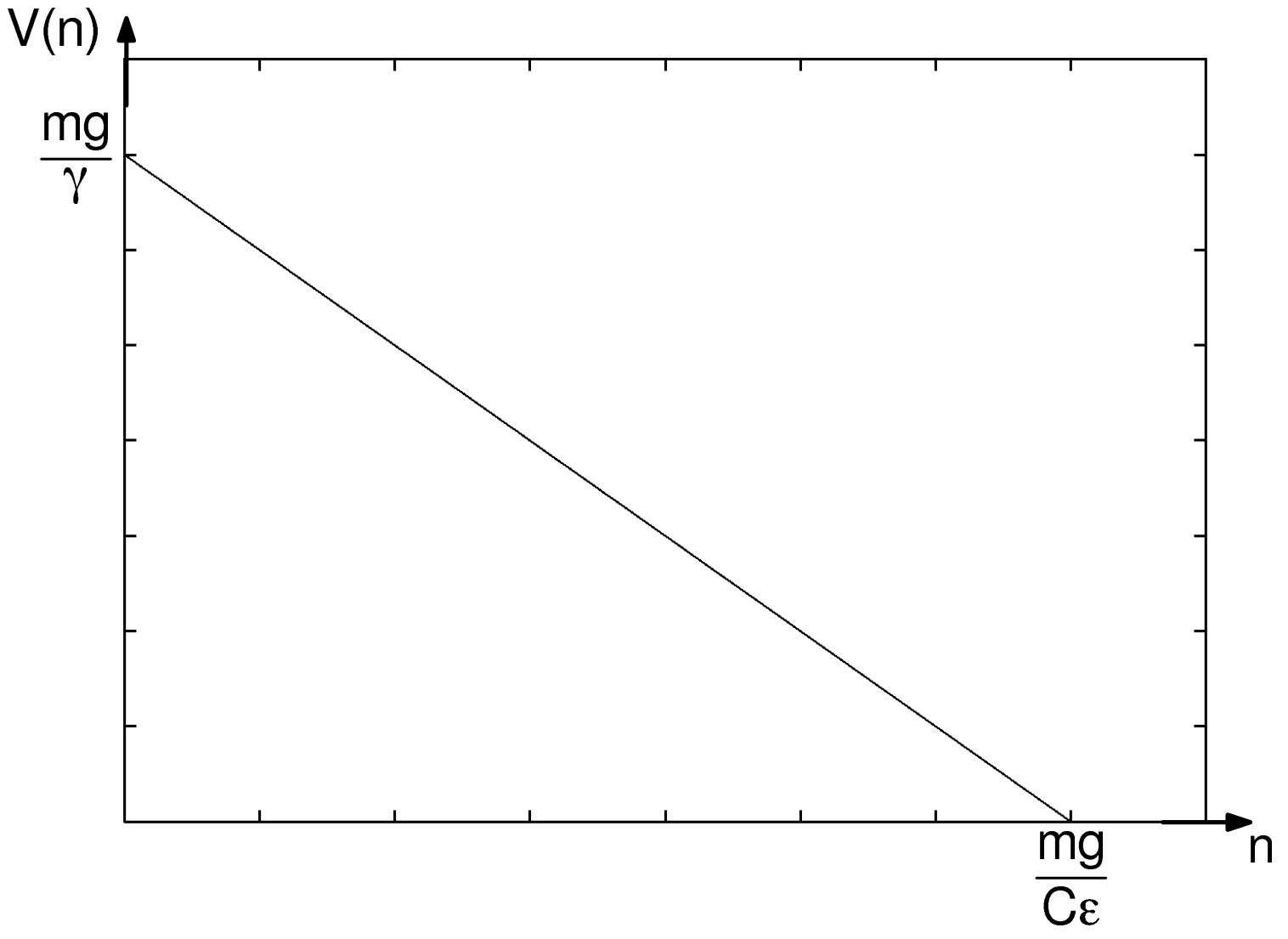

Bild 3 V(n) aus F(n) berechnet, für konstantes C

Das tatsächliche V(n) eines Verkehrsflusses muß experimentell bei homogenen Verhältnissen ermitelt werden. Homogene Verkehrsflüsse beobachtet man i.a. bei sehr geringen Dichten; Meßwerte für größere Dichten kann man z.B. in Tunneln erhalten. Einige Beispiele aus der Literatur seien im folgenden aufgeführt:

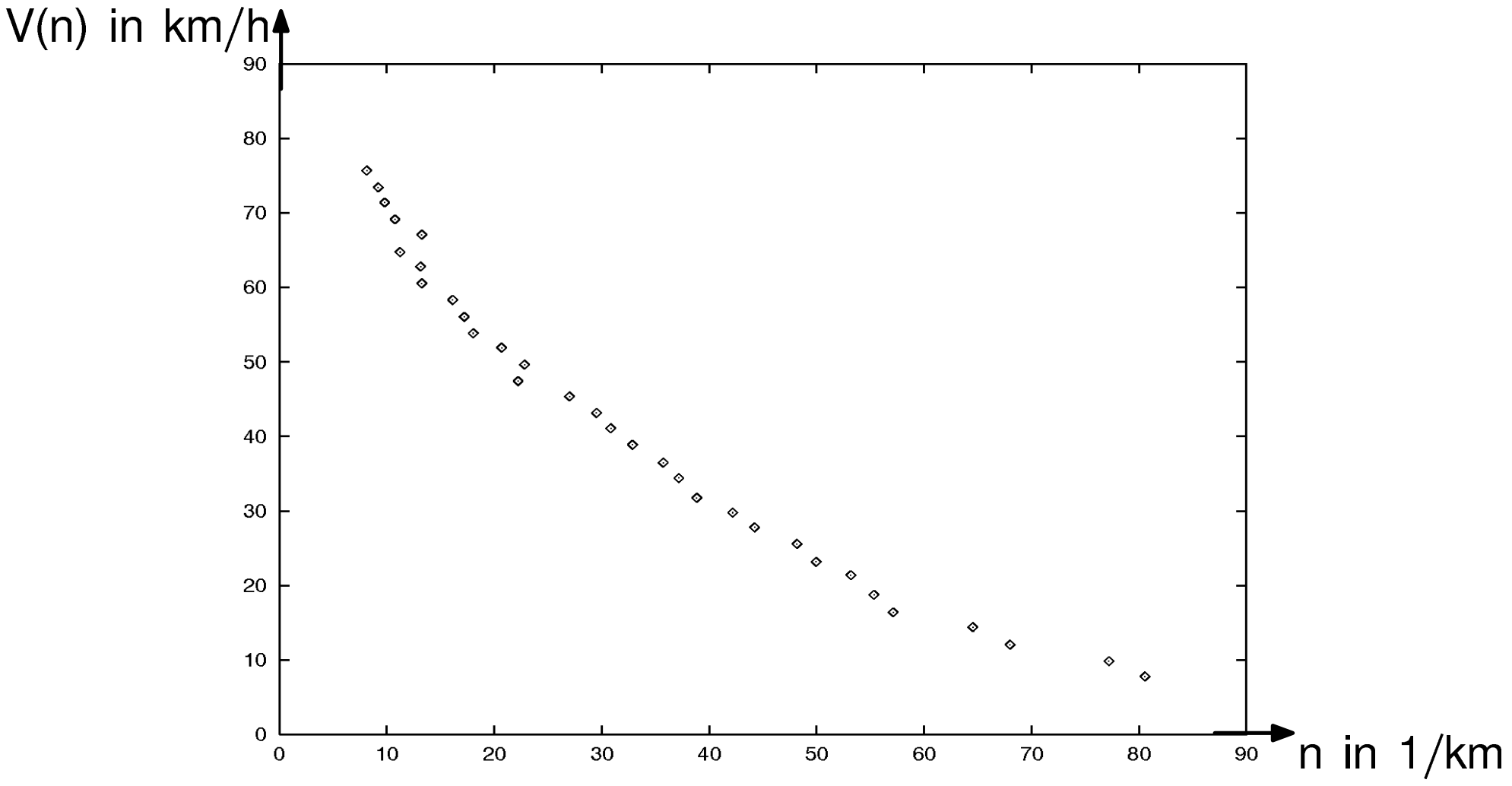

Bild 4 Experimentell ermittelte V(n)-Werte für einen Verkehrsfluß im Holland-Tunnel, aus [39]

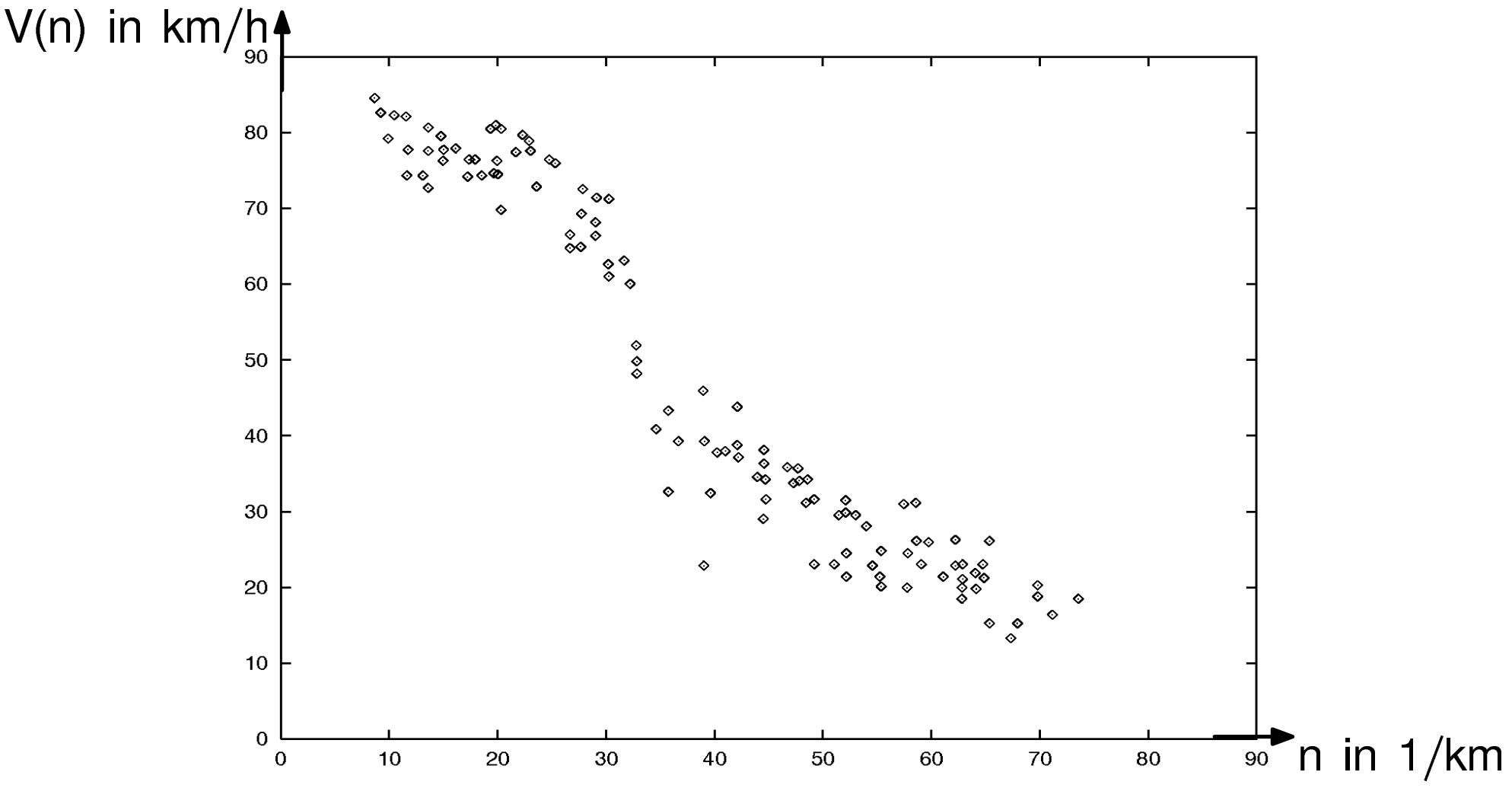

Bild 5 Experimentell ermittelte V(n)-Werte für einen Verkehrsfluß auf einem Freeway, aus [50]

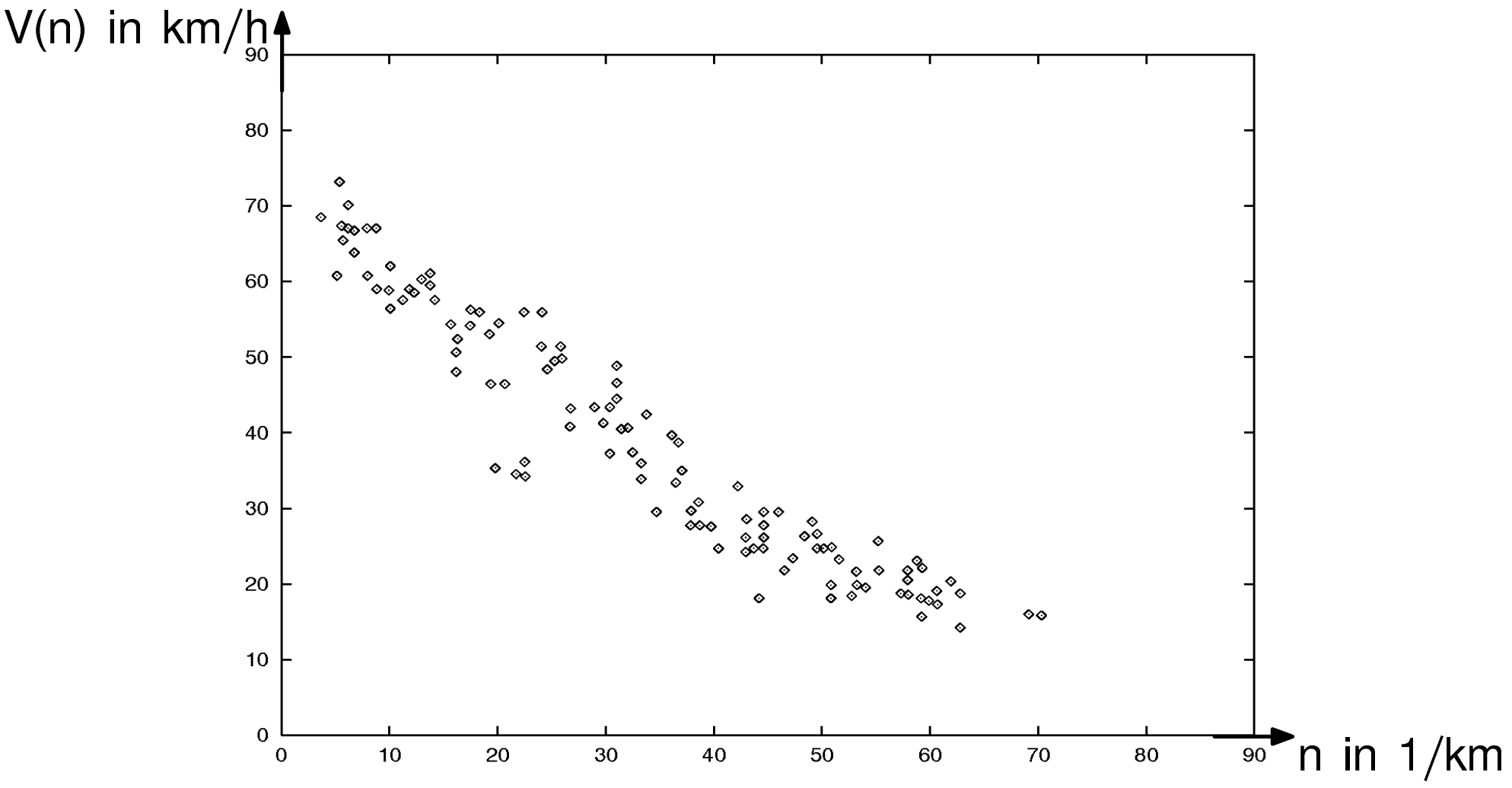

Bild 6 Experimentell ermittelte V(n)-Werte für einen Verkehrsfluß in einem Tunnel, aus [50]

Für die Simulationsrechnungen in [35] wurde für V(n) folgende Kurve verwendet:

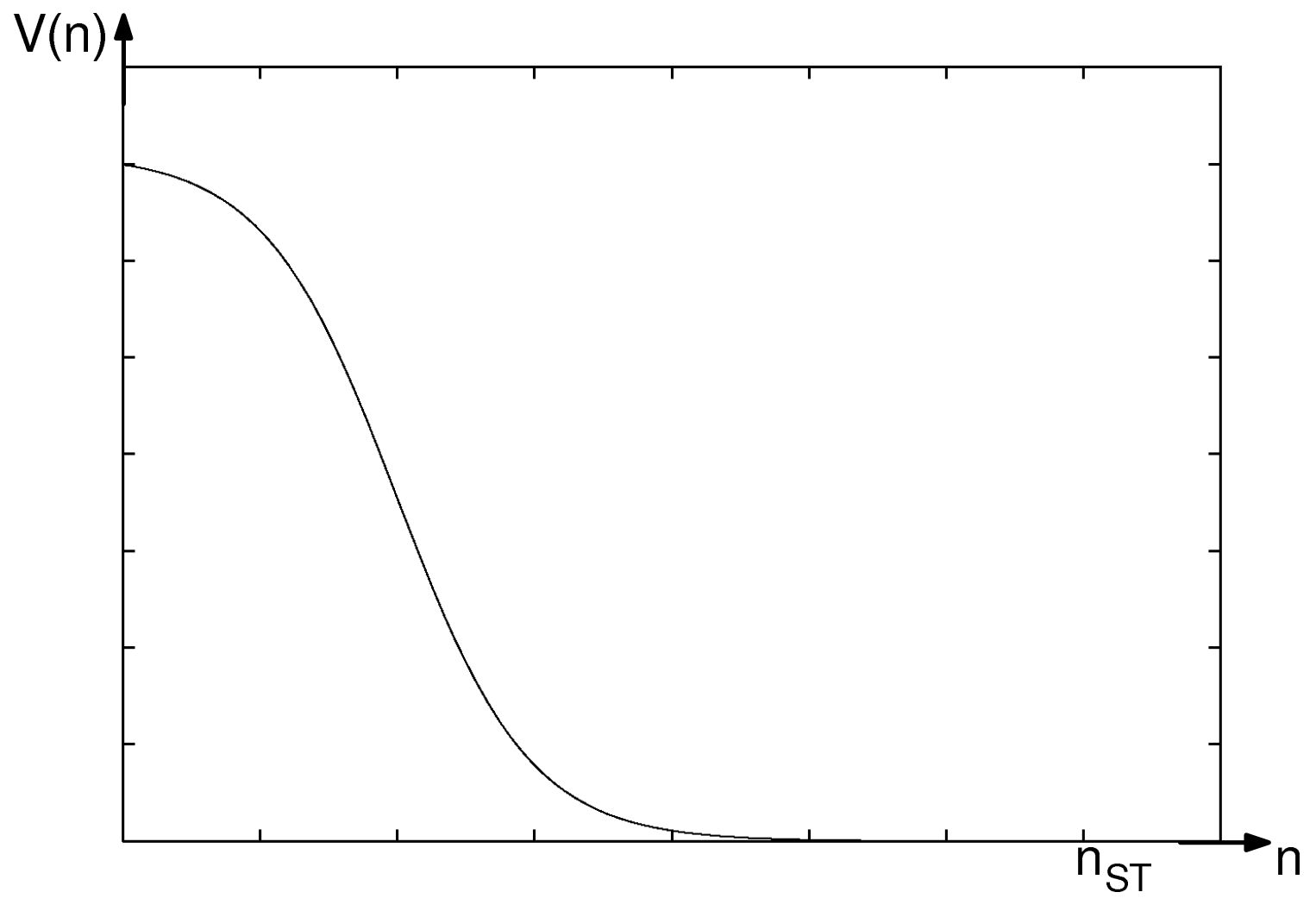

Bild 7 V(n) aus [35]

$n_{ST}$ ist die maximal mögliche Dichte der Autos, d.h. die Dichte, die dem Zustand Stoßstange an Stoßstange entspricht. Man kann diese Form der V(n)-Kurve aus der dichteabhängigen Kraft $F(n,T_B)$ erhalten, indem man den Querschnitt C dichteabhängig macht, d.h. C = C(n). Das würde folgendem Verlauf von C(n) entsprechen:

$$C(n) = \frac{1}{\varepsilon n}(mg - \gamma V(n)) \qquad (61)$$

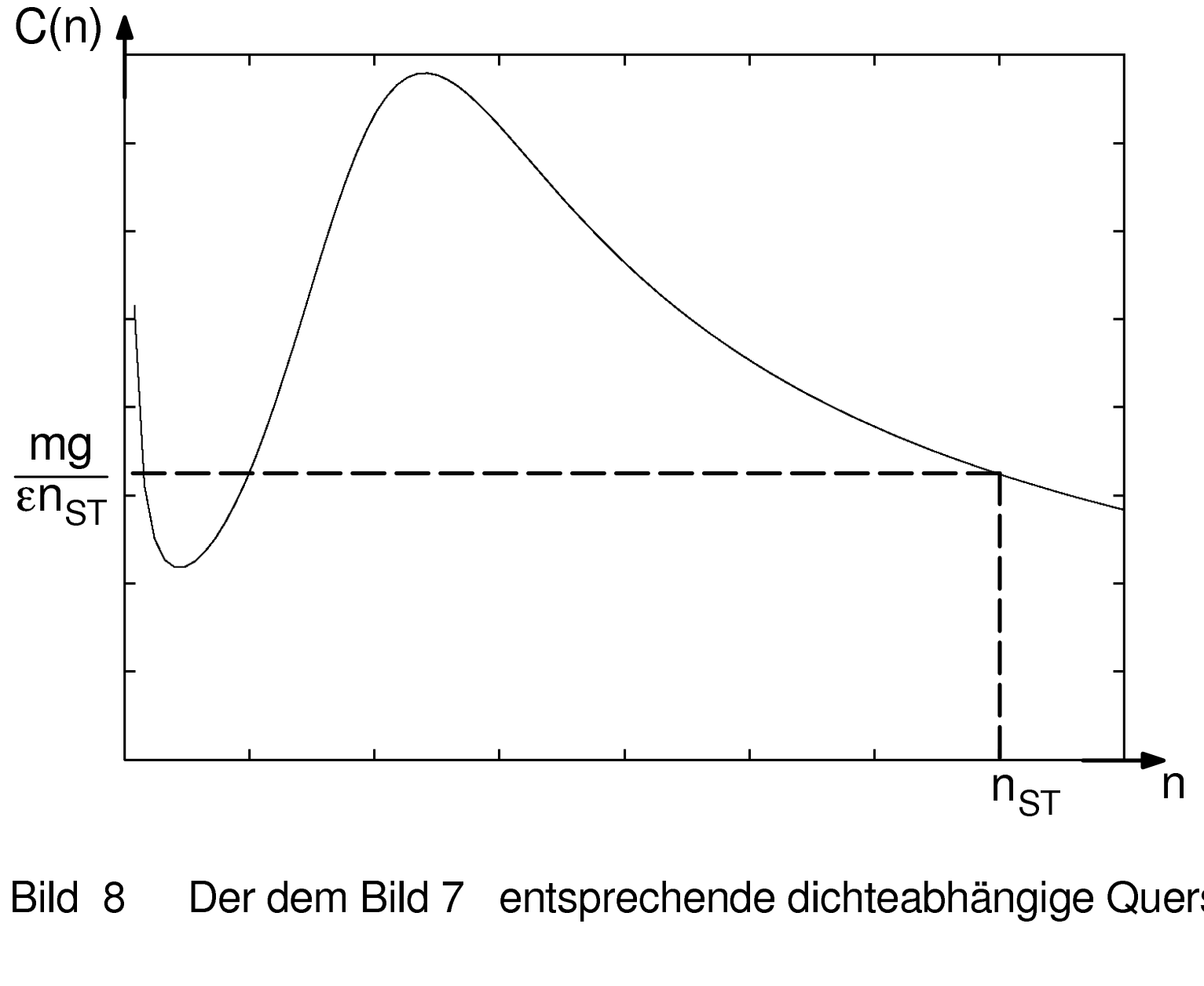

Bild 8 Der dem Bild 7 entsprechende dichteabhängige Querschnitt C(n)

In dieser Arbeit soll aber der Wirkungsquerschnitt C als Konstante behandelt werden. Im Falle granularer Flüsse ist der tatsächliche Verlauf von V(n) nicht bekannt und im Falle der Verkehrsflüsse soll der lineare Zusammenhang aus Bild 3 als erste Näherung der experimentell ermittelten Zusammenhänge aus Bild 4 - 6 aufgefaßt werden.

## 1.6 Lineare Stabilitätsuntersuchungen

### 1.6.1 Stabilitätsanalyse der Burgers-Gleichung

Nun soll untersucht werden, ob sich die homogene Lösung der Burgers-Gleichung stabil verhält. Dazu setze man:

$$n(x,t) = n_h + \delta n(x,t) \tag{62}$$

d.h. der homogene Zustand $n_h$ wird durch kleine Fluktuationen $\delta n$ gestört. Man setze (62) in die Burgers-Gleichung ein, berücksichtige nur lineare Terme in $\delta n$ und erhält:

$$\frac{\partial}{\partial t}(\delta n) + \frac{1}{\gamma}\frac{\partial}{\partial x}\left[mg\frac{\partial}{\partial x}(\delta n) - 2C\varepsilon n_h\frac{\partial}{\partial x}(\delta n)\right] = \frac{\varepsilon}{\gamma}\frac{\partial^2}{\partial x^2}(\delta n) \tag{63}$$

Die Störung sei als Welle angesetzt:

$$\delta n(x,t) = \delta n_0 \exp[-\alpha t + ikx] \tag{64}$$

und erfülle die als periodisch angesetzten Randbedingungen:

$$\delta n(0,t) = \delta n(L,t) \tag{65}$$

Die Randbedingungen sollen periodisch sein, weil das den Vergleich mit Simulationsrechnungen erleichtert.

Aus (65) folgt, daß k nur folgende Werte annehmen darf:

$$k = \frac{2\pi z}{L} \qquad \text{wobei } z = \pm 1, \pm 2, \ldots \tag{66}$$

(64) in (63) ergibt:

$$\alpha\gamma + 2ikC\varepsilon n_h - ikmg - \varepsilon k^2 = 0 \tag{67}$$

Einsetzen von $\alpha = \lambda + i\nu \quad (\lambda, \nu \in \mathbb{R})$ und Zerlegung in Real- und Imaginärteil liefert:

$$\lambda\gamma - \varepsilon k^2 = 0 \tag{68}$$

$$\nu\gamma + 2kC\varepsilon n_h - kmg = 0 \tag{69}$$

Aus (68) erhält man:

$$\lambda = \frac{\varepsilon k^2}{\gamma} \tag{70}$$

$\lambda>0$ würde bedeuten, daß die Fluktuationen mit der Zeit exponentiell abklingen, $\lambda<0$ bedeutet Anwachsen der Störungen. $\lambda$ kann nach (70) niemals negativ werden, d.h. die homogene Lösung der Burgers-Gleichung ist in jedem Falle stabil.

### 1.6.2 Stabilitätsanalyse der zu Kerner/Konhäuser analogen Gleichungen

Im vorigen Abschnitt wurde gezeigt, daß die homogene Lösung der Burgers-Gleichung in keinem Falle instabil werden kann. Eine Beschreibung von Verklumpungserscheinungen ist mit ihr also nicht möglich. Es soll nun eine Stabilitätsanalyse für die beiden hydrodynamischen Gleichungen (22) und (55) durchgeführt werden, wobei in (55) $\gamma \gg m\frac{\partial u}{\partial x}$ angenommen werden soll:

$$\frac{\partial n}{\partial t} + \frac{\partial}{\partial x}[nu] = 0 \tag{71}$$

$$\frac{\partial u}{\partial t} + u\frac{\partial u}{\partial x} = \frac{1}{m}\left[F(n, T_B) - \gamma u - \frac{\varepsilon}{n}\frac{\partial n}{\partial x} + \frac{\varepsilon m}{\gamma}\frac{\partial^2 u}{\partial x^2}\right] \tag{72}$$

Wieder stört man den homogenen Zustand von Dichte und mittlerer Geschwindigkeit durch kleine Fluktuationen $\delta$.

$$n(x,t) = n_h + \delta n(x,t) \tag{73}$$

$$u(x,t) = u_h + \delta u(x,t) \tag{74}$$

Quadratische Terme in den Fluktuationen werden wieder vernachlässigt. (73) und (74) in (71) eingesetzt, liefert:

$$\frac{\partial}{\partial t}(\delta n) + n_h\frac{\partial}{\partial x}(\delta u) + u_h\frac{\partial}{\partial x}(\delta n) = 0 \tag{75}$$

Nun werden (73) und (74) in (72) eingesetzt. Man beachte dabei, daß in erster Näherung

$$F(n_h + \delta n, T_B) = F(n_h, T_B) + \zeta(n_h)\delta n \tag{76}$$

gilt, wobei

$$\zeta(n_h) = \frac{\partial F(n, T_B)}{\partial n}\Big|_{n=n_h} \tag{77}$$

gesetzt wurde. Es ergibt sich also:

$$\frac{\partial}{\partial t}(\delta u) + u_h\frac{\partial}{\partial x}(\delta u) = \frac{1}{m}\left[F(n_h, T_B) - \gamma u_h + \zeta(n_h)\delta n - \gamma\delta u - \frac{\varepsilon}{n_h}\frac{\partial}{\partial x}(\delta n) + \frac{\varepsilon m}{\gamma}\frac{\partial^2}{\partial x^2}(\delta u)\right] \tag{78}$$

Für den homogenen Zustand gilt nun aber gerade:

$$F(n_h, T_B) - \gamma u_h = 0 \tag{79}$$

Die Störungen seien wieder als Wellen angesetzt:

$$\delta n(x,t) = \delta n_0 \exp[-\alpha t + ikx] \tag{80}$$

$$\delta u(x,t) = \delta u_0 \exp[-\alpha t + ikx] \tag{81}$$

und sollen die als periodisch angesetzten Randbedingungen erfüllen:

$$\delta n(0,t) = \delta n(L,t) \tag{82}$$

$$\delta u(0,t) = \delta u(L,t) \tag{83}$$

Aus (82) und (83) folgt, daß k nur folgende Werte annehmen darf:

$$k = \frac{2\pi z}{L} \qquad \text{wobei } z = \pm 1, \pm 2, \ldots \tag{84}$$

Durch Einsetzen von (80) und (81) in (75) bzw. (78) erhält man:

$$-\alpha\delta u + ikn_h\delta u + iku_h\delta n = 0 \tag{85}$$

$$-\alpha\delta u + iku_h\delta u = \frac{1}{m}\left[\zeta(n_h)\delta n - \gamma\delta u - \frac{ik\varepsilon}{n_h}\delta n - \frac{\varepsilon m k^2}{\gamma}\delta u\right] \tag{86}$$

(85) und (86) bilden ein homogenes Gleichungssystem in $\delta n$, $\delta u$. Damit es nichttriviale Lösungen gibt, muß die Koeffizientendeterminante verschwinden, was folgende Gleichung ergibt:

$$\left(\alpha - iku_h - \frac{\gamma}{2m} - \frac{\varepsilon k^2}{2\gamma}\right)^2 - \frac{\gamma^2}{4m^2} + \frac{\varepsilon k^2}{2m} - \frac{\varepsilon^2 k^4}{4\gamma^2} - \frac{ikC\varepsilon n_h}{m} = 0 \tag{87}$$

Setzt man nun $\alpha = \lambda + i\nu \quad (\lambda, \nu \in \mathbb{R})$, ergeben sich folgende zwei Gleichungen für den

Real- und Imaginärteil:

$$\lambda^2 - \nu^2 - \frac{\lambda\varepsilon k^2}{\gamma} - \frac{\lambda\gamma}{m} + 2\nu k u_h + k^2\left(\frac{\varepsilon}{m} - u_h^2\right) = 0 \tag{88}$$

$$2\lambda\nu - 2\lambda k u_h - \frac{\nu\varepsilon k^2}{\gamma} - \frac{\nu\gamma}{m} + k\left(\frac{\varepsilon k^2 u_h}{\gamma} + \frac{\gamma u_h}{m} + \frac{n_h\zeta(n_h)}{m}\right) = 0 \tag{89}$$

Aus der Gleichung für den Realteil folgt:

$$\nu_{1/2} = ku_h \pm \sqrt{\frac{\varepsilon k^2}{m} + \lambda^2 - \frac{\lambda\varepsilon k^2}{\gamma} - \frac{\lambda\gamma}{m}} \tag{90}$$

Sinn dieser Betrachtungen ist es, die kritischen Dichten auszurechnen, bei denen das Teilchensystem vom stabilen zum instabilen Zustand wechselt. Ein negatives $\lambda$ entspricht dabei dem instabilen Zustand, ein positives $\lambda$ entspricht dem stabilen Zustand. Beim Wechsel von stabil zu instabil hat $\lambda$ also sehr kleine Beträge, so daß man mit der Abkürzung:

$$c_0 = \sqrt{\frac{\varepsilon}{m}} \tag{91}$$

erhält:

$$\nu_{1/2} = k(u_h \pm c_0) \tag{92}$$

Das wird nun in die Gleichung für den Imaginärteil eingesetzt:

$$\lambda = \frac{\varepsilon k^2}{2\gamma} + \frac{\gamma}{2m} \mp \frac{n_h \zeta(n_h)}{2c_0 m} \tag{93}$$

$\varsigma(n_h)$ ist die Ableitung der Kraft $F(n,T_B)$ (Analogon zur maximalen sicheren Geschwindigkeit V(n)) nach der Dichte und kann deshalb nie positiv werden. Eine Erklärung dafür läßt sich sehr schön im Bild der Verkehrsflüsse geben (siehe auch [34]). Wenn die Dichte steigt, muß auch die Geschwindigkeit V(n) abnehmen. Denn steigende Dichte bedeutet immer kleinerer Abstand zwischen den Fahrzeugen. Um die Sicherheit auch für einen kleineren Abstand zwischen den Autos zu gewährleisten, muß der Fahrer seine Geschwindigkeit reduzieren. V(n) ist also immer eine monoton fallende Funktion der Dichte n, d.h. ihre Ableitung kann nie positiv werden (siehe Bild 3 und 7 ).

Es wird nun nach negativen Werten für $\lambda$, d.h. nach instabilen Zuständen gesucht. Aus den eben dargelegten Betrachtungen folgt, daß das Minuszeichen in (93) niemals negative $\lambda$ liefern kann. $\nu_1 = k(u_h+c_0)$ entfällt also als Lösung. Durch Einsetzen von (84) erhält man somit als Bedingung für Instabilität:

$$-\left[\gamma + \frac{n_h}{c_0}\zeta(n_h)\right]\frac{\gamma}{\varepsilon m} > \left(\frac{2\pi z}{L}\right)^2 \tag{94}$$

Der Übergang vom stabilen zum instabilen Zustand findet bei einer kritischen Dichte $n_c$ statt. Die kleinste kritische Dichte ergibt sich für $z^2=1$. Es gilt also:

$$-\left[\gamma + \frac{n_c}{c_0}\zeta(n_c)\right]\frac{\gamma}{\varepsilon m} = \left(\frac{2\pi}{L}\right)^2 \tag{95}$$

Es sei bemerkt, daß es für ein allgemeines $\varsigma(n)$ auch mehrere $n_c$ geben kann. Das bedeutet, daß es dann bestimmte Dichtebereiche gibt, in denen sich das System stabil verhält und in allen anderen Dichtebereichen verhält es sich instabil. Man betrachte nun den konkreten Fall:

$$F(n, T_B) = mg - C\varepsilon n \tag{96}$$

Es folgt daraus:

$$\zeta(n_c) = -C\varepsilon \tag{97}$$

(97) in (95) eingesetzt, liefert:

$$n_c = \frac{\gamma}{\sqrt{\varepsilon m}\,C} + \frac{\sqrt{\varepsilon m}}{\gamma C}\left(\frac{2\pi}{L}\right)^2 \tag{98}$$

Bei festem $\varepsilon$, m, C und L (Werte aus Unterkapitel 2.2) und variablem Reibungskoeffizienten $\gamma$ erhält man das folgende Bifurkations-Diagramm:

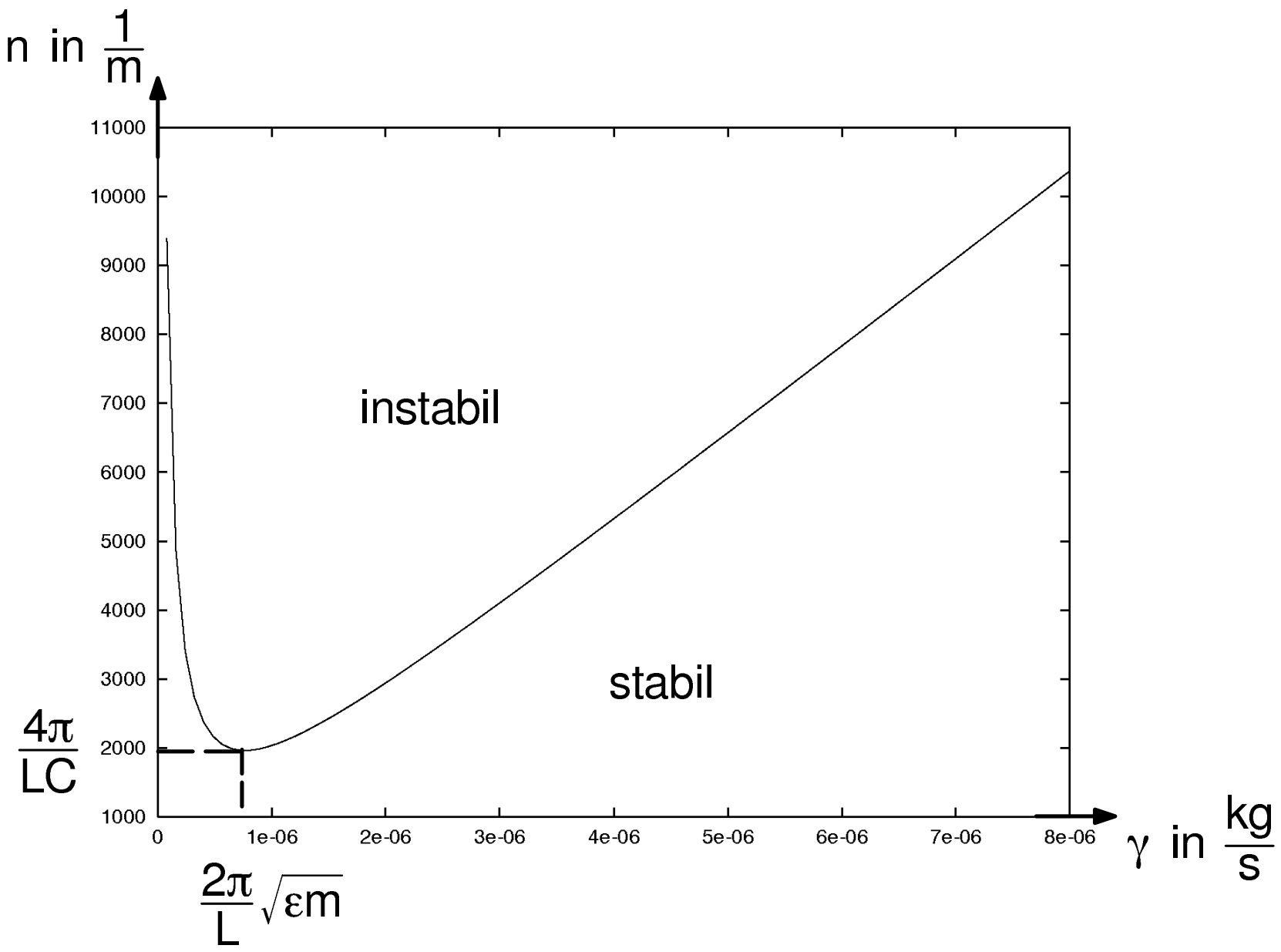

Bild 9 Bifurkations-Diagramm für die kritische Dichte als Funktion des Reibungskoeffizienten

Im Grenzfall eines sehr langen Rohres, genauer für

$$L \gg \frac{2\pi}{\gamma}\sqrt{\varepsilon m} \tag{99}$$

kann der zweite Term vernachlässigt werden. Es ergibt sich:

$$n_C = \frac{\gamma}{\sqrt{\varepsilon m}\,C} \tag{100}$$

Mit der Einführung der Bremslänge

$$L_B = \frac{\sqrt{\varepsilon m}}{\gamma} \tag{101}$$

wird daraus

$$n_C = \frac{1}{L_B C} \tag{102}$$

Für alle Dichten kleiner $n_C$ hat man es mit einem stabilen Verhalten zu tun, bei Dichten größer $n_C$ mit Instabilität.

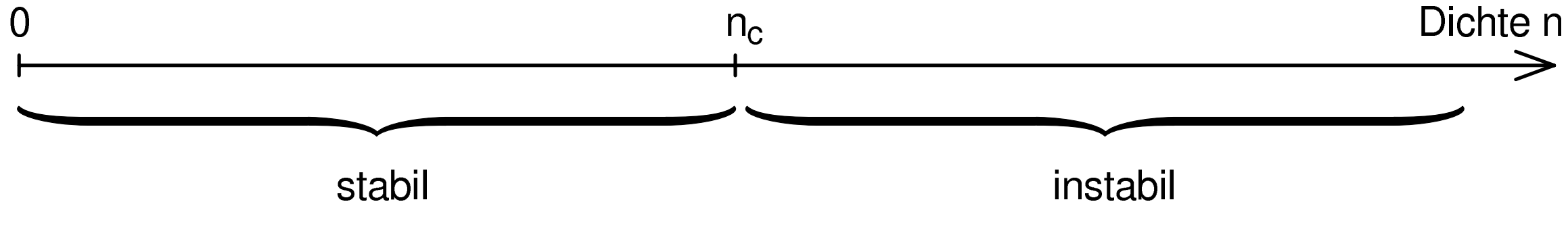

Eine Klumpen- bzw. Staubildung läßt sich also vermeiden, falls:

– alle Teilchen sich mit annähernd gleicher Geschwindigkeit bewegen (kleines $\varepsilon$),

– die Wechselwirkung der Teilchen untereinander gering ist (kleines C),

– die Teilchen sich nicht so träge verhalten (kleines m) und

– die Reibung groß bzw. die Relaxationszeit klein ist (großes $\gamma$).

Anhand Gleichung (100) läßt sich nun auch verstehen, warum die homogene Lösung der Burgers-Gleichung niemals instabil sein kann. Zu ihrer Herleitung wurde der Grenzfall $\gamma$ gegen unendlich angenommen, was zur Folge hat, daß auch $n_c$ gegen unendlich geht. Somit sind Instabilitäten nicht mehr möglich.

Als letztes soll der Frage nachgegangen werden, mit welcher Geschwindigkeit sich die Fluktuationen bewegen. Mit $\alpha=\lambda+i\nu$ können (80), (81) so geschrieben werden:

$$\delta n \sim \delta u \sim \exp(-\alpha t + ikx) = \exp\left(-\lambda t - ik\left(\frac{\nu}{k}t - x\right)\right) \tag{103}$$

Die Fluktuationen bewegen sich also mit der Geschwindigkeit

$$v_F = \frac{\nu}{k} \tag{104}$$

Diese Größe gilt es jetzt zu berechnen. Mit der Abkürzung

$$y = \alpha - iku_h - \frac{\gamma}{2m} - \frac{\varepsilon k^2}{2\gamma} \tag{105}$$

und

$$y = a + ib \qquad (a, b \in \mathbb{R}) \tag{106}$$

d.h.

$$a = \lambda - \frac{\gamma}{2m} - \frac{\varepsilon k^2}{2\gamma} \tag{107}$$

$$b = \nu - ku_h \tag{108}$$

läßt sich (87) wie folgt in Real- und Imaginärteil zerlegen:

$$a^2 - b^2 + \frac{\varepsilon}{2m}k^2 - \frac{\gamma^2}{4m^2} - \left(\frac{\varepsilon k^2}{2\gamma}\right)^2 = 0 \tag{109}$$

$$2ab - \frac{C\varepsilon n_h}{m}k = 0 \tag{110}$$

Aus (108) und (110) folgt:

$$b = \frac{C\varepsilon n_h k}{2am} = \nu - ku_h \tag{111}$$

was nach $\nu$ umgestellt wird.

$$\nu = \frac{C\varepsilon n_h k}{2am} + ku_h \tag{112}$$

Einsetzen von (15) und Divison durch k ergibt:

$$v_F = \frac{mg}{\gamma} - \frac{C\varepsilon n_h}{\gamma}\left(1 - \frac{\gamma}{2am}\right) \tag{113}$$

Wiederum Einsetzen von (107) liefert:

$$v_F = \frac{mg}{\gamma} - \frac{C\varepsilon n_h}{\gamma}\left[1 - \frac{\gamma}{2m\lambda - \gamma - \frac{m\varepsilon k^2}{\gamma}}\right] \tag{114}$$

Das Interesse soll dem kritischen Punkt gelten. Dort ist $\lambda$=0, so daß sich unter Benutzung der Bremslänge folgendes ergibt:

$$v_F = \frac{mg}{\gamma} - \frac{C\varepsilon n_h}{\gamma}\left(1 + \frac{1}{1 + L_B^2 k^2}\right) \tag{115}$$

Für kleine $L_B k$ oder bei Vernachlässigung von $\frac{\varepsilon m}{\gamma}\frac{\partial^2 u}{\partial x^2}$ in (72) ergäbe sich:

$$v_F = \frac{mg}{\gamma} - \frac{2C\varepsilon n_h}{\gamma} \tag{116}$$

(116) kann nun sehr einfach interpretiert werden:

Die homogene Stromdichte $j_h$ ergibt sich mit (15) zu:

$$j_h = u_h n_h = \frac{mgn_h}{\gamma} - \frac{C\varepsilon n_h^2}{\gamma} \tag{117}$$

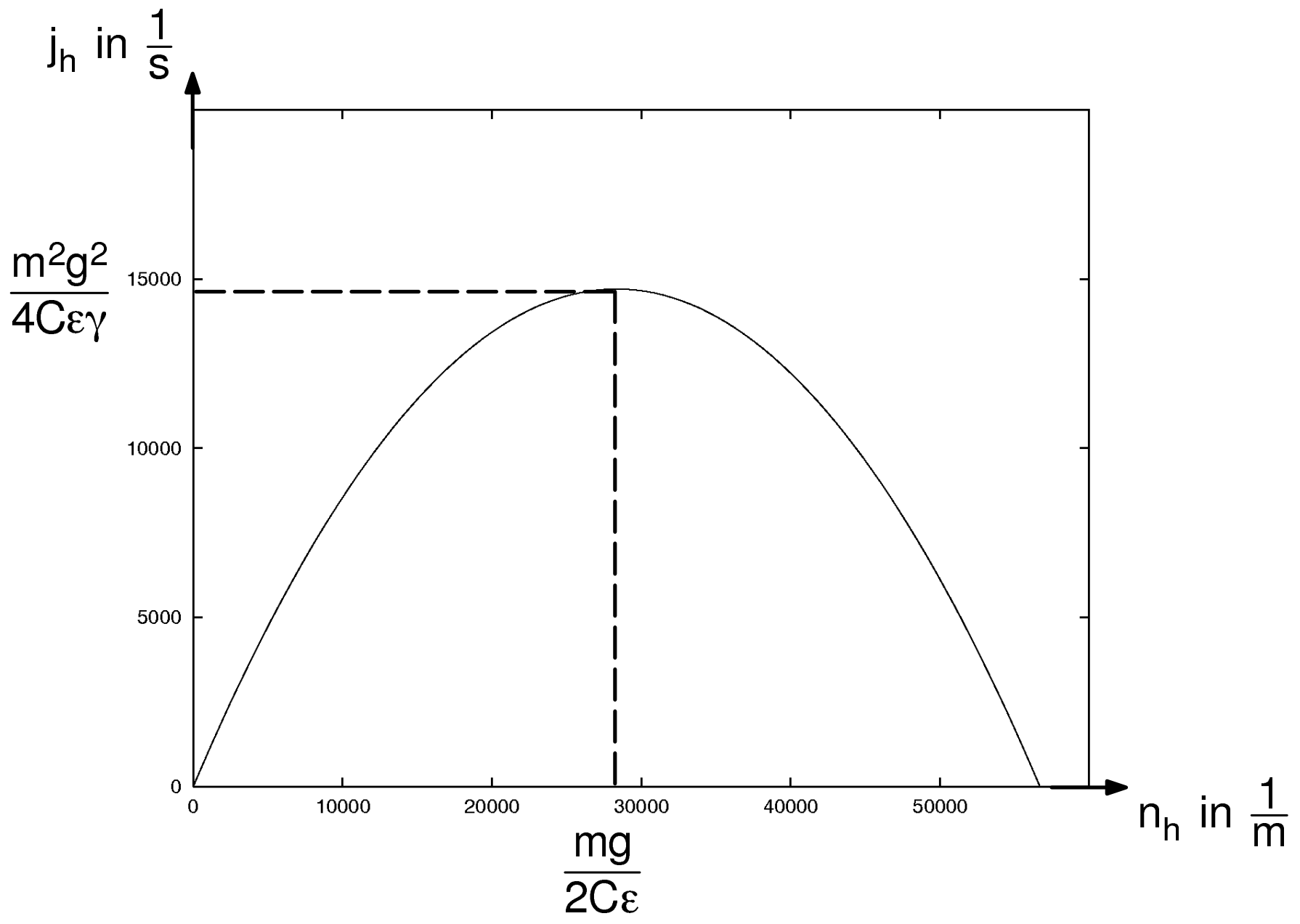

Bild 10 Homogene Stromdichte als Funktion der homogenen Dichte (Zahlenwerte entsprechen dem Parametersatz aus 2.2)

Daraus folgt:

$$\frac{\partial j_h}{\partial n_h} = \frac{mg}{\gamma} - \frac{2C\varepsilon n_h}{\gamma} = v_F \qquad (118)$$

Sind die Fluktuationen gerade am Beginn ihres Wachstumsprozesses (kritischer Punkt $\lambda$=0), bewegen sie sich in Richtung Schwerkraft, falls $\frac{\partial j_h}{\partial n_h}$ positiv ist, und sie bewegen sich rückwärts, falls $\frac{\partial j_h}{\partial n_h}$ negativ ist.

### 1.6.3 Stabilitätsanalyse für temperaturunabhängige Kraft

Die zu Kerner und Konhäuser analogen Gleichungen wurden unter den Voraussetzungen $F(n,T) \equiv F(n,T_B)$ und $\gamma \gg \frac{\partial T}{\partial t},\ u\frac{\partial T}{\partial x}$ erhalten. Es soll nun von Interesse sein, was sich am Stabilitätsverhalten des granularen Systems ändert, wenn man auf die letzte Voraussetzung verzichtet. Es muß also eine Stabilitätsanalyse für das Differentialgleichungssystem (22), (24) und (26) durchgeführt werden. Die Temperatur der Brownschen Teilchen sei dabei im Gleichgewicht mit dem Bad, d.h.:

$$F(n,T) \equiv F(n) = mg - C\varepsilon n(x,t) \qquad (119)$$

Wie gewohnt werden die Störungen wie folgt angesetzt:

$$n(x,t) = n_h + \delta n(x,t) \tag{120}$$

$$u(x,t) = u_h + \delta u(x,t) \tag{121}$$

$$T(x,t) = T_h + \delta T(x,t) \tag{122}$$

wobei gilt:

$$u_h = \frac{mg - Ck_B T_h n_h}{\gamma} \tag{123}$$

$$T_h = T_B = \frac{\varepsilon}{k_B} \tag{124}$$

Dies in (22), (24) und (26) eingesetzt, liefert:

$$\frac{\partial}{\partial t}(\delta n) + n_h \frac{\partial}{\partial x}(\delta u) + u_h \frac{\partial}{\partial x}(\delta n) = 0 \tag{125}$$

$$\frac{\partial}{\partial t}(\delta u) + u_h \frac{\partial}{\partial x}(\delta u) = \frac{1}{m}\left[\zeta(n_h)\delta n - \gamma \delta u - k_B \frac{\partial}{\partial x}(\delta T) - \frac{\varepsilon}{n_h}\frac{\partial}{\partial x}(\delta n)\right] \tag{126}$$

$$\frac{\partial}{\partial t}(\delta T) + u_h \frac{\partial}{\partial x}(\delta T) = -\frac{2\gamma}{m}\delta T - \frac{2\varepsilon}{k_B}\frac{\partial}{\partial x}(\delta u) \tag{127}$$

wobei

$$\zeta(n_h) = \frac{\partial F(n, T_B)}{\partial n}\Big|_{n=n_h} \tag{128}$$

Die Störungen seien wieder als Wellen angesetzt:

$$\delta n(x,t) = \delta n_0 \exp[-\alpha t + ikx] \tag{129}$$

$$\delta u(x,t) = \delta u_0 \exp[-\alpha t + ikx] \tag{130}$$

$$\delta T(x,t) = \delta T_0 \exp[-\alpha t + ikx] \tag{131}$$

und sollen die als periodisch angesetzten Randbedingungen erfüllen:

$$\delta n(0,t) = \delta n(L,t) \tag{132}$$

$$\delta u(0,t) = \delta u(L,t) \tag{133}$$

$$\delta T(0,t) = \delta T(L,t) \tag{134}$$

Es folgt, daß k nur folgende Werte annehmen darf:

$$k = \frac{2\pi z}{L} \qquad \text{wobei } z = \pm 1, \pm 2, \ldots \tag{135}$$

Durch Einsetzen von (129), (130), (131) in (125), (126), (127) erhält man:

$$-\alpha\delta n + ikn_h\delta u + iku_h\delta n = 0 \qquad (136)$$

$$-\alpha\delta u + iku_h\delta u = \frac{1}{m}\left[\zeta(n_h)\delta n - \gamma\delta u - ikk_B\delta T - ik\frac{\varepsilon}{n_h}\delta n\right] \qquad (137)$$

$$-\alpha\delta T + iku_h\delta T = -\frac{2\gamma}{m}\delta T - 2ik\frac{\varepsilon}{k_B}\delta u \qquad (138)$$

(136), (137) und (138) bilden ein homogenes Gleichungssystem in $\delta n$, $\delta u$, $\delta T$. Damit es nichttriviale Lösungen gibt, muß die Koeffizientendeterminante verschwinden. Dies führt auf die folgende Gleichung für $\alpha$:

$$\alpha^3 + r\alpha^2 + s\alpha + t = 0 \qquad (139)$$

wobei die komplexen Koeffizienten r, s, t für die spezielle Kraft

$$F(n, T_B) = mg - C\varepsilon n \qquad (140)$$

also

$$\zeta(n_h) = -C\varepsilon \qquad (141)$$

wie folgt lauten:

$$r = -3\left(\frac{\gamma}{m} + iku_h\right) \qquad (142)$$

$$s = \frac{2\gamma^2}{m^2} + \frac{3\varepsilon k^2}{m} - 3k^2u_h^2 + i\left(\frac{6\gamma ku_h}{m} - \frac{C\varepsilon n_h k}{m}\right) \qquad (143)$$

$$t = \frac{3\gamma k^2u_h^2}{m} - \frac{C\varepsilon n_h k^2 u_h}{m} - \frac{2\varepsilon\gamma k^2}{m^2} + i\left(k^3u_h^3 - \frac{2\gamma^2 ku_h}{m^2} - \frac{3\varepsilon k^3 u_h}{m} + \frac{2C\varepsilon n_h\gamma k}{m^2}\right) \qquad (144)$$

Durch die Ersetzung

$$y = \alpha + \frac{r}{3} \qquad (145)$$

gelangt man zu der reduzierten Gleichung:

$$y^3 + py + q = 0 \qquad (146)$$

Die Koeffizienten p und q lauten dabei:

$$p = \frac{3s - r^2}{3} = \frac{3\varepsilon k^2}{m} - \frac{\gamma^2}{m^2} - i\frac{C\varepsilon n_h k}{m} \qquad (147)$$

$$q = \frac{2r^3}{27} - \frac{rs}{3} + t = \frac{\varepsilon\gamma k^2}{m^2} + i\frac{C\varepsilon n_h\gamma k}{m^2} \qquad (148)$$

Zur Lösung der kubischen Gleichung (146) benutzt man die Cardanische Formel. Es ergeben sich drei Lösungen für $\alpha$:

$$\alpha_1(n_h, z) = -\frac{r}{3} + u + v \tag{149}$$

$$\alpha_2(n_h, z) = -\frac{r}{3} - \frac{u+v}{2} + i\sqrt{3}\frac{u-v}{2} \tag{150}$$

$$\alpha_3(n_h, z) = -\frac{r}{3} - \frac{u+v}{2} - i\sqrt{3}\frac{u-v}{2} \tag{151}$$

wobei folgende Abkürzungen benutzt wurden:

$$u = \sqrt[3]{-\frac{q}{2} + \sqrt{D}} \tag{152}$$

$$v = -\frac{p}{3u} \tag{153}$$

$$D = \left(\frac{p}{3}\right)^3 + \left(\frac{q}{2}\right)^2 \tag{154}$$

Beim Wurzelziehen in (152) braucht man nur mit dem Hauptwert zu rechnen. Die anderen Lösungen der Wurzel sind bereits in den jeweils anderen beiden Lösungen von $\alpha$ enthalten.

Es ist nun von Interesse, für welche z und $n_h$ der Realteil von $\alpha$ kleiner Null wird, denn das ist das Kriterium dafür, daß die homogene Lösung instabil wird. Für die kleinste mögliche Wellenzahl mit z=1 sei hier der Realteil von $\alpha$ als Funktion von $n_h$ graphisch dargestellt. Für die Parameter $\varepsilon$, C, g, m, $\gamma$ und L wurden die Werte aus dem Kapitel "Simulationen" verwendet.

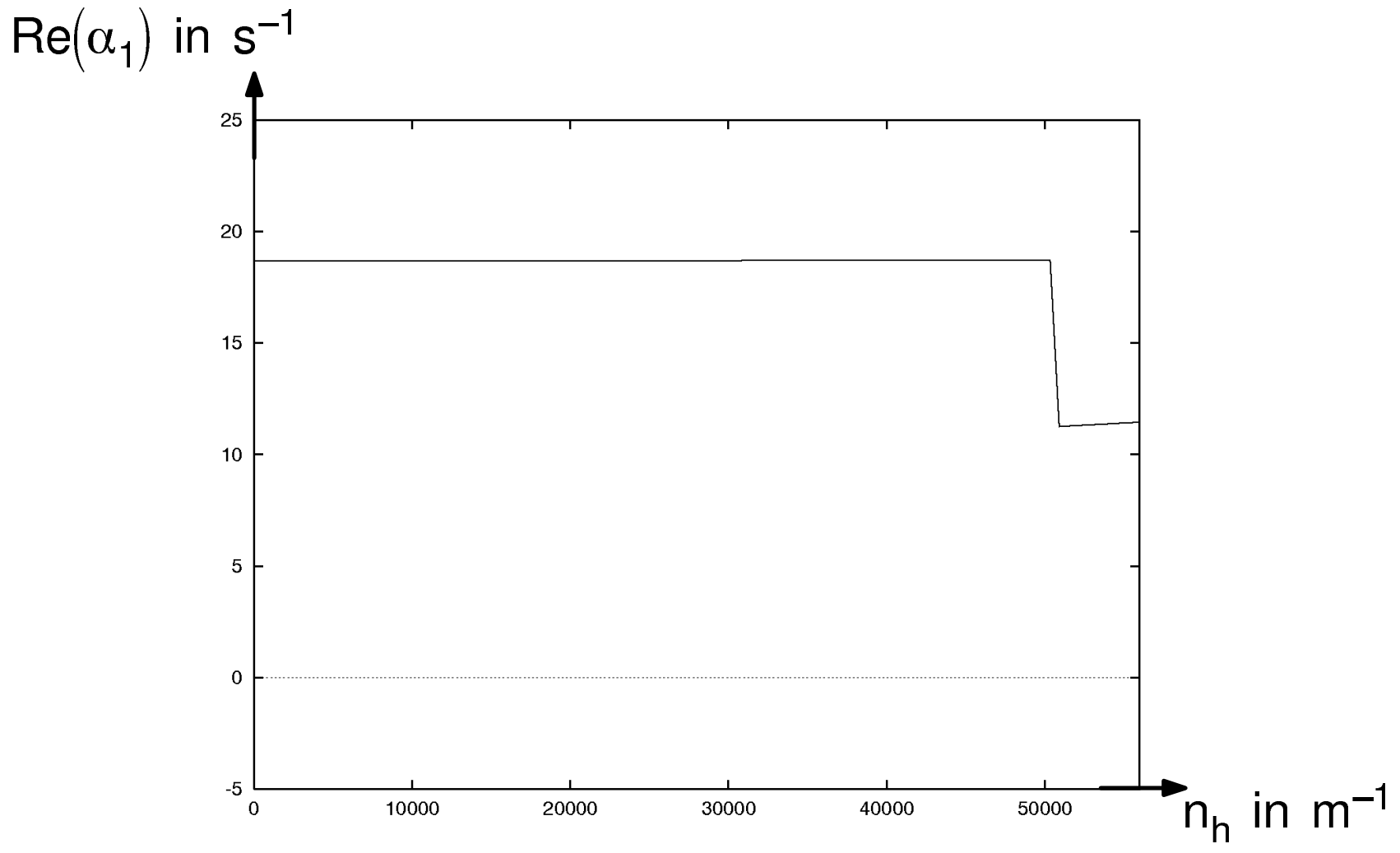

Bild 11 Realteil von $\alpha_1$ als Funktion der homogenen Dichte

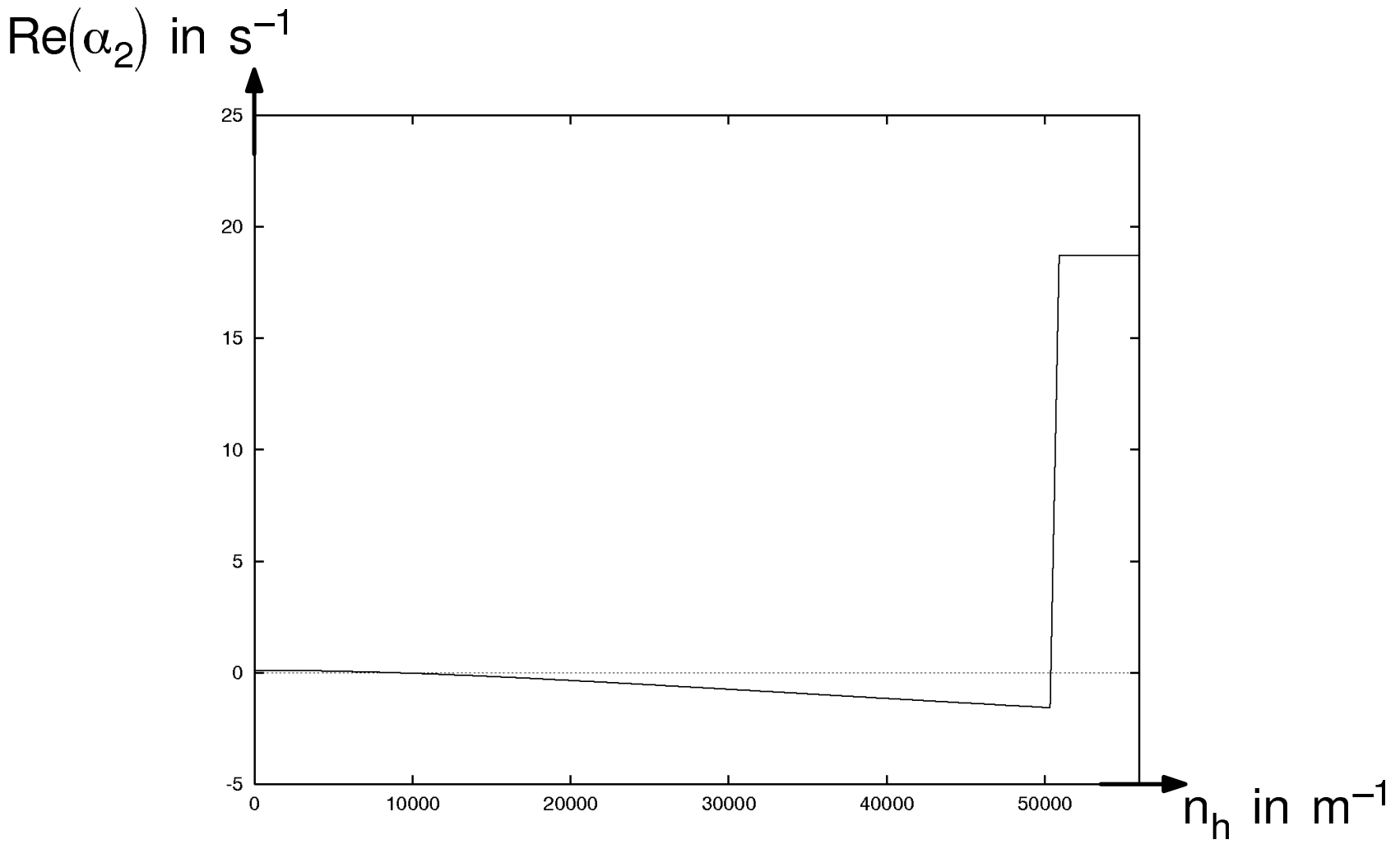

Bild 12 Realteil von $\alpha_2$ als Funktion der homogenen Dichte

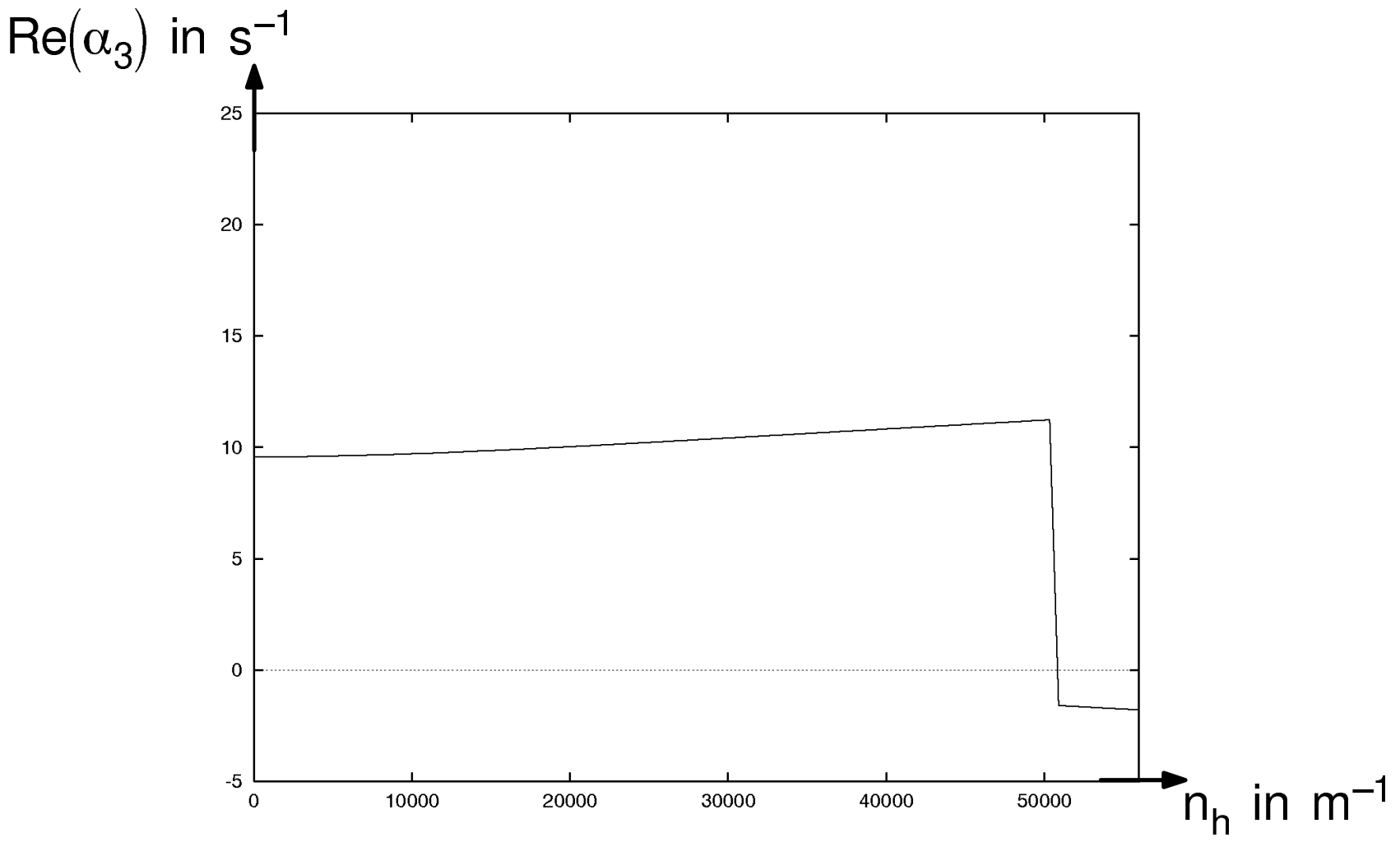

Bild 13 Realteil von $\alpha_3$ als Funktion der homogenen Dichte

Es ist deutlich zu erkennen, daß der Realteil von $\alpha_1$ niemals negativ wird. Der Realteil von $\alpha_2$ ist für Dichten zwischen $9100 m^{-1}$ und $50400 m^{-1}$ negativ, der Realteil von $\alpha_3$ für Dichten größer $50400 m^{-1}$. Im Falle z=1 erhält man also eine kritische Dichte $n_c = 9100 m^{-1}$. Für $n_h < n_c$ ist die homogene Lösung stabil, für $n_h > n_c$ instabil. Auf diese Art und Weise kann man für jedes z eine kritische Dichte $n_c$ erhalten. Wie $n_c$ von der Wellenzahl k bzw. von z abhängt, wird im folgenden Bild gezeigt:

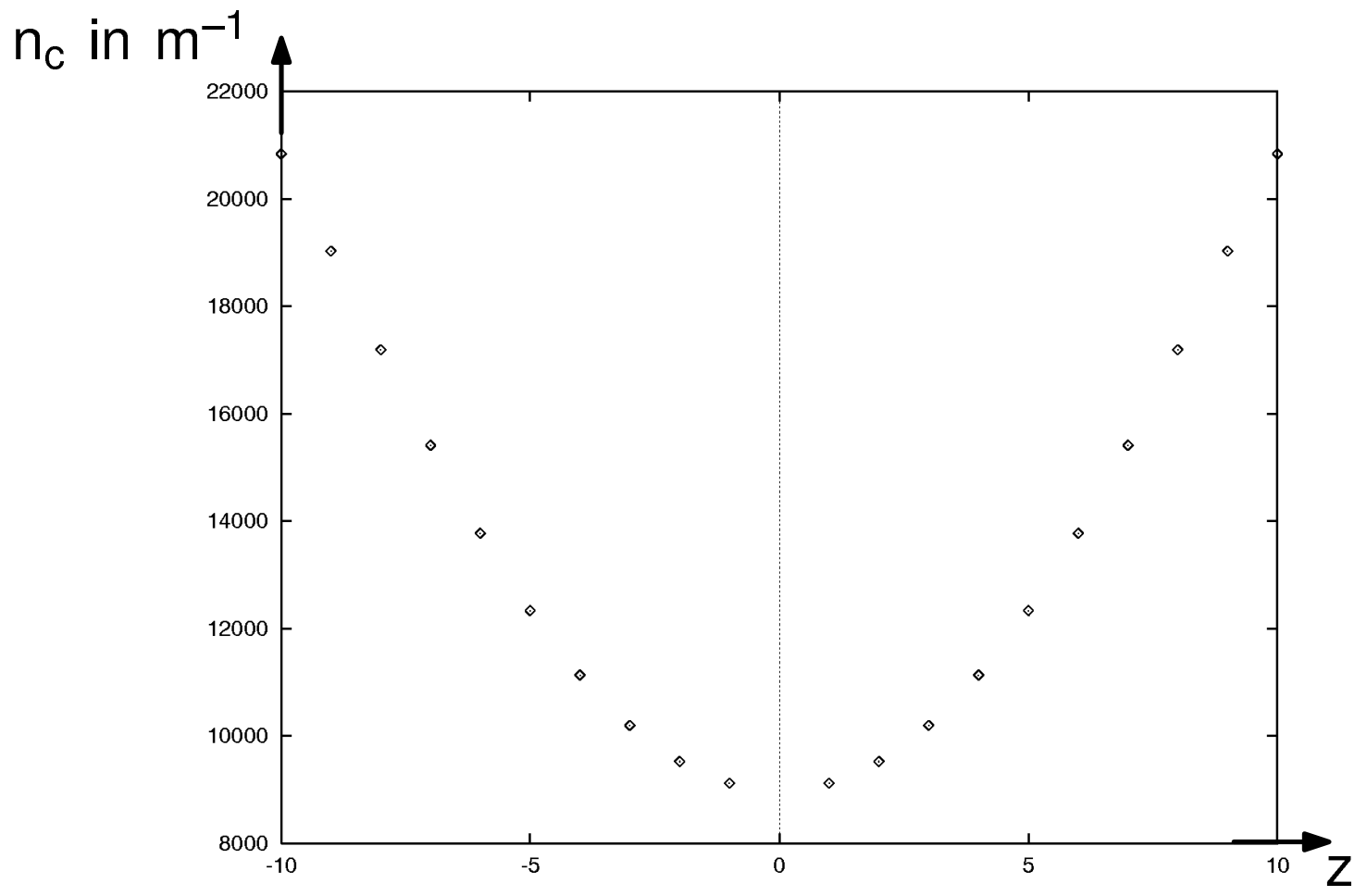

Bild 14 Kritische Dichte als Funktion von z

Am "instabilsten" sind die Moden z=±1. Für diese ergibt sich im Falle unendlich langer Rohre:

$$n_c = \frac{\gamma}{\sqrt{\varepsilon m}\, C} = \frac{1}{L_B C} \qquad (155)$$

was das gleiche Resultat wie bei den zu Kerner/Konhäuser analogen Gleichungen ist (siehe (102)).

### 1.6.4 Stabilitätsanalyse für temperaturabhängige Kraft

Für den Fall, daß der Kraftansatz auch Nichtgleichgewichtstemperaturen für die Teilchen zuläßt, d.h. für:

$$F(n, T) = mg - Ck_B T(x, t) n(x, t) \qquad (156)$$

sei hier die Stabilität der homogenen Lösung untersucht. Um den Leser nicht mit den fast identischen Rechnungen zu langweilen, seien nur die wichtigsten Resultate angegeben. Es sei bemerkt, daß durch Einführung einer Variablen a die Formeln für temperaturunabhängige und temperaturabhängige Kraft völlig identisch sind. Man setze

$$a = 1 \quad \text{für temperaturunabhängige Kraft und} \qquad (157)$$

$$a = 3 \quad \text{für temperaturabhängige Kraft.} \qquad (158)$$

Es ergibt sich also:

$$\alpha^3 + r\alpha^2 + s\alpha + t = 0 \qquad (159)$$

$$r = -3\left(\frac{\gamma}{m} + iku_h\right) \qquad (160)$$

$$s = \frac{2\gamma^2}{m^2} + \frac{3\varepsilon k^2}{m} - 3k^2u_h^2 + i\left(\frac{6\gamma k u_h}{m} - \frac{aC\varepsilon n_h k}{m}\right) \tag{161}$$

$$t = \frac{3\gamma k^2 u_h^2}{m} - \frac{aC\varepsilon n_h k^2 u_h}{m} - \frac{2\varepsilon\gamma k^2}{m^2} + i\left(k^3u_h^3 - \frac{2\gamma^2 k u_h}{m^2} - \frac{3\varepsilon k^3 u_h}{m} + \frac{2C\varepsilon n_h \gamma k}{m^2}\right) \tag{162}$$

$$y = \alpha + \frac{r}{3} \tag{163}$$

$$y^3 + py + q = 0 \tag{164}$$

$$p = \frac{3s - r^2}{3} = \frac{3\varepsilon k^2}{m} - \frac{\gamma^2}{m^2} - i\frac{aC\varepsilon n_h k}{m} \tag{165}$$

$$q = \frac{2r^3}{27} - \frac{rs}{3} + t = \frac{\varepsilon\gamma k^2}{m^2} + i(2-a)\frac{C\varepsilon n_h \gamma k}{m^2} \tag{166}$$

$$\alpha_1(n_h, z) = -\frac{r}{3} + u + v \tag{167}$$

$$\alpha_2(n_h, z) = -\frac{r}{3} - \frac{u+v}{2} + i\sqrt{3}\frac{u-v}{2} \tag{168}$$

$$\alpha_3(n_h, z) = -\frac{r}{3} - \frac{u+v}{2} - i\sqrt{3}\frac{u-v}{2} \tag{169}$$

$$u = \sqrt[3]{-\frac{q}{2} + \sqrt{D}} \tag{170}$$

$$v = -\frac{p}{3u} \tag{171}$$

$$D = \left(\frac{p}{3}\right)^3 + \left(\frac{q}{2}\right)^2 \tag{172}$$

Hier wieder die graphische Veranschaulichung der Realteile der $\alpha_i$:

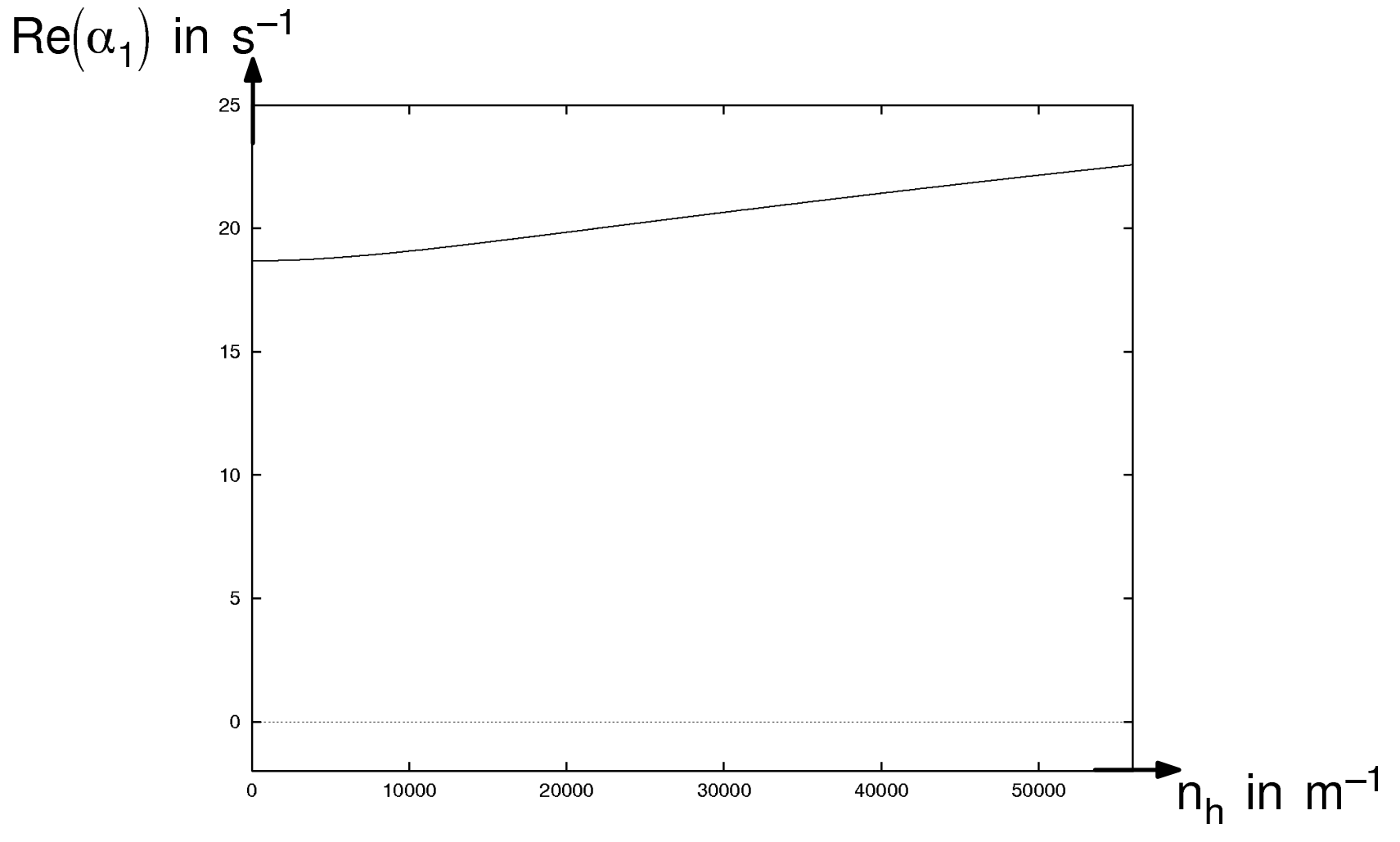

Bild 15 Realteil von $\alpha_1$ als Funktion der homogenen Dichte

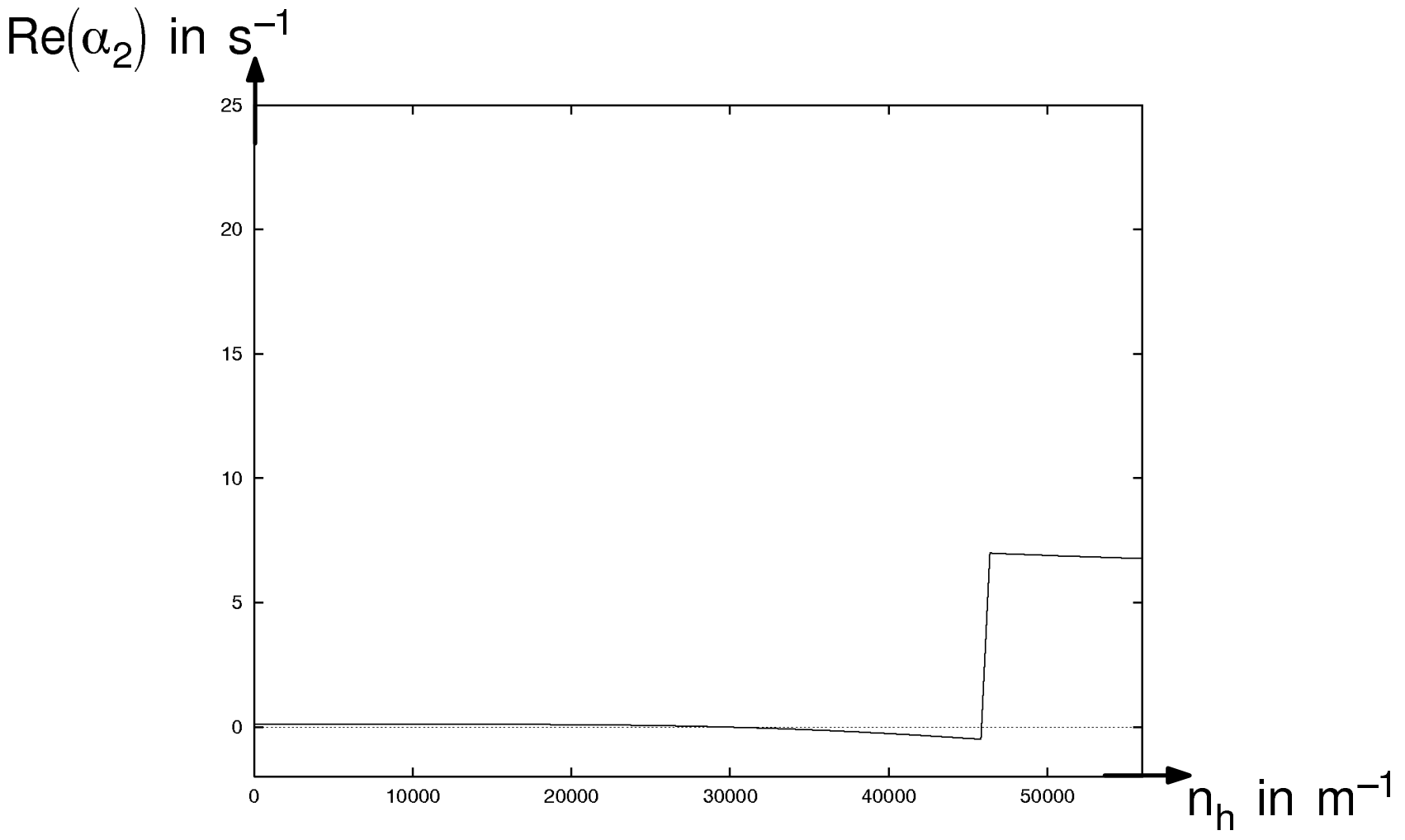

Bild 16 Realteil von $\alpha_2$ als Funktion der homogenen Dichte

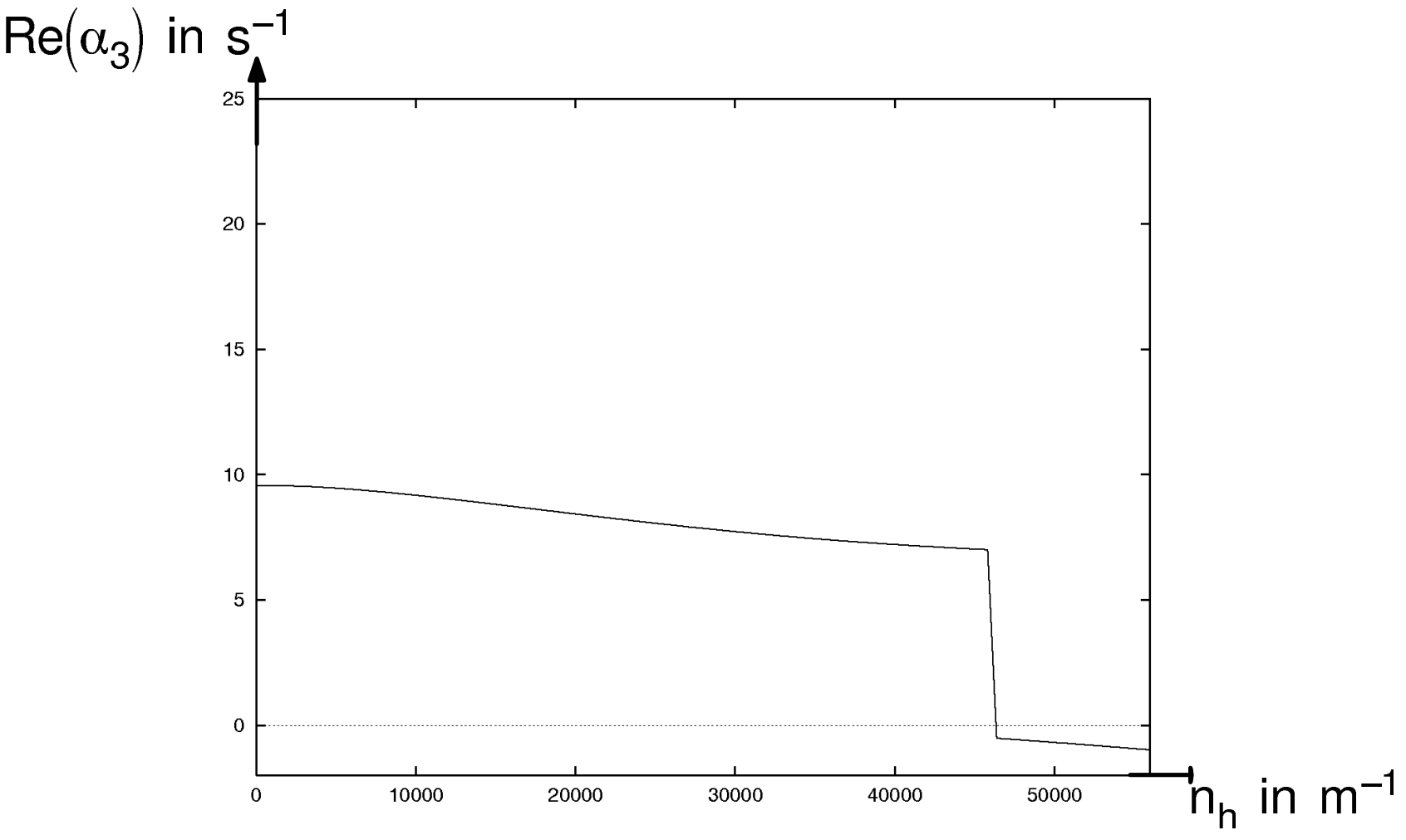

Bild 17 Realteil von $\alpha_3$ als Funktion der homogenen Dichte

Wie $n_c$ im Falle a=3 von der Wellenzahl k bzw. von z abhängt, wird im folgenden Bild gezeigt:

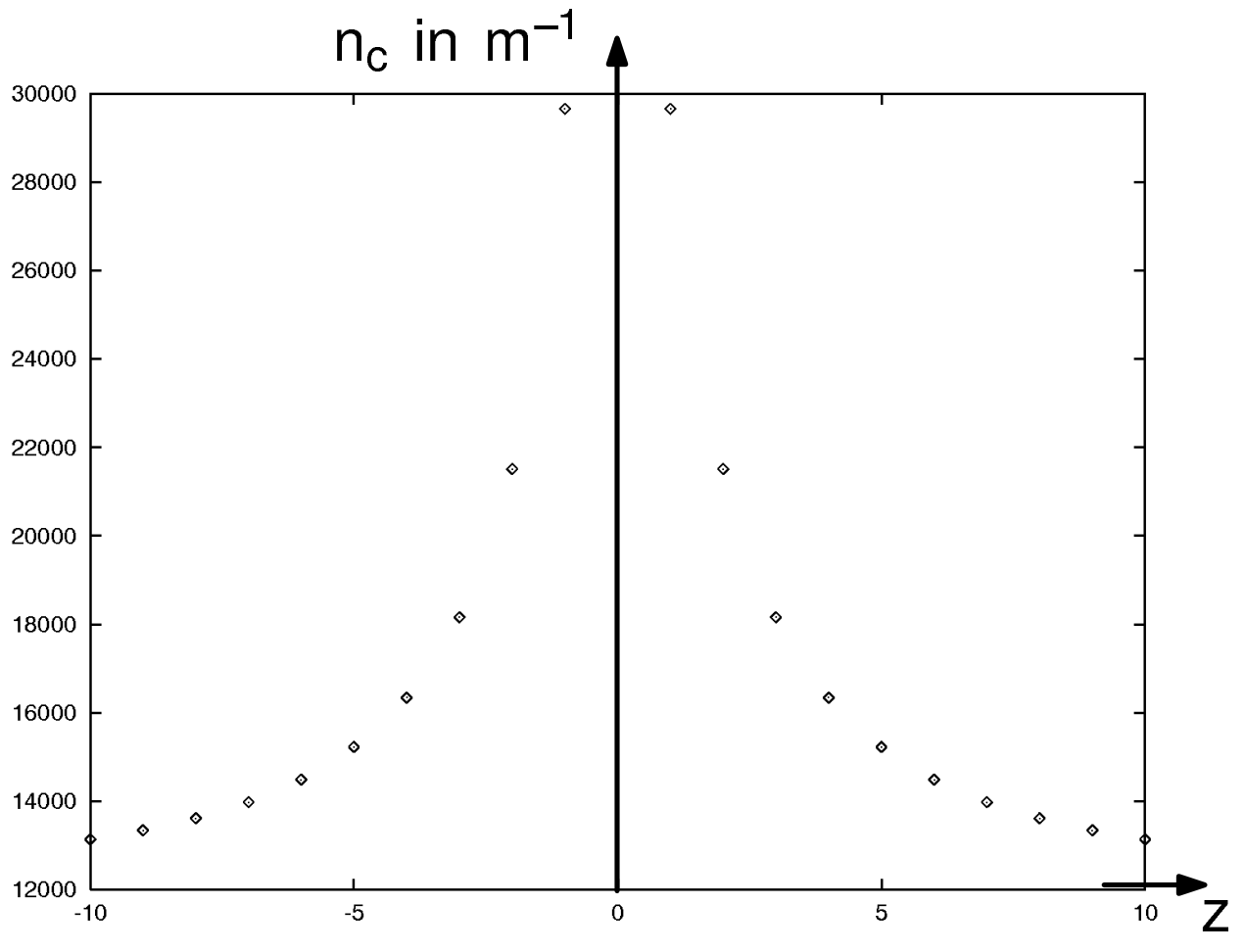

Bild 18 Kritische Dichte als Funktion von z

Diesmal sind die Moden mit großem $z^2$ am "instabilsten".

### 1.6.5 Zusammenfassung der Stabilitätsanalysen

– Die homogene Lösung der Burgers-Gleichung ist immer stabil. Somit ist sie zur Beschreibung von Verklumpungserscheinungen nicht geeignet.

– Für ein unendlich langes Rohr liefern die zu Kerner/Konhäuser analogen Gleichungen dieselbe kritische Dichte wie im Falle der temperaturunabhängigen Kraft.

– Im Falle der temperaturunabhängigen Kraft wird die kritische Dichte mit wachsender Wellenzahl größer.

– Im Falle der temperaturabhängigen Kraft wird die kritische Dichte mit wachsender Wellenzahl kleiner.

– Für den gewählten Parametersatz liefert die temperaturunabhängige Kraft als kleinste mögliche kritische Dichte $n_c \approx 9000 m^{-1}$ und die temperaturabhängige Kraft $n_c \approx 12000 m^{-1}$.

# 2 Simulationen

## 2.1 Algorithmus

Die im vorigen Kapitel gemachten Aussagen der Theorie sollen nun durch Computersimulationen überprüft werden. Dazu werden direkt die Langevin-Gleichungen (36) und (37) wie folgt simuliert:

$$x_i(t+\Delta t) = x_i(t) + v_i(t)\Delta t \tag{173}$$

$$v_i(t+\Delta t) = v_i(t) + \left(\frac{F(n,T)}{m} - \frac{\gamma v_i(t)}{m}\right)\Delta t - \frac{\sqrt{2\varepsilon\gamma}\,\xi_i}{m}\sqrt{\Delta t} \tag{174}$$

Der Fluktuationsterm muß mit $\sqrt{\Delta t}$ multipliziert werden, um Unabhängigkeit von der Größe des Zeitschritts $\Delta t$ zu erreichen. Dieser Fakt wird verständlich, wenn man bedenkt, daß die Summe zweier gaußverteilter Zufallszahlen mit den Standardabweichungen $\sigma_1$ und $\sigma_2$ eine gaußverteilte Zufallszahl mit der Standardabweichung $\sqrt{\sigma_1 + \sigma_2}$ ergibt.

Folgende Dinge sind bei den Simulationen zu beachten:

– Der Zeitschritt muß klein gegen die Relaxationszeit sein.

$$\Delta t \ll \frac{m}{\gamma} \tag{175}$$

– Zur Berechnung der Dichte n, der mittleren Geschwindigkeit u, der Temperatur T und der Geschwindigkeitsverteilung w wird der Wertebereich der jeweiligen Variable in M äquidistante Boxen unterteilt. Dabei muß einerseits $M \gg 1$ gelten, um charakteristische Inhomogenitäten des Systems noch auflösen zu können, aber andererseits darf M auch nicht zu groß gewählt werden, damit noch genügend viele Teilchen in einer Box liegen.

– Soll die kritische Dichte $n_c$ aus Gleichung (155) bestimmte werden, darf die Rohrlänge L nicht zu klein sein.

$$L \gg \frac{2\pi}{\gamma}\sqrt{\varepsilon m} \tag{176}$$

## 2.2 Bestimmung der Parameter

Das Ziel soll es sein, die Parameter der Simulation so realitätsnah wie möglich festzulegen. Im Unterkapitel 3.4 werden Experimente mit Sand in einem dünnem Rohr des Innendurch-messers 6mm vorgestellt. Es werden dort die Werte für m, $\gamma$, $n_{ST}$ und $n_c$ abgeschätzt. Weiterhin ist der Wert von g bekannt.

$$m = 7,4 * 10^{-7}kg \tag{177}$$

$$\gamma = 7 * 10^{-6}\frac{kg}{s} \tag{178}$$

$$n_{ST} = 56000\frac{1}{m} \tag{179}$$

$$n_c = 9000\frac{1}{m} \tag{180}$$

$$g = 9,81\frac{m}{s^2} \tag{181}$$

Es fehlen noch die Werte für $\varepsilon$ und C. Dazu folgende Überlegungen:

Das Verhältnis von kritischer Dichte zur Staudichte soll mit a bezeichnet werden, das Verhältnis von Nullstelle der V(n)-Kurve (siehe Bild 3 ) zur Staudichte soll mit b bezeichnet werden. Mit den angegebenen Werten für $n_{ST}$ und $n_c$ ergibt sich a = 0,16. Da der tatsächliche Verlauf der V(n)-Kurve für Sand nicht bekannt ist, soll als erste Abschätzung b = 1 gesetzt werden, d.h. $n_{ST} = \frac{mg}{C\varepsilon}$. Mit diesen Annahmen erhält man mit (155):

$$\varepsilon = 2,0 * 10^{-8}Nm \tag{182}$$

$$C = 6,4 * 10^{-3} \tag{183}$$

Aus (175) folgt nun $\Delta t \ll 0,11s$ und aus (176) folgt $L \gg 0,12m$. Es werden folgende Festlegungen getroffen:

$$\Delta t = 0.01s \tag{184}$$

$$L = 1m \tag{185}$$

$$M = 100 \tag{186}$$

Damit liegen alle Simulationsparameter fest.

## 2.3 Homogene Anfangsbedingungen, temperaturunabhängige Kraft

In diesem Unterkapitel soll das geschlossene Rohr untersucht werden. Es werden also periodische Randbedingungen realisiert, d.h. ein Teilchen, das ein Rohrende verläßt, wird am anderen Rohrende wieder eingefügt. Die Kraft wird als temperaturunabhängig angesetzt, d.h.:

$$F(n,T) \equiv F(n,T_B) = mg - C\varepsilon n(x,t) \tag{187}$$

### 2.3.1 Stabiles Regime

Als erstes soll die homogene Anfangsbedingung

$$n(x,0) = 8000\frac{1}{m} \qquad \forall x \in [0,L] \tag{188}$$

$$u(x,0) = 0 \qquad \forall x \in [0,L] \tag{189}$$

untersucht werden. Da n(x,0) < $n_c$ ist, erwartet man, daß sich das System stabil verhält, also die Dichte im wesentlichen für alle Zeiten erhalten bleibt und die mittlere Geschwindigkeit den stationären Zustand (19) annimmt.

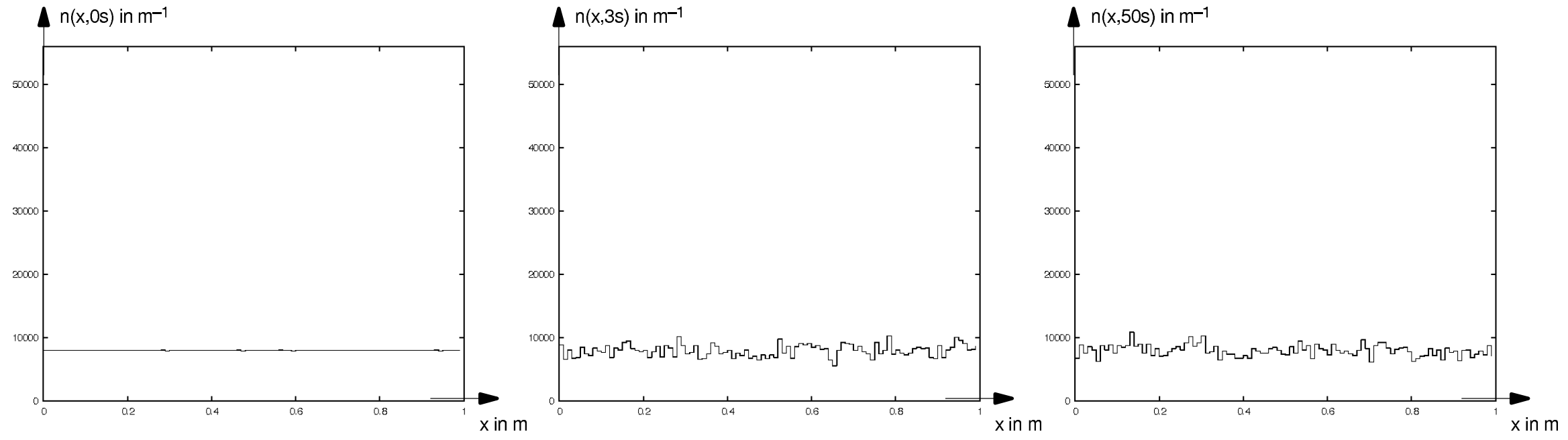

Bild 19 Dichte eines stabilen Systems zur Zeit 0s, 3s und 50s

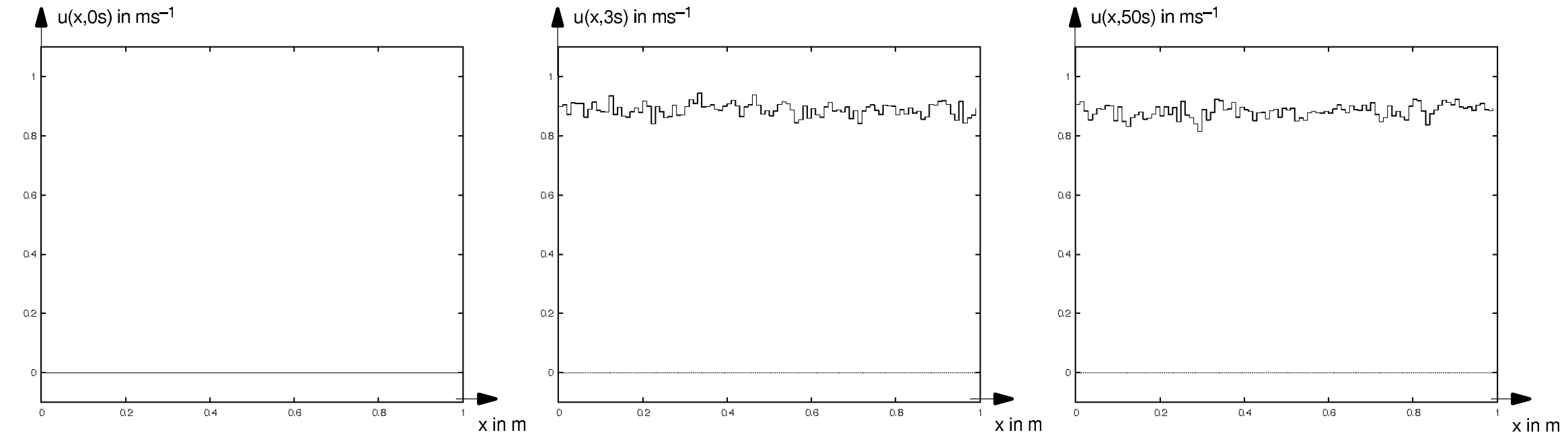

Bild 20 Mittlere Geschwindigkeit eines stabilen Systems zur Zeit 0s, 3s und 50s

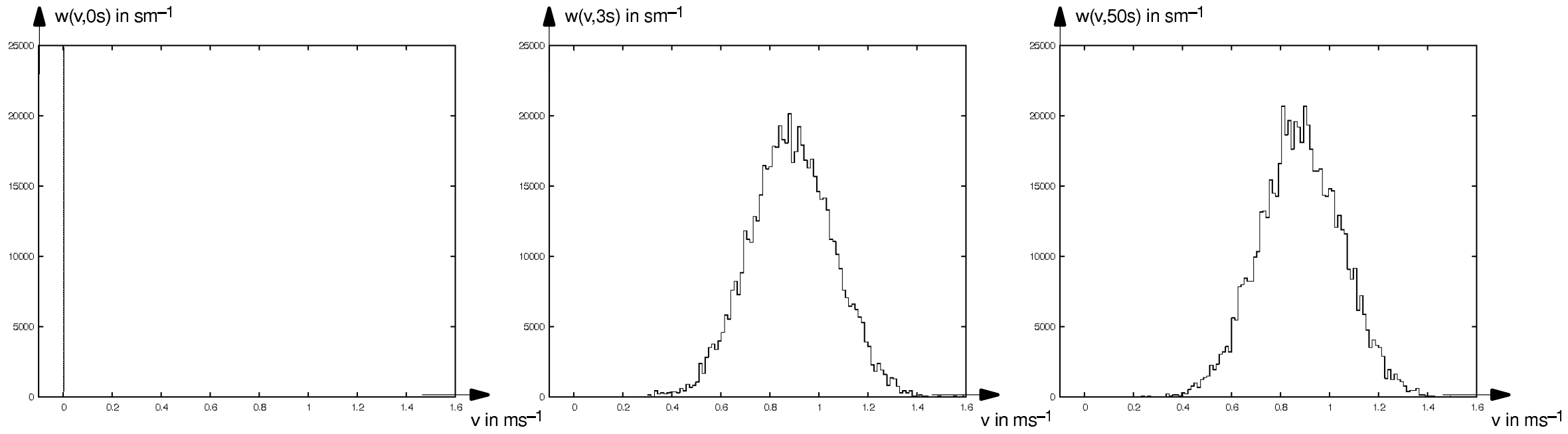

Bild 21 Geschwindigkeitsverteilung eines stabilen Systems zur Zeit 0s, 3s und 50s

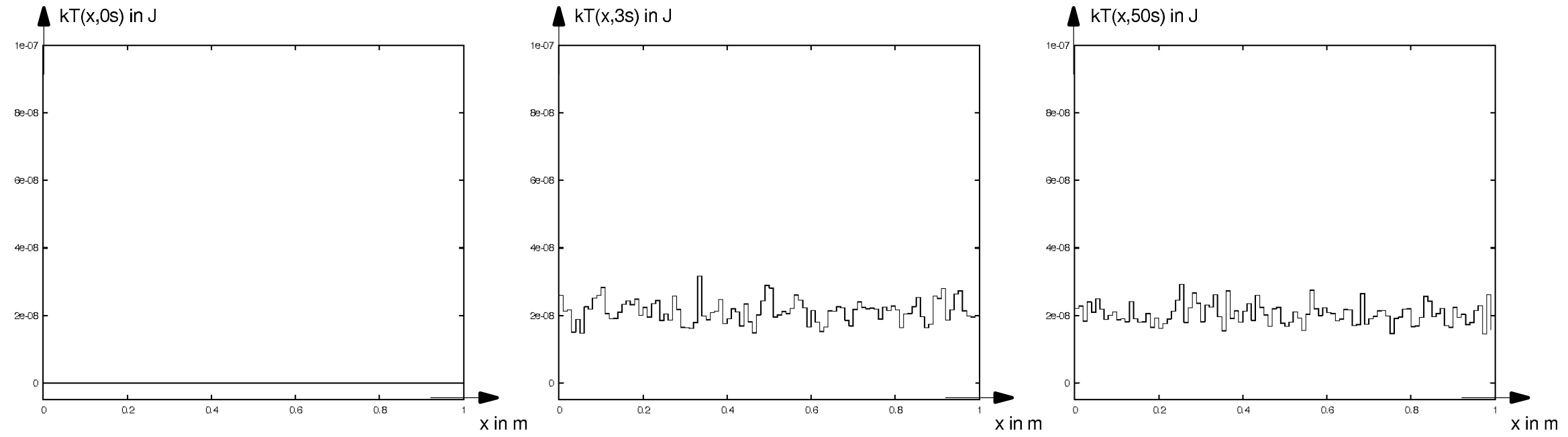

Bild 22 Mittlere thermische Energie kT eines stabilen Systems zur Zeit 0s, 3s und 50s

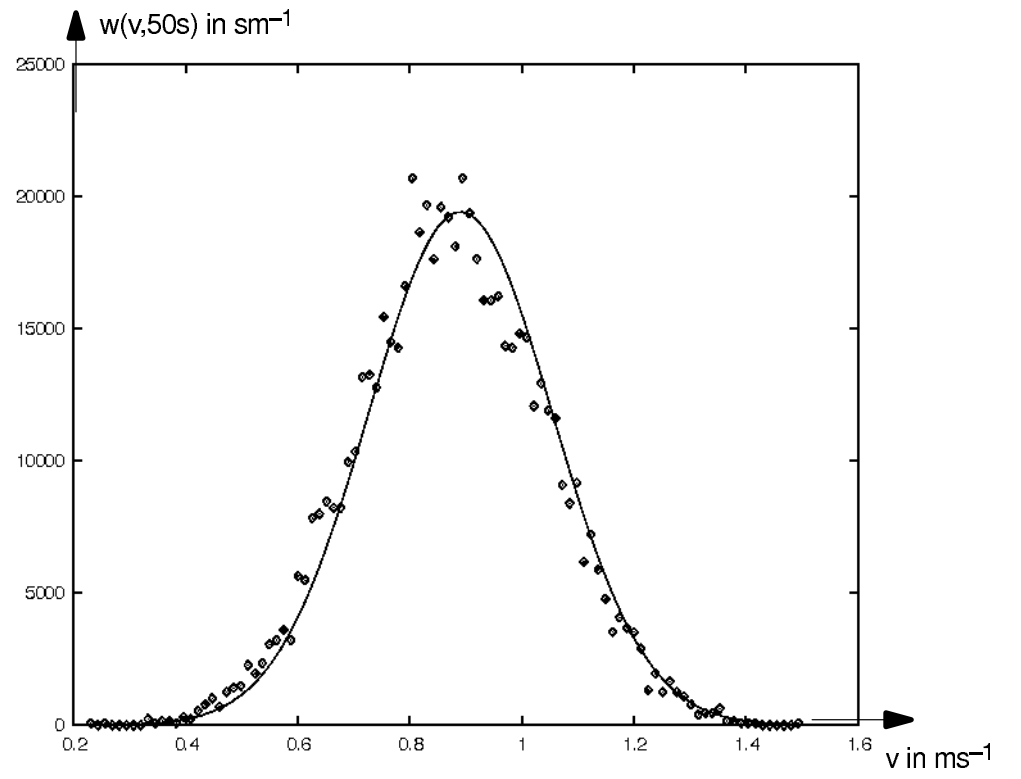

Bild 23 Vergleich der Geschwindigkeitsverteilung eines homogenen, stabilen Systems aus Simulation (Punkte) und Gleichung (19) der Theorie (durchgezogene Kurve) zur Zeit 50s

Die Bilder 19 - 23 zeigen deutliche Übereinstimmung zwischen Theorie und Simulation. Insbesondere stellt sich sehr "rasch" die mittlere Geschwindigkeit aus (15) ein und die mittlere thermische Energie der Teilchen weicht nirgends stark von $\varepsilon$ ab.

### 2.3.2 Instabiles Regime

In diesem Abschnitt soll gezeigt werden, daß bei homogenen Anfangsdichten $n_h > n_c = 9000 m^{-1}$ Klumpenbildung auftritt. Daß dies bereits deutlich für $n_h = 11000 m^{-1}$ geschieht, zeigt das folgende Bild.

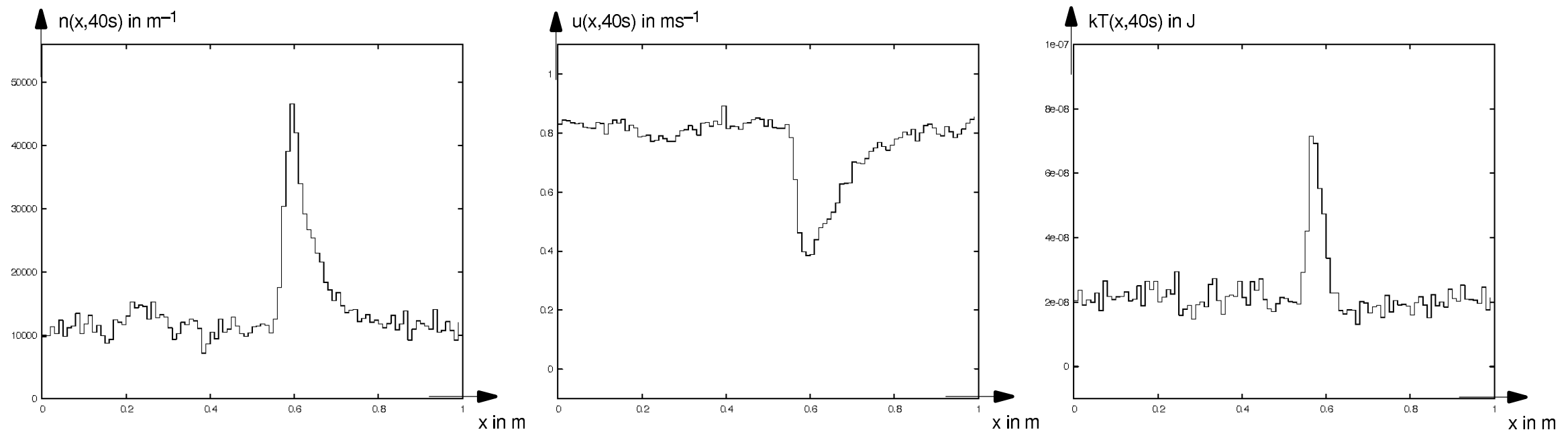

Bild 24 Dichte, mittlere Geschwindigkeit und mittlere thermische Energie eines instabilen Systems zur Zeit 40s

Aus pädagogischen Gründen soll die ausführliche graphische Darstellung jedoch für eine noch größere Anfangsdichte erfolgen. Denn dann ist das Verschmelzen von zwei Klumpen besser zu erkennen. Es werden also die folgenden Anfangsbedingungen betrachtet:

$$n(x,0) = 14000\frac{1}{m} \qquad \forall x \in [0,L] \tag{190}$$

$$u(x,0) = 0 \qquad \forall x \in [0,L] \tag{191}$$

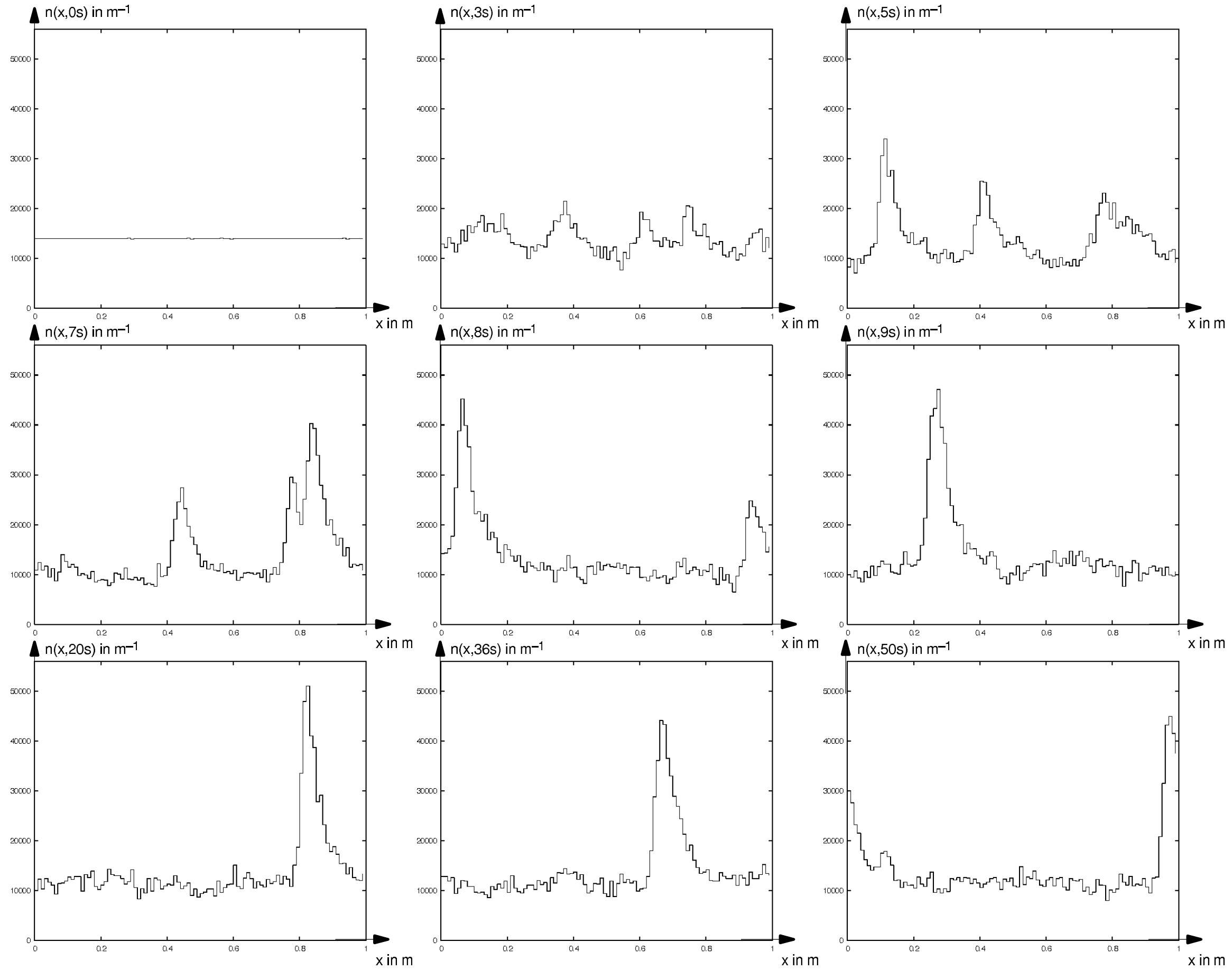

Bild 25 Dichte eines instabilen Systems zur Zeit 0s, 3s, 5s, 7s, 8s, 9s, 20s, 36s und 50s

Bild 25 zeigt deutlich, wie sich im Laufe der Zeit kleine Fluktuationen aufschaukeln und makroskopische Klumpen bilden. Unterschiedlich große Klumpen bewegen sich auch mit unterschiedlicher Geschwindigkeit, so daß es zur Verschmelzung zweier ineinanderlaufender Klumpen kommt. Aufgrund der periodischen Randbedingungen bleibt im Grenzfall unendlich großer Zeiten nur ein einziger Klumpen bestehen, der sich annähernd forminvariant mit konstanter Geschwindigkeit fortbewegt.

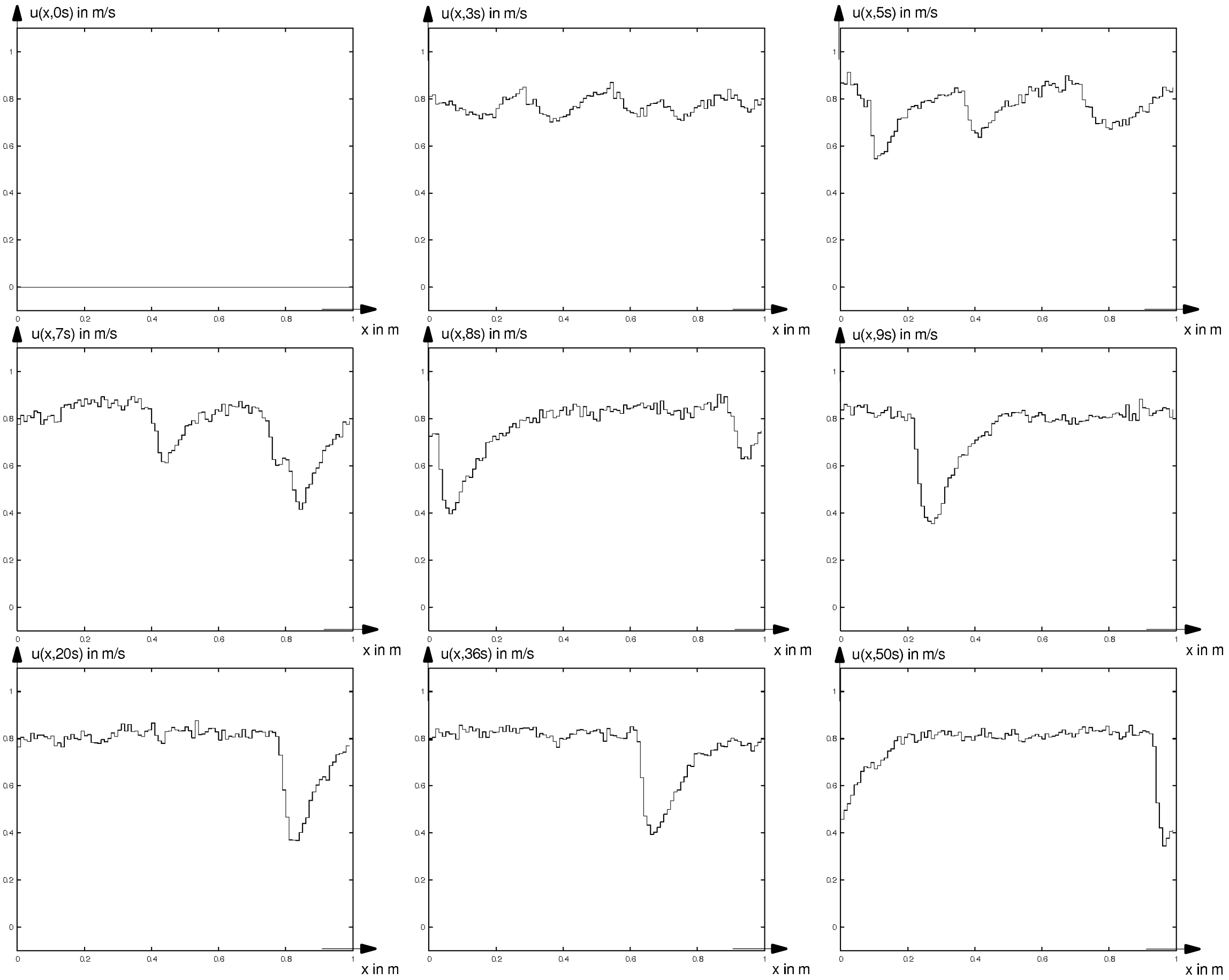

Bild 26 Mittlere Geschwindigkeit eines instabilen Systems zur Zeit 0s, 3s, 5s, 7s, 8s, 9s, 20s, 36s und 50s

Bild 26 zeigt die zu Bild 25 analogen Momentanaufnahmen der mittleren Geschwindigkeit u(x,t). Es ist deutlich zu sehen, daß ein Klumpen durch eine Gebiet mit hoher Dichte und niedriger mittlerer Geschwindigkeit ausgezeichnet ist, während zwischen zwei Klumpen annähernd freie Bewegung stattfindet, d.h. geringe Dichte bei großer Geschwindigkeit.

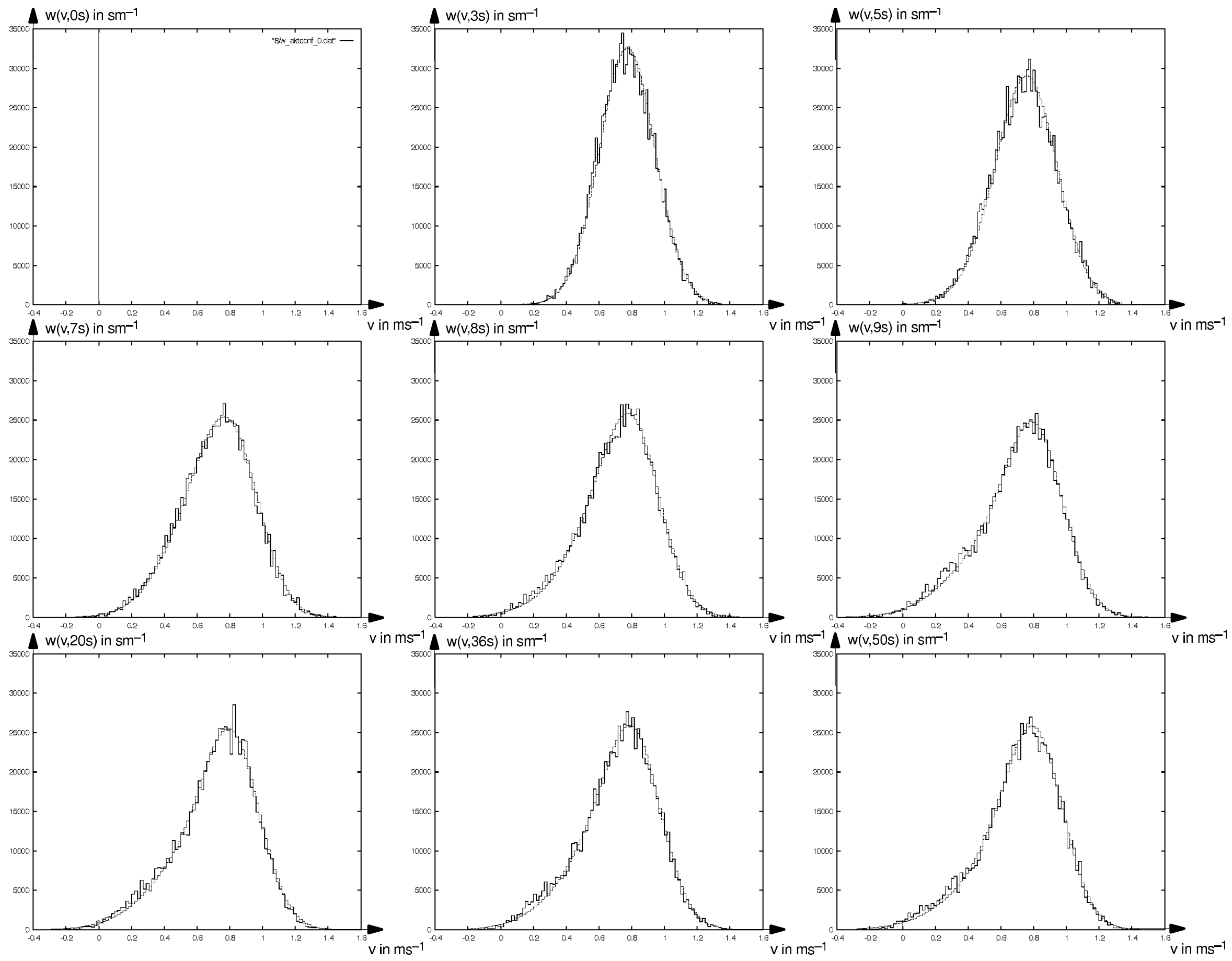

Bild 27 Geschwindigkeitsverteilung eines instabilen Systems zur Zeit 0s, 3s, 5s, 7s, 8s, 9s, 20s, 36s und 50s

Die Simulationen haben gezeigt, daß die Geschwindigkeitsverteilung w(v,t) im Grenzfall unendlicher Zeiten auch für instabile Systeme einer stationären Lösung entgegenstrebt, die jedoch keine Gaußglocke mehr ist. Bei genauer Betrachtung erkennt man, daß in jedem Diagramm von Bild 27 eigentlich zwei Kurven eingezeichnet sind, eine relativ glatte und eine etwas "zerhacktere". Die etwas "zerhacktere" Kurve ist einfach durch Abzählen der Teilchen in der entsprechenden Geschwindigkeitsbox entstanden. Für die glatte Kurve wurde angenommen, daß in jeder Box ein lokales Gleichgewicht herrscht, also jede Box ihre eigene Gaußglocke (19) besitzt. Die Überlagerung aller dieser Gaußglocken ergibt die glatte Kurve, was in Formeln so aussieht:

$$w(v) = \sum_{i=0}^{M-1} \sqrt{\frac{m}{2\pi k_B T_i}}\, n_i \frac{L}{M} \exp\left[ -\frac{m}{2k_B T_i}(v - u_i)^2 \right] \tag{192}$$

wobei $n_i$, $u_i$, $T_i$ die Dichte, mittlere Geschwindigkeit und Temperatur in der i-ten Box sind. Da beide Kurven übereinstimmen, kann die Annahme des lokalen Gleichgewichts (25) als bestätigt angesehen werden.

Für große Zeiten, d.h. wenn nur noch ein einzelner Klumpen übrig geblieben ist, kann die Geschwindigkeitsverteilung auch aus der Summe von nur zwei Gaußglocken erhalten werden. Die eine der beiden Gaußglocken stellt dabei die Geschwindigkeitsverteilung nur der Teilchen des Klumpens dar, die andere Gaußglocke ist die Geschwindigkeitsverteilung nur der Teilchen außerhalb des Klumpens.

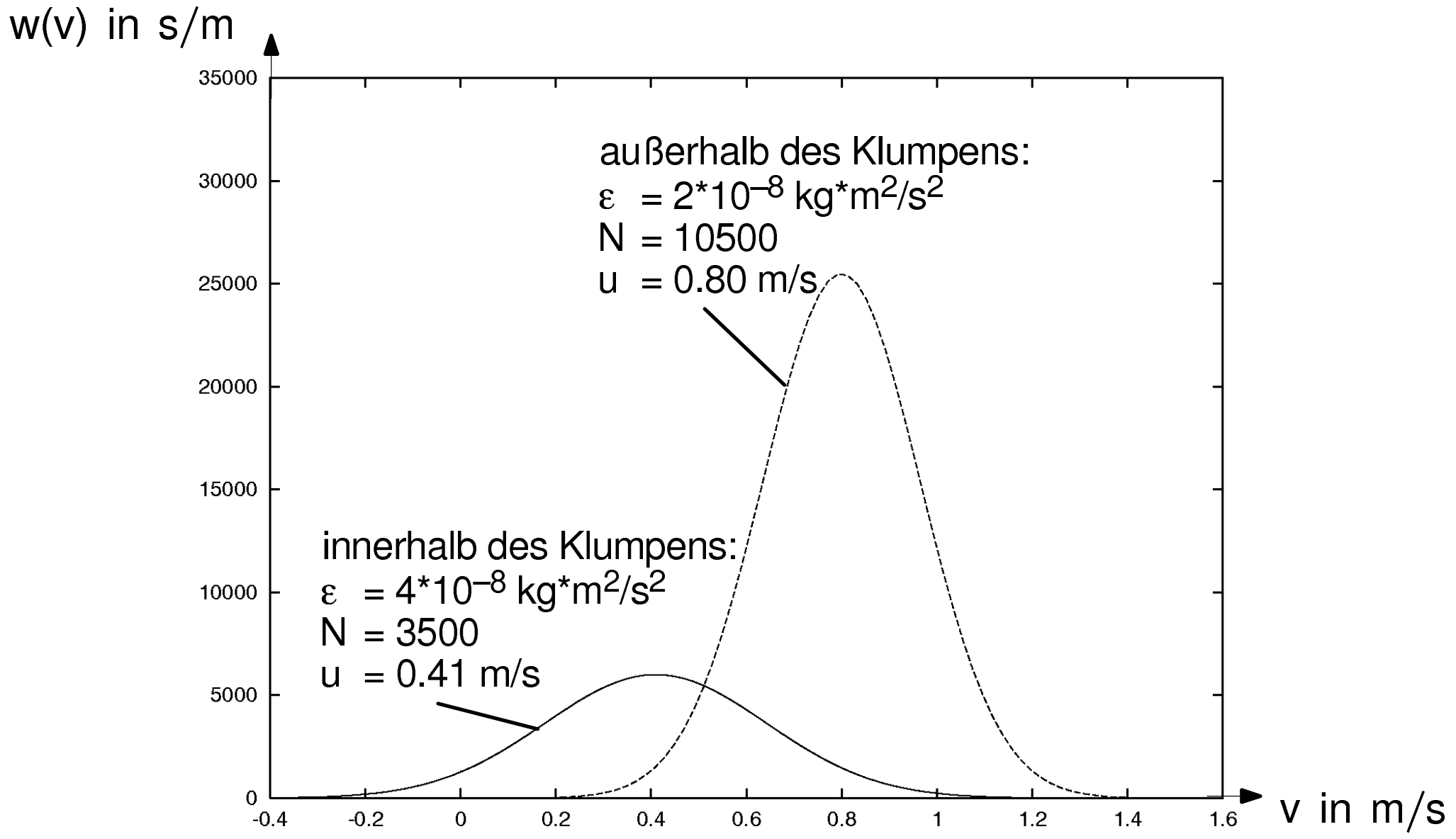

Bild 28 Unterschiedliche Geschwindigkeitsverteilungen innerhalb und außerhalb des Klumpens

Das folgende Bild vergleicht die Summe dieser beiden Gaußglocken mit der aus der Simulation erhaltenen Geschwindigkeitsverteilung.

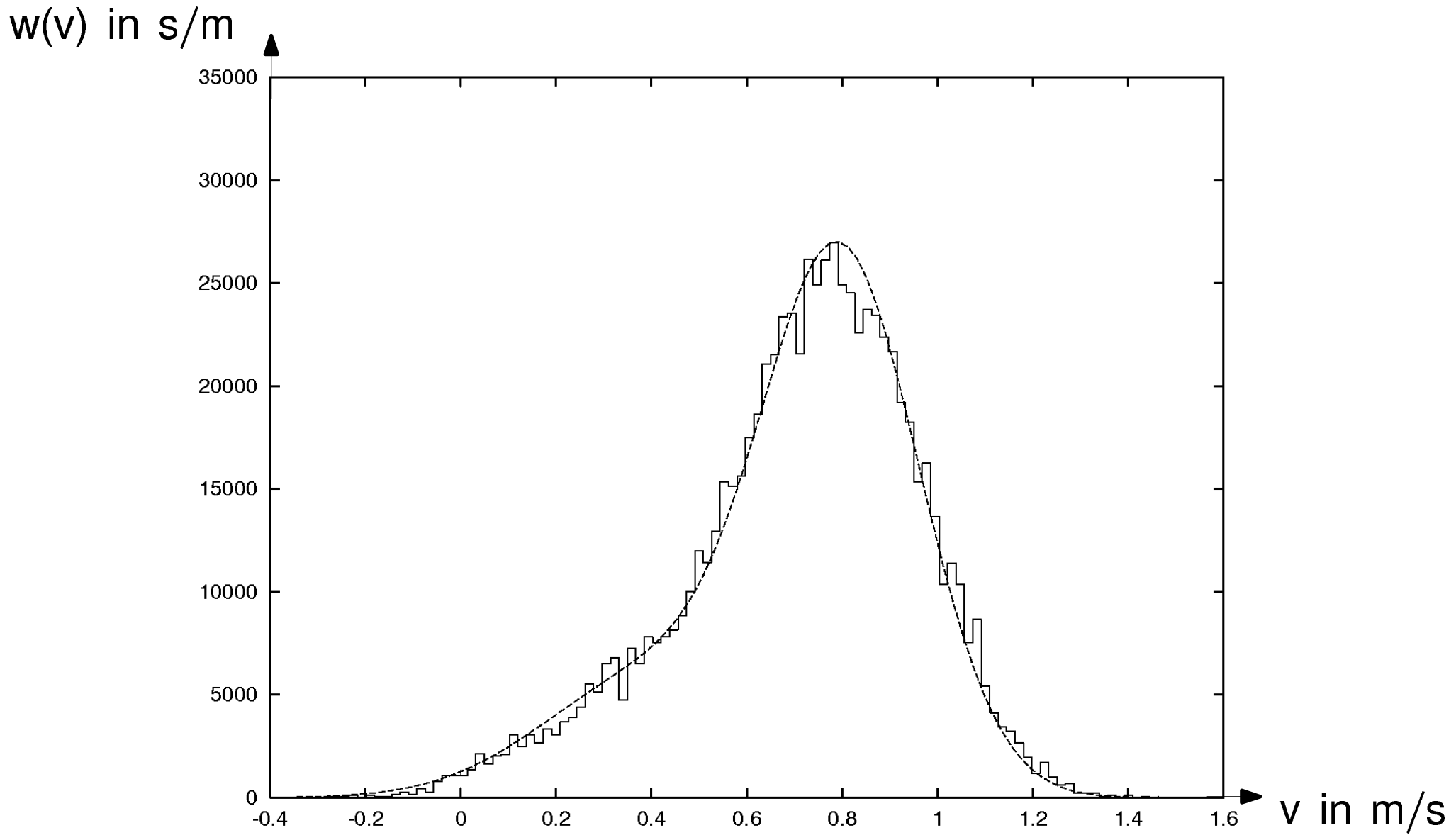

Bild 29 Summe der Gaußglocken aus Bild 28 und Geschwindigkeitsverteilung aus Simulation

Experimentelle Daten für w(v) sind z.B. in [39] zu finden:

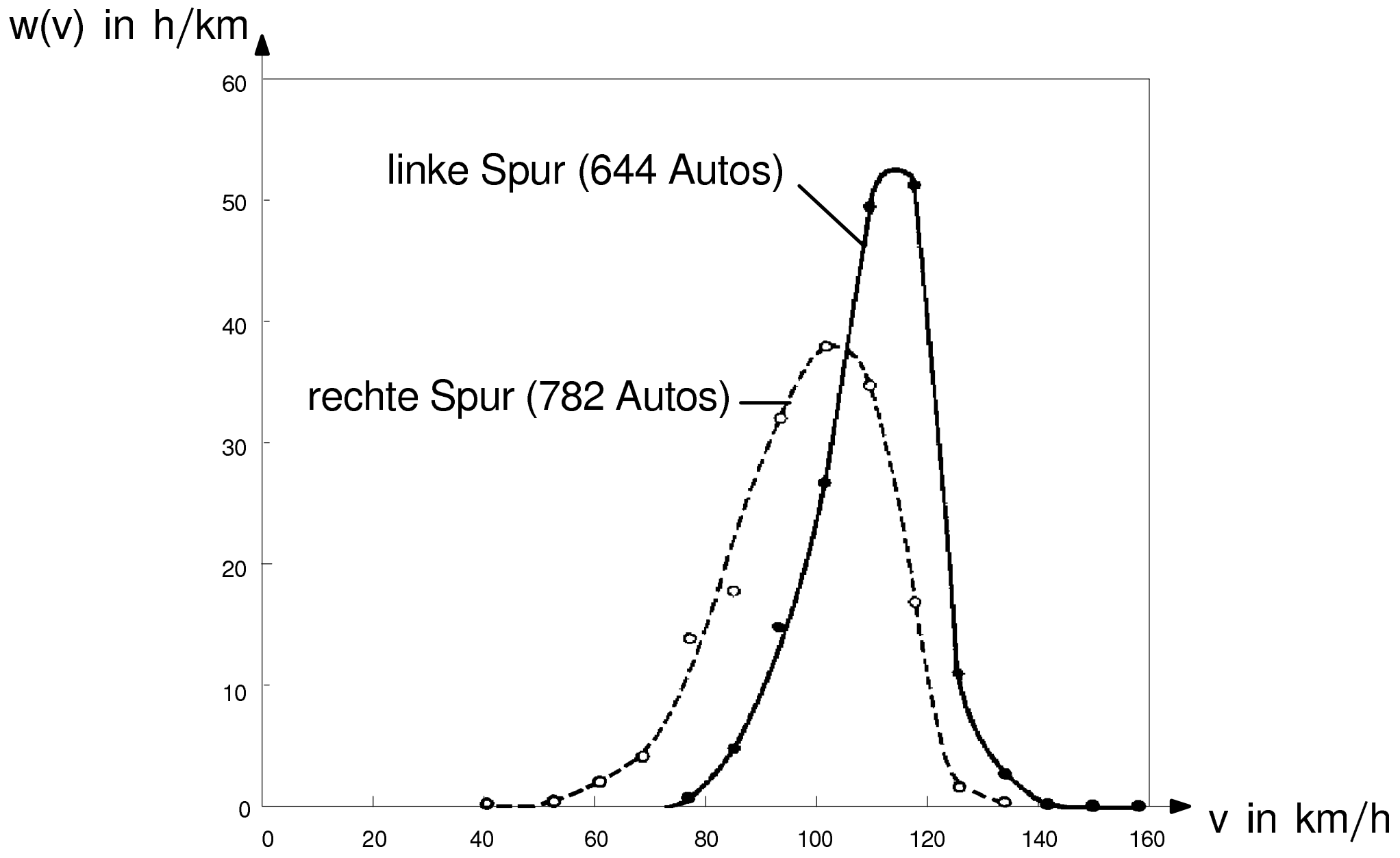

Bild 30 Experimentell ermittelte w(v)-Werte für einen Verkehrsfluß auf dem Lansing-Highway, aus [39]

Es ist zu erkennen, daß die Geschwindigkeitsverteilung je mehr einer Gaußglocke ähnelt, je geringer die Wechselwirkung zwischen den Autos ist (vergleiche linke Spur in Bild 30 und Bild 21 (50s) bei homogenen Verhältnissen). Mit Zunahme der Wechselwirkung wird w(v) immer asymmetrischer (vergleiche rechte Spur in Bild 30 und Bild 27 (50s) bei inhomogenen Verhältnissen).

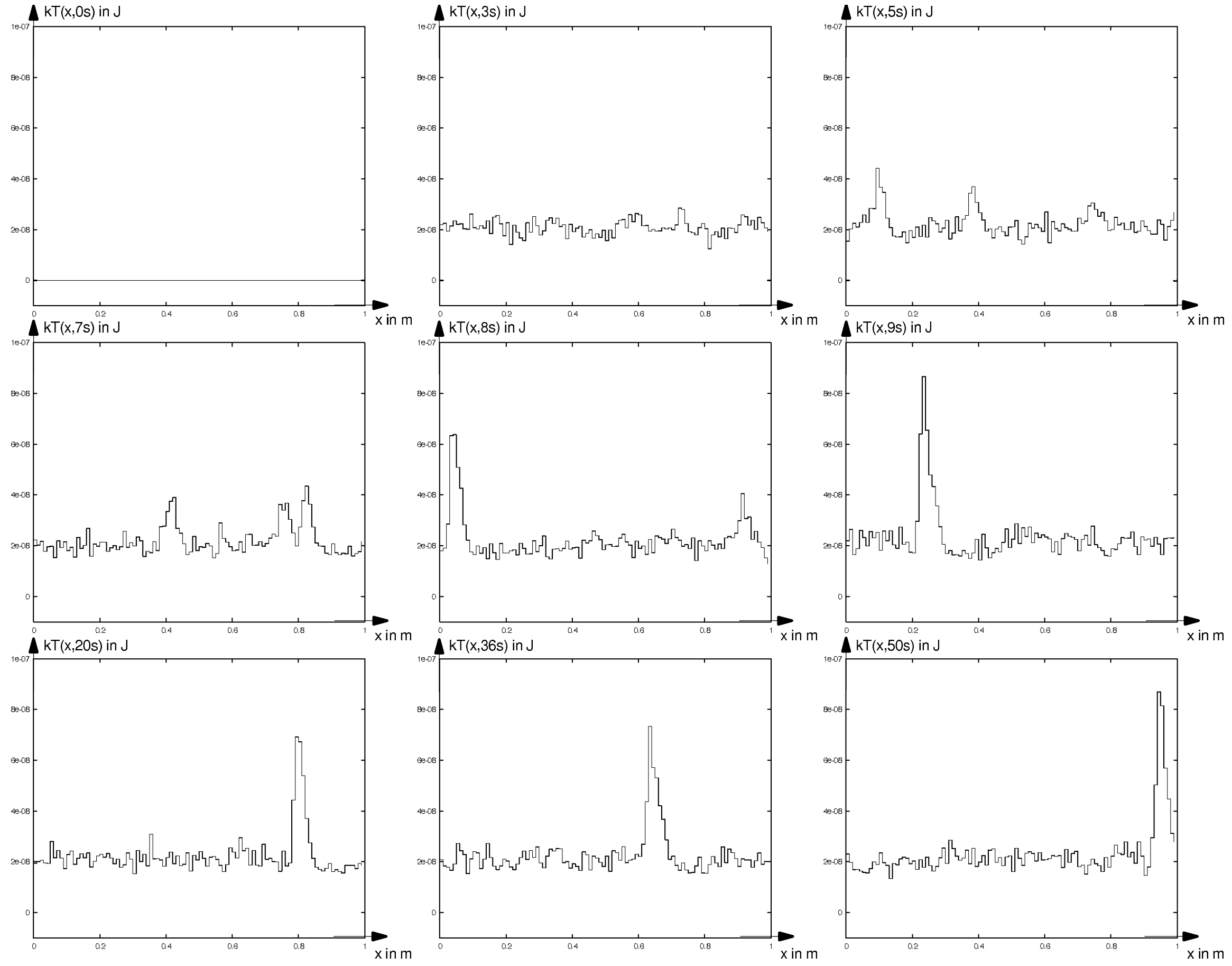

Bild 31 Mittlere thermische Energie kT eines instabilen Systems zur Zeit 0s, 3s, 5s, 7s, 8s, 9s, 20s, 36s und 50s

Obwohl in diesem Unterkapitel die thermische Energie in der Kraft F(n,T) als konstant gleich $\varepsilon$ angenommen wurde ($T \equiv T_B$), läßt Bild 31 deutliche Abweichungen von $\varepsilon$ erkennen. Das könnte ein Indiz dafür sein, daß das Modell mit temperaturabhängiger Kraft der Realität näher kommt. Bei genauem Hinsehen erkennt man, daß die thermische Energie dort deutlich größer als $\varepsilon$ ist, wo der Gradient der mittleren Geschwindigkeit negativ ist und etwas kleiner als $\varepsilon$ ist, wo der Gradient der mittleren Geschwindigkeit positiv ist (siehe Bild 26 ). Diese Tatsache bestätigt die Gültigkeit der Gleichung (54) für den Fall großer Zeiten

$$k_B T = \frac{\varepsilon}{1 + \frac{m}{\gamma}\frac{\partial u}{\partial x}} \qquad (193)$$

Daß die thermische Energie deutlich größer, aber nur etwas kleiner als $\varepsilon$ sein kann, liegt daran, daß der Gradient der mittleren Geschwindigkeit bei negativem Vorzeichen betragsmäßig sehr groß ist, aber bei positivem Vorzeichen betragsmäßig relativ klein ist (siehe Bild 26 ).

## 2.4 Homogene Anfangsbedingungen, temperaturabhängige Kraft

In diesem Unterkapitel soll wieder das geschlossene Rohr mit periodischen Randbedingungen untersucht werden. Die Kraft wird aber diesmal als temperaturabhängig angesetzt, d.h.:

$$F(n, T) = mg - Ck_BT(x, t)n(x, t) \tag{194}$$

Die Parameter C und $\varepsilon$ wurden mittels (155) aus der gemessenen kritischen Dichte $n_c \approx 9000 m^{-1}$ ermittelt und sollen auch weiterhin so verwendet werden. Da die Gleichung (155) aber nur für temperaturunabhängige Kraft gilt, ergibt sich für temperaturabhängige Kraft und beibehaltenem C und $\varepsilon$ eine andere kritische Dichte entsprechend Bild 18 . Die kleinste kritische Dichte ($n_c \approx 12000 m^{-1}$) erhält man für $z^2 \to \infty$, was aber in realen System wegen der nicht unendlich kleinen Teilchengröße unmöglich ist.

### 2.4.1 Stabiles Regime

Als erstes soll die homogene Anfangsbedingung

$$n(x, 0) = 11000 \frac{1}{m} \qquad \forall x \in [0, L] \tag{195}$$

$$u(x, 0) = 0 \qquad \forall x \in [0, L] \tag{196}$$

untersucht werden. Da $n(x,0) < n_c \approx 12000 m^{-1}$ ist, erwartet man, daß sich das System stabil verhält, also die Dichte im wesentlichen für alle Zeiten erhalten bleibt und die mittlere Geschwindigkeit den stationären Zustand (19) annimmt.

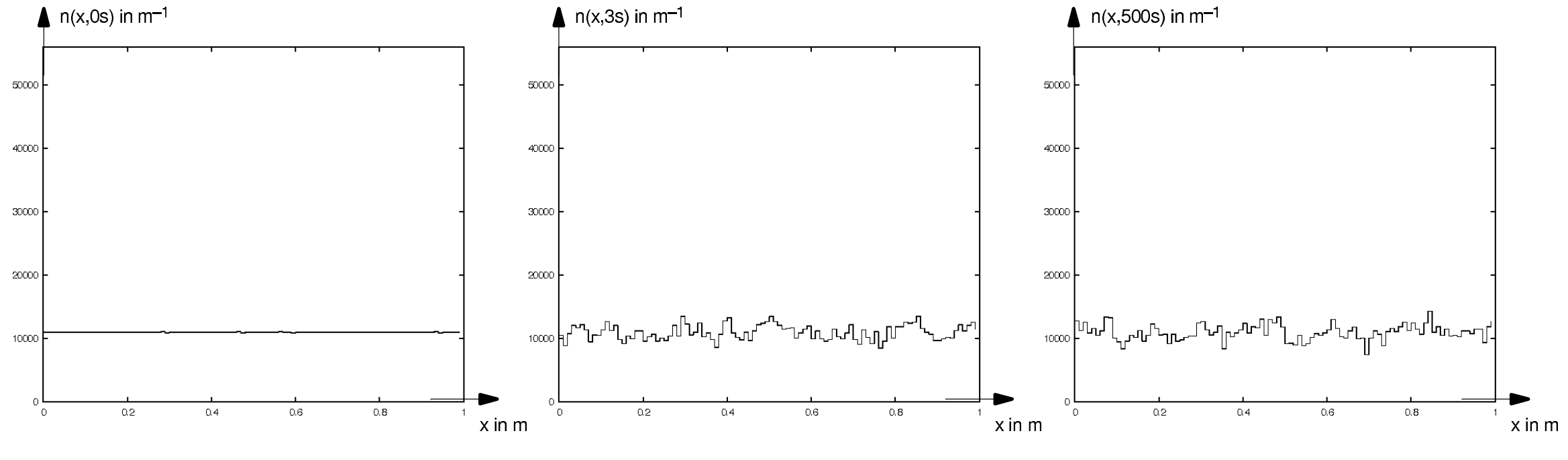

Bild 32 Dichte eines stabilen Systems zur Zeit 0s, 3s und 500s

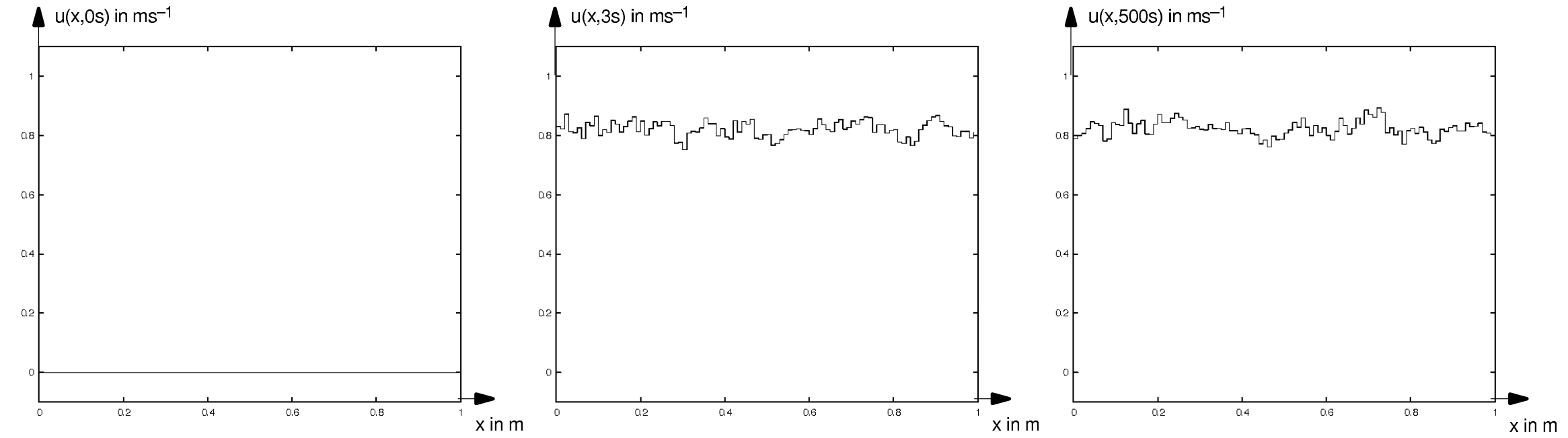

Bild 33 Mittlere Geschwindigkeit eines stabilen Systems zur Zeit 0s, 3s und 500s

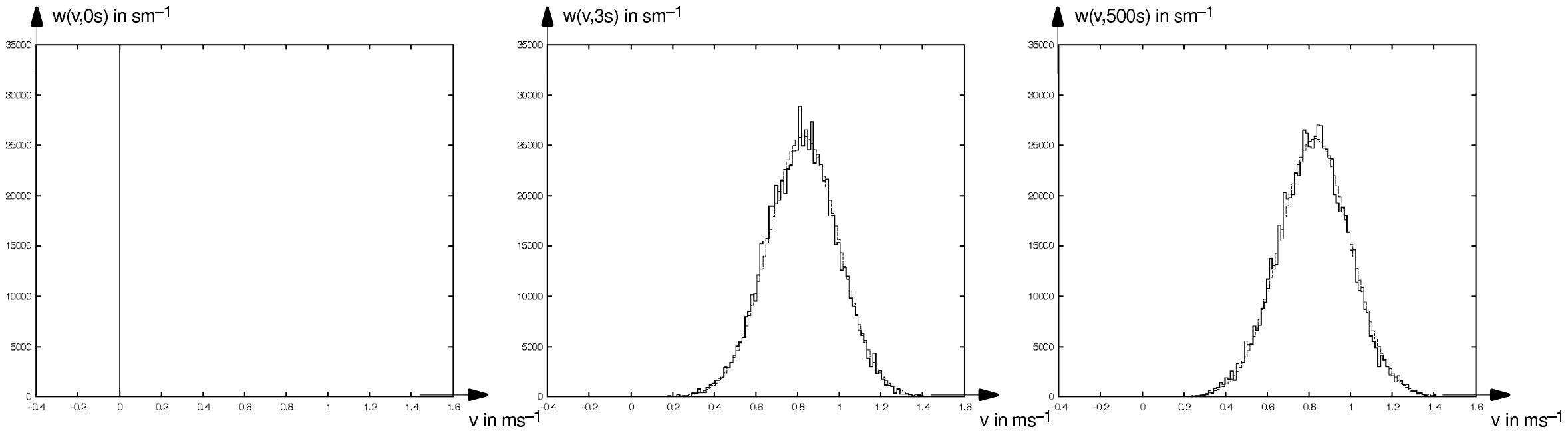

Bild 34 Geschwindigkeitsverteilung eines stabilen Systems zur Zeit 0s, 3s und 500s

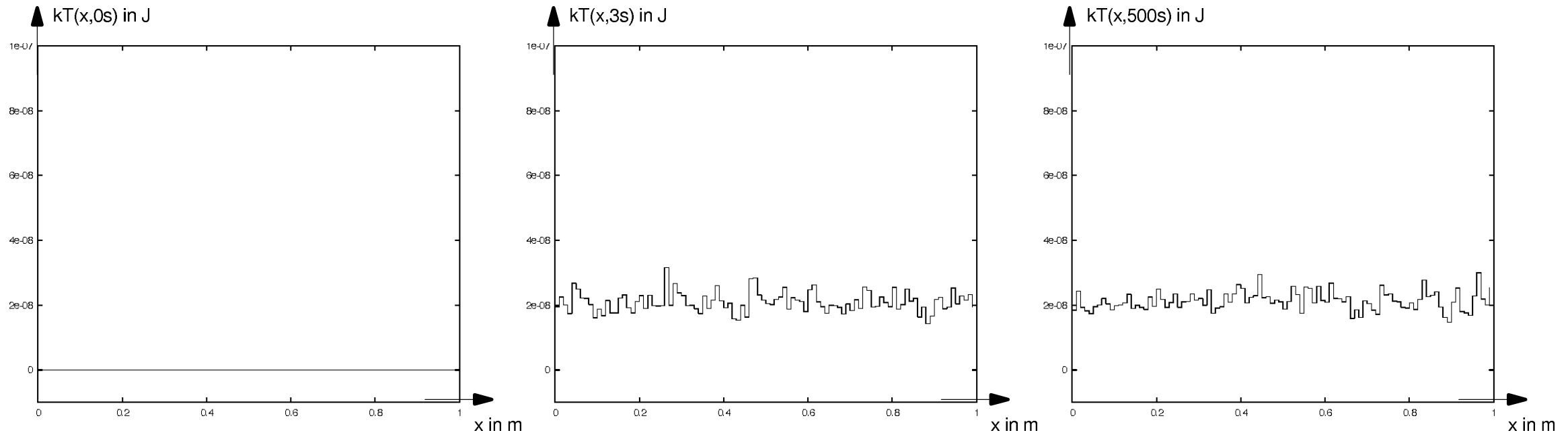

Bild 35 Mittlere thermische Energie kT eines stabilen Systems zur Zeit 0s, 3s und 500s

Die Bilder 32 - 35 zeigen deutliche Übereinstimmung zwischen Theorie und Simulation. Es sei noch einmal explizit darauf hingewiesen, daß bei unverändertem Parametersatz die Anfangsdichte $n_h$=11000$m^{-1}$ im Falle der temperaturunabhängigen Kraft zu einer Klumpenbildung führt, aber im Falle der temperaturabhängigen Kraft die Verhältnisse homogen bleiben.

### 2.4.2 Instabiles Regime

Jetzt sollen die folgenden Anfangsbedingungen betrachtet werden:

$$n(x,0) = 14000\frac{1}{m} \quad \forall x \in [0,L] \tag{197}$$

$$u(x,0) = 0 \quad \forall x \in [0,L] \tag{198}$$

Diesmal ist $n(x,0) > n_c$, d.h. es ist ein instabiles Verhalten zu erwarten. Die Simulation ergab folgende Bilder:

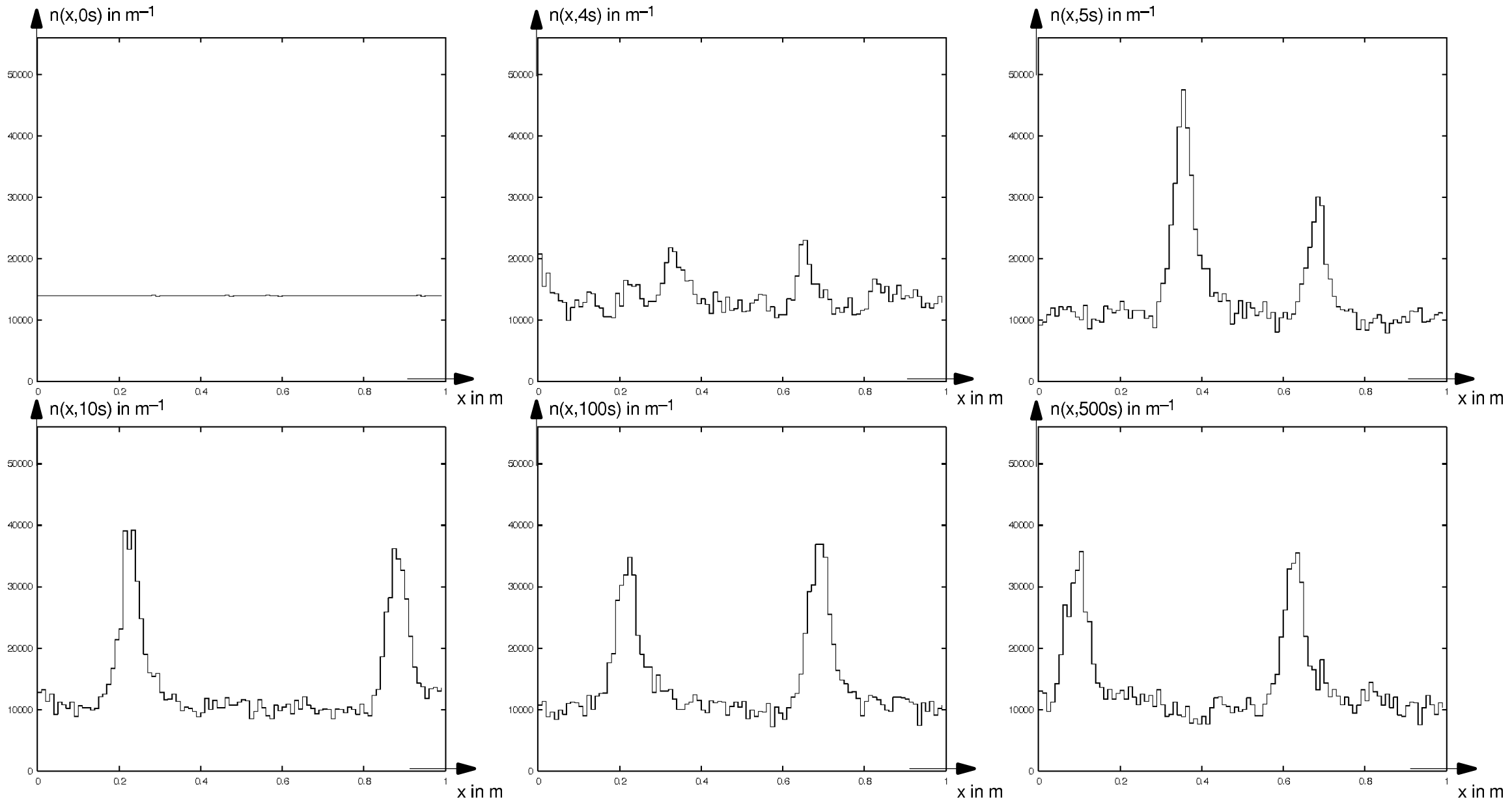

Bild 36 Dichte eines instabilen Systems zur Zeit 0s, 4s, 5s, 10s, 100s und 500s

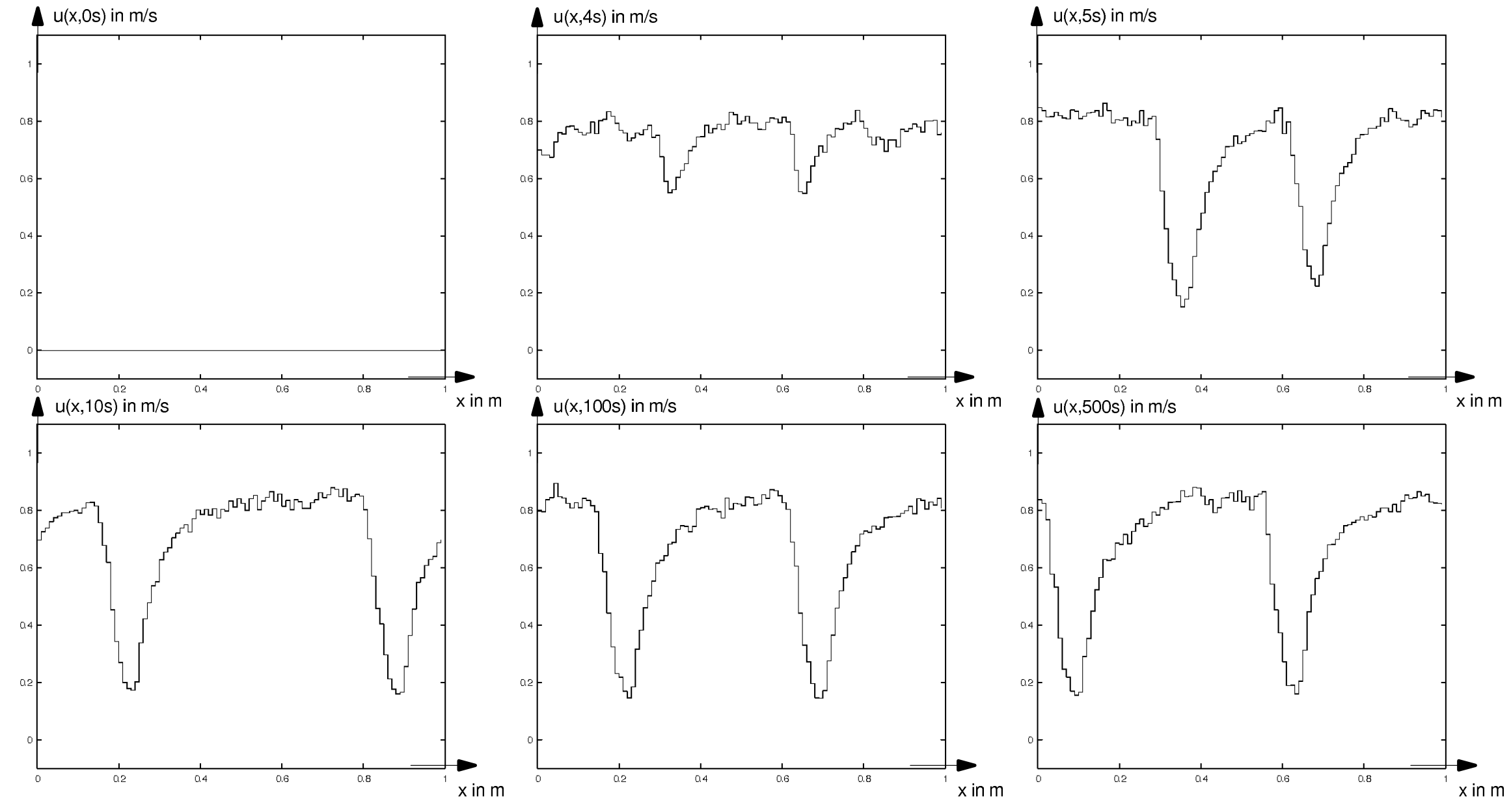

Bild 37 Mittlere Geschwindigkeit eines instabilen Systems zur Zeit 0s, 4s, 5s, 10s, 100s und 500s

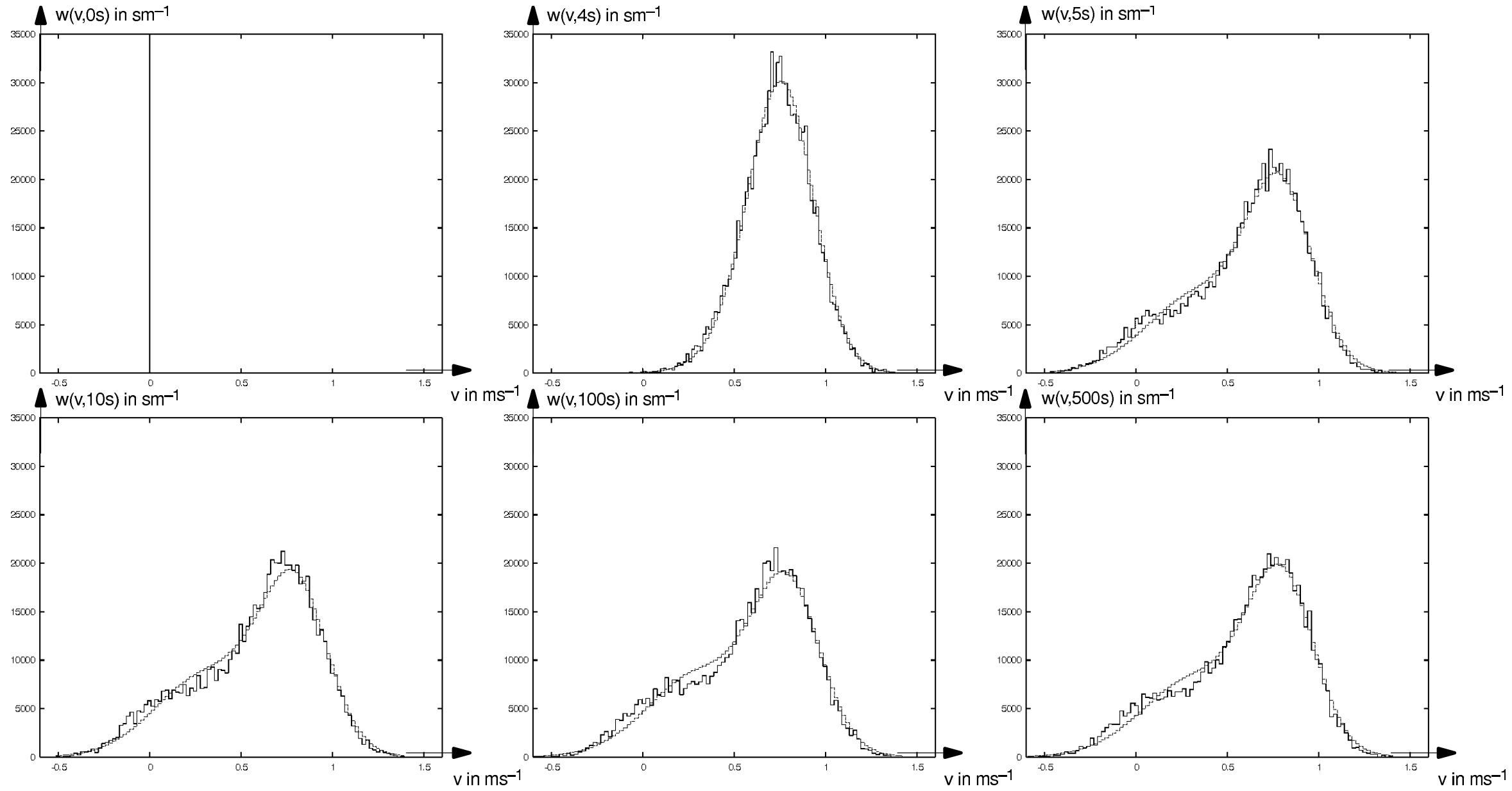

Bild 38 Geschwindigkeitsverteilung eines instabilen Systems zur Zeit 0s, 4s, 5s, 10s, 100s und 500s

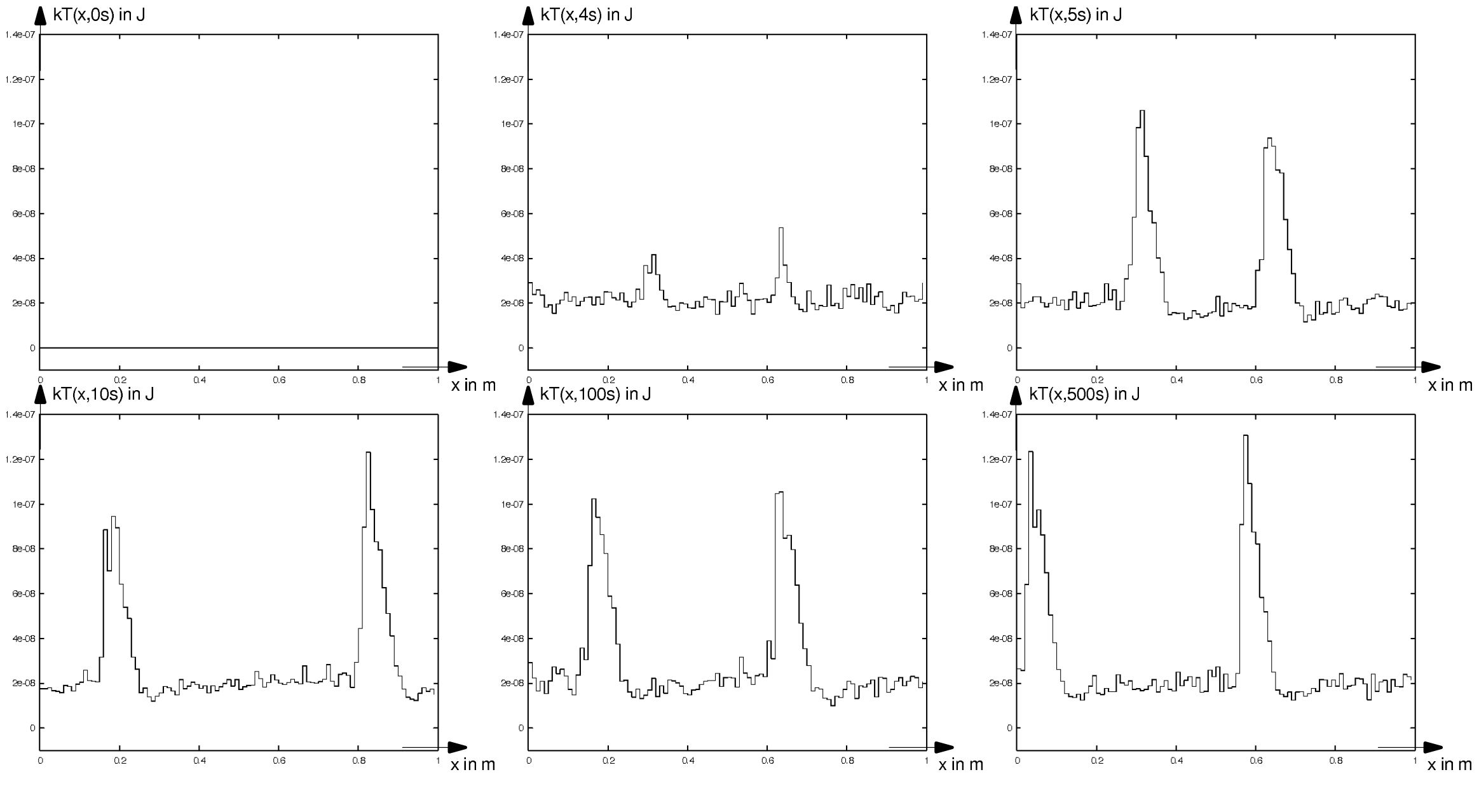

Bild 39 Mittlere thermische Energie kT eines instabilen Systems zur Zeit 0s, 4s, 5s, 10s, 100s und 500s

Wie erwartet kommt es zu Verklumpungserscheinungen. Dieselbe Simulationsrechnung wurde mit anderen Zufallszahlen $\xi(t)$ mehrmals durchgeführt. Für große Zeiten (t=500s) blieb manchmal ein großer Klumpen übrig, meistens zwei und manchmal auch drei. Dies ist ein Unterschied zur Simulation der temperaturunabhängigen Kraft, wo immer genau ein Klumpen übrig blieb. Die Ursache dafür liegt in der unterschiedlichen Monotonie der $n_c(k)$-Kurven (vergleiche Bild 14 mit Bild 18 ). Bild 18 läßt erkennen, daß die kritische Dichte mit zunehmender Wellenzahl sinkt, somit wird die Herausbildung mehrerer Klumpen gefördert. Da für immer größer werdende Dichten auch Terme höherer Ordnung, die in der angeführten linearen Stabilitätsanalyse nicht berücksichtigt wurden, wichtig werden, kommt es nicht zur Ausbildung unendlich vieler Klumpen, sondern nur einiger weniger. Die Ausbildung unendlich vieler Klumpen ist in der Simulation wegen der Teilchenzahlerhaltung und der Boxenunterteilung bei der Dichtebestimmung sowieso nicht möglich.

# 3 Experimente

## 3.1 Experimenteller Aufbau

In wieweit die vorgeschlagenen Langevin-Gleichungen das reale Verhalten granularer Ströme wiedergeben, läßt sich nur durch das Experiment zeigen. Es wird folgender Versuchsaufbau vorgeschlagen.

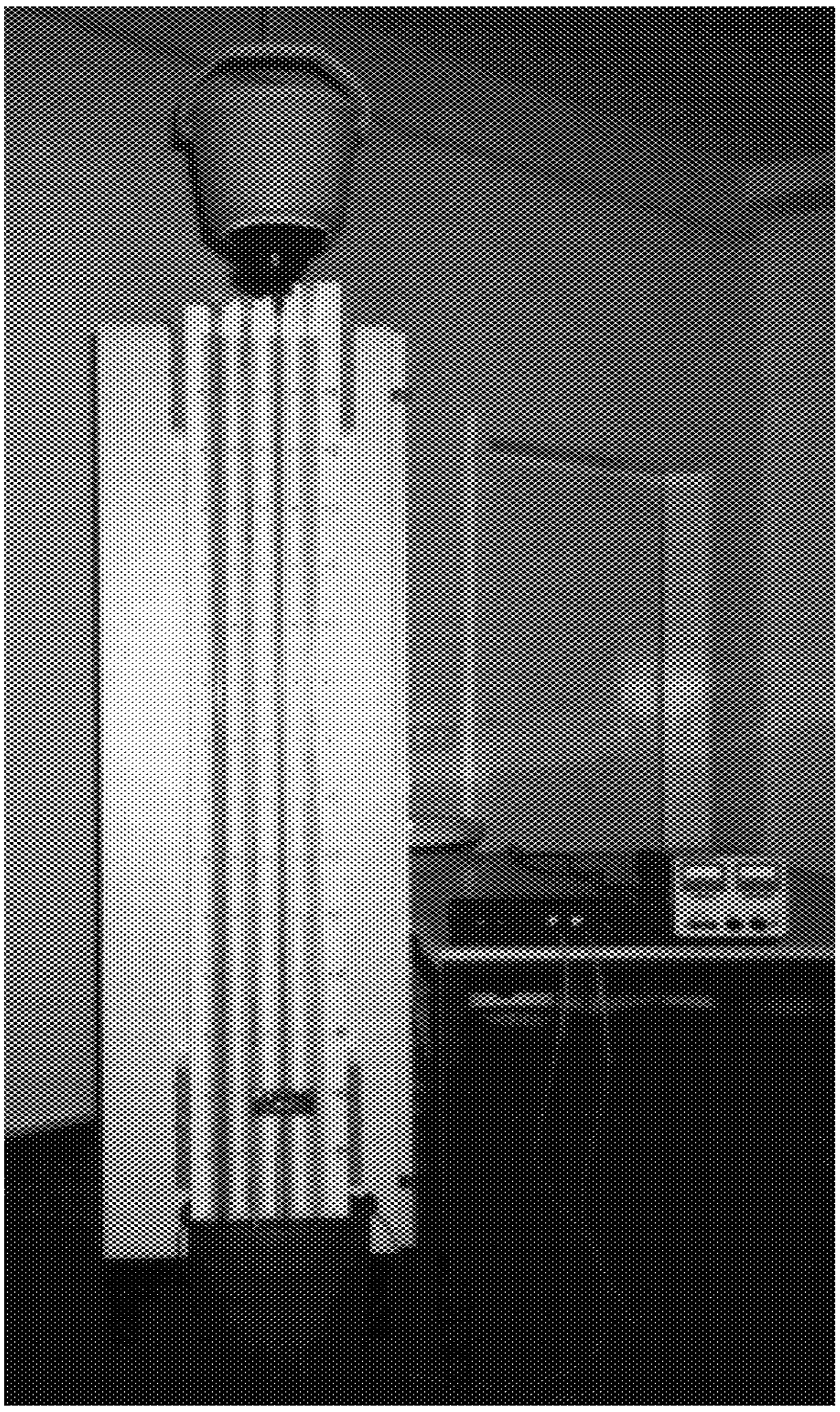

Bild 40 Experimentelle Anordnung

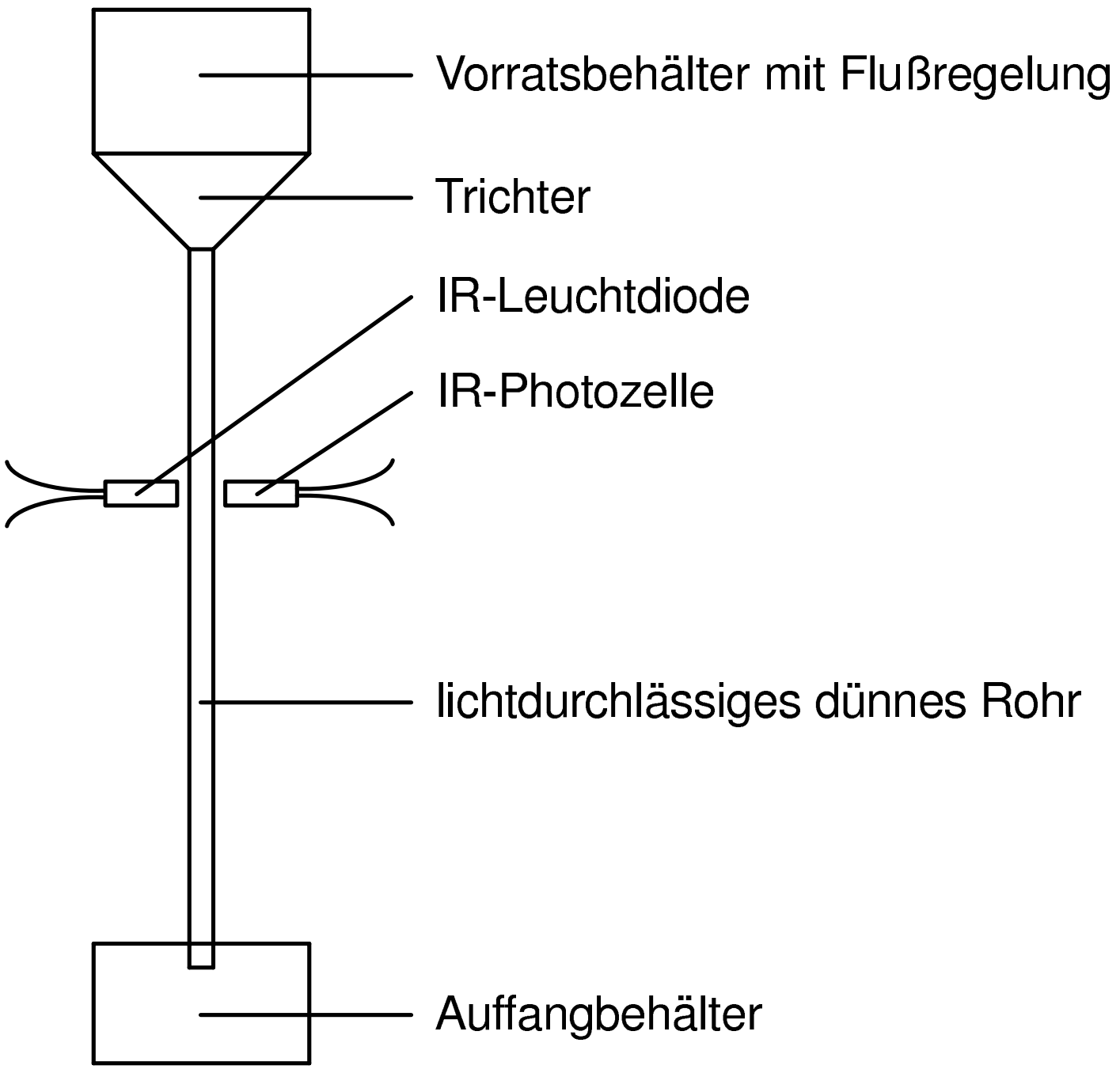

Bild 41 Experimenteller Aufbau (schematisch)

Als granulares Medium wurde Sand in 3 verschiedenen Korngrößen verwendet. Diese 3 Korngrößen sollen im folgenden als feine, mittlere bzw. grobe Körnung bezeichnet werden. Über einen Trichter gelangt der Sand aus einem Vorratsbehälter in eines der senkrecht angebrachten Rohre. Der Durchmesser des Rohres muß sehr viel kleiner als seine Länge sein ($d << L$). Die Dichte des Sandes im Rohr soll mittels Photozelle bestimmt werden. Aus diesem Grunde muß das Rohrmaterial lichtdurchlässig sein. Zum Einsatz kamen 4 Gummischläuche mit den Innendurchmessern 4mm, 6mm, 8mm und 10mm. Zwei längs der Schläuche angebrachte Schienen verhindern ein Verbiegen und Vibrieren der Rohre. Die Entfernung zwischen Trichter und Photozelle kann in 10cm-Schritten variiert werden.

## 3.2 Verwendete Meßtechnik

Wie oben erwähnt, wird die Dichte des Sandes an einer bestimmten Rohrstelle mit einer Photozelle gemessen. Es wurde der Infrarot-Phototransistor SFH 309 in Verbindung mit der Infrarot-Leuchtdiode SFH 409 eingesetzt. Das Signal des Phototransistors kann entweder direkt (Kanal 1) oder nach Verstärkung durch die folgende Schaltung (Kanal 0) einer Meßkarte zugeführt werden. Durch die Verstärkerschaltung hat man die Möglichkeit selbst kleinste Dichteunterschiede aufzulösen. So ist es z.B. möglich, in einem sehr, sehr dünnen Sandfluß einzelne Sandkörner zu unterscheiden.

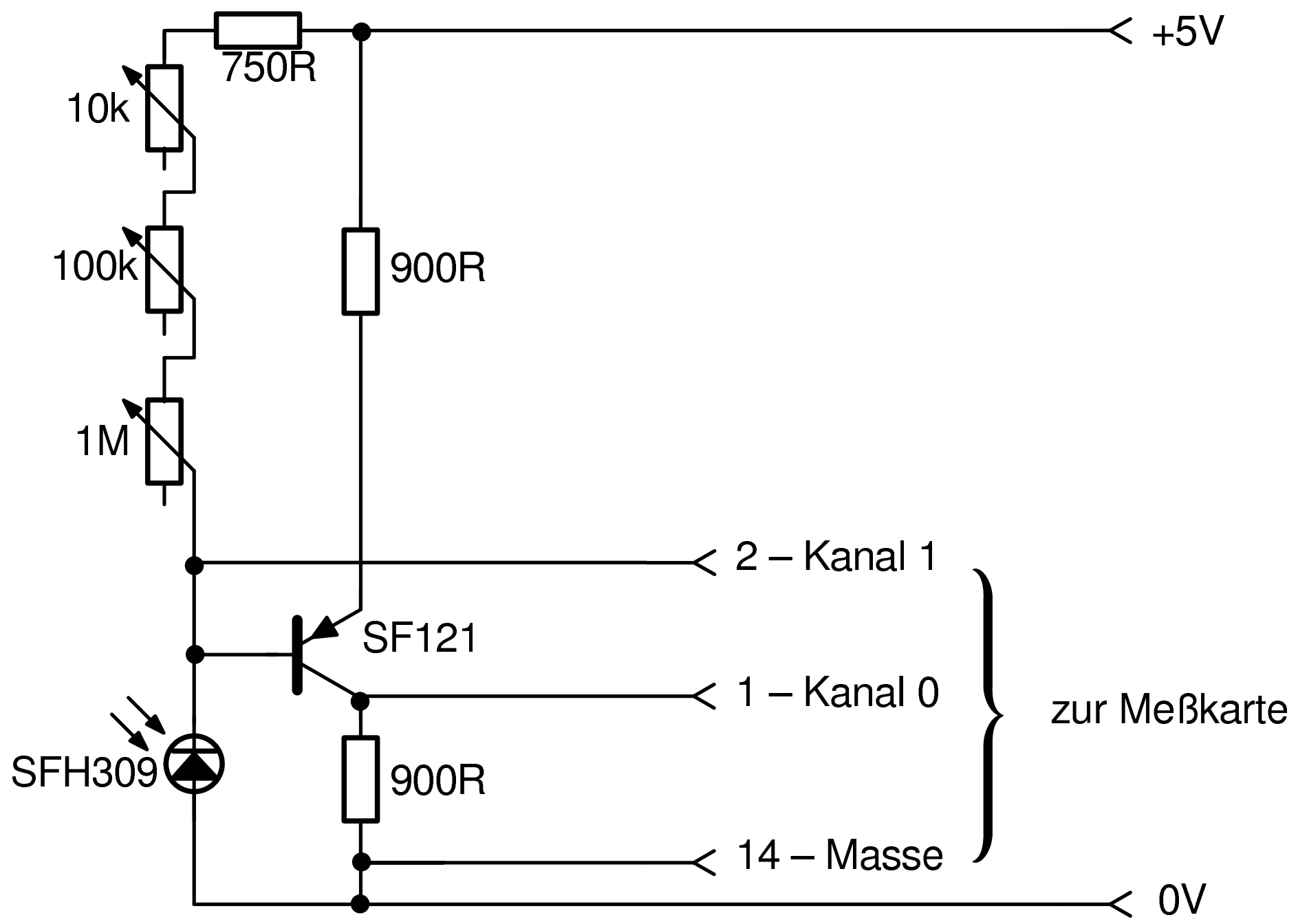

Bild 42 Verstärkerschaltung für das Signal des Phototransistors

Als Meßkarte wurde eine KFA 8I verwendet; diese besitzt 8 AD-Wandler, welche mit einer Frequenz von bis zu 80kHz abgefragt werden können und eine Auflösung von 12bit besitzen. Der Einsatz einer Meßkarte ermöglicht die Erfassung und Weiterverarbeitung der Daten durch einen Rechner.

Der Zusammenhang zwischen Spannung der Schaltung und Sanddichte im Rohr ist streng monoton, aber nicht notwendig linear. Die drei Potentiometer dienen der genauen Arbeitspunkteinstellung.

## 3.3 Suche nach Verklumpungserscheinungen

In der Literatur sind einige Experimente zum Sandfluß im dünnen Rohr zu finden, siehe z.B. [8], [13]. Ein Bild von Pöschel [8] sei hier gezeigt.

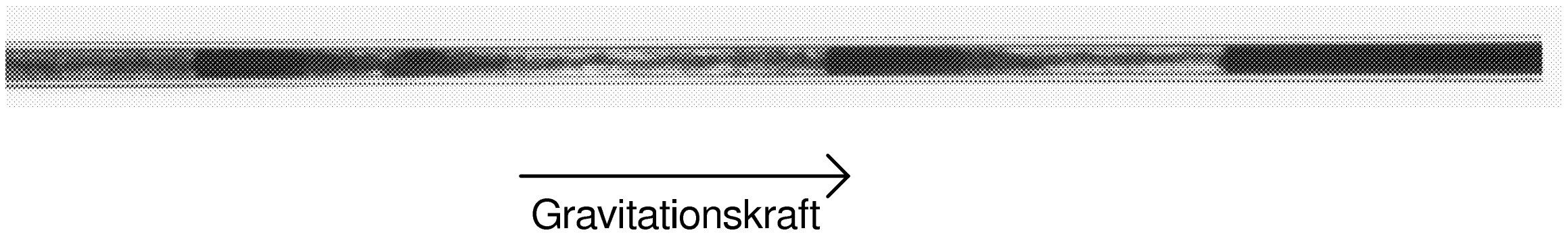

Bild 43 Beispiel für Sandfluß im Rohr aus [8]

Nicht jede Kombination von Sandkörnigkeit und Rohrdurchmesser führt zu Verklumpungserscheinungen. Deshalb muß als erstes eine geeignete Kombination gefunden werden. Es hat sich gezeigt, daß die deutlichsten Verklumpungserscheinungen bei der Verwendung einer mittleren Körnung des Sandes und eines Rohrinnendurchmessers von 6mm auftreten. Bemerkt werden muß allerdings, daß auch bei dieser Kombination die Verklumpungserscheinungen nicht ganz so deutlich ausgeprägt waren wie in Bild 43 . Das hat vermutlich seine Ursache darin, daß der durch den Trichter am oberen Rohrende ermöglichte Sandfluß nicht groß genug ist, so daß die im Rohr erreichte mittlere Dichte nicht deutlich genug über der kritischen Dichte liegt. Andere Ursachen könnten z.B. sein, daß die Innenwand des Rohres nicht rauh genug ist oder daß das Rohr nicht lang genug ist. Später wird sich nämlich zeigen, daß die Klumpen mit wachsendem Abstand vom Trichter immer seltener, aber dafür intensiver werden.

## 3.4 Bestimmung der Parameter

In diesem Kapitel sollen verschiedene Parameter eines granularen Flusses ermittelt werden. Alle Angaben beziehen sich auf eine mittlere Körnung des Sandes und einen Rohrinnendurchmesser von 6mm. Aufgrund der stark eingeschränkten experimentellen Möglichkeiten im Rahmen dieser Arbeit, soll kein Anspruch auf sehr genaue Messungen erhoben werden. Es geht vielmehr darum, eine grobe Abschätzung der Parameter zu erhalten, um damit möglichst realitätsnahe Simulationsrechnungen durchführen zu können.

### 3.4.1 Bestimmung der mittleren Masse m eines Sandkorns

Es wurden 1806 Sandkörner abgezählt und anschließend gewogen. Es ergab sich:

$$m = \frac{(1,34 \pm 0,02)\mathrm{g}}{1806 \pm 100} = (7,4 \pm 0,6) * 10^{-7}\mathrm{kg} \tag{199}$$

### 3.4.2 Bestimmung der Staudichte $n_{ST}$

Eine gewisse Menge Sand wurde gewogen, eine Volumenbestimmung erfolgte mit einem Meßzylinder. Es ergab sich folgende Massendichte:

$$\varrho_{ST} = \frac{(473 \pm 4)\mathrm{g}}{(300 \pm 10)\mathrm{cm}^3} = (1580 \pm 70)\frac{\mathrm{kg}}{\mathrm{m}^3} \tag{200}$$

Es hat sich gezeigt, daß die Herstellerangabe von 6mm für den Rohrinnendurchmesser sehr ungenau ist. Das Rohrvolumen V soll deshalb nicht aus der Rohrgeometrie, sondern wie folgt bestimmt werden. Das Rohr wird auf einer Länge von L=1,50m mit Sand gefüllt, anschließend wird der Sand gewogen. Man erhält:

$$V = \frac{m}{\varrho_{ST}} = \frac{(62 \pm 2)g}{(1,58 \pm 0,07)\frac{g}{cm^3}} = (39 \pm 3)cm^3 \qquad (201)$$

Für den Rohrquerschnitt A ergibt sich:

$$A = \frac{V}{L} = \frac{(39 \pm 3)cm^3}{(150 \pm 1)cm} = (2,6 \pm 0,2) * 10^{-5}m^2 \qquad (202)$$

Mit Kenntnis der Teilchenmasse m und des Rohrquerschnittes A läßt sich die dreidimensionale Massendichte in eine eindimensionale Teilchenzahldichte umrechnen:

$$n_{ST} = \frac{\varrho_{ST} * A}{m} = (56000 \pm 13000)\frac{1}{m} \qquad (203)$$

### 3.4.3 Bestimmung des Reibungskoeffizienten $\gamma$

In der Langevin-Gleichung (37) kann der letzte Term für sehr kleine Dichten vernachlässigt werden, also:

$$m\dot{v}_i = mg - \gamma v_i + \sqrt{2\varepsilon\gamma}\,\xi_i(t) \qquad (204)$$

Durch Mittelung von (204) erhält man:

$$m\langle\dot{v}\rangle = mg - \gamma\langle v\rangle \qquad (205)$$

Im stationären Falle ist $\langle\dot{v}\rangle = 0$ , es ist also:

$$\gamma = \frac{mg}{\langle v\rangle} \qquad (206)$$

$\langle v\rangle$ läßt sich als Quotient aus Teilchenstrom J und Teilchendichte $\varrho$ ermitteln. Man realisiere dazu einen sehr dünnen Fluß, damit die Bedingung kleiner Dichten erfüllt ist und messe, wieviel Masse Sand in einer bestimmten Zeit das Rohr durchfließt:

$$J = \frac{\Delta m}{\Delta t} = \frac{(523 \pm 2)g}{(254 \pm 1)s} = (2,06 \pm 0,02)\frac{g}{s} \qquad (207)$$

Bei unverändertem Fluß J werden oberes und unteres Rohrende gleichzeitig verschlossen. Der im Rohr verbleibende Sand wird gewogen, die Rohrlänge L ist bekannt. Es ergibt sich:

$$\varrho = \frac{m}{L} = \frac{(2,8 \pm 0,6)g}{(1,50 \pm 0,01)m} = (1,9 \pm 0,4)\frac{g}{m} \qquad (208)$$

Daraus folgt:

$$\langle v\rangle = \frac{J}{\varrho} = (1,1 \pm 0,3)\frac{m}{s} \qquad (209)$$

Gleichung (209) gilt nur als Näherung, da sich der Sand im oberen Rohrteil noch nicht mit der stationären Geschwindigkeit, sondern beschleunigt bewegt.

Schließlich ergibt sich der Reibungskoeffizient zu:

$$\gamma = (7 \pm 2) * 10^{-6} \frac{kg}{s} \qquad (210)$$

### 3.4.4 Bestimmung der kritischen Dichte $n_C$

Es wird der maximale Teilchenstrom eingestellt, der gerade noch keine Klumpenbildung zuläßt. Wieder werden oberes und unteres Rohrende gleichzeitig verschlossen und die Sandmenge im Rohr gewogen.

$$\varrho_C = (250 \pm 60) \frac{kg}{m^3} \qquad (211)$$

Das entspricht einer Teilchenzahldichte von:

$$n_C = (9000 \pm 4000) \frac{1}{m} \qquad (212)$$

## 3.5 Messung der zeitlichen Abstände zweier aufeinanderfolgender Klumpen

Die Verklumpung tritt nicht überall im Rohr auf, sondern es ist so, daß für kleine Abstände vom Trichter (Einschüttstelle) der Sandfluß annähernd laminar ist und eine Verklumpung erst in einiger Entfernung vom Trichter einsetzt. Deshalb wird die Photozelle möglichst weit unten am Rohr plaziert. Die Samplefrequenz, mit der die Spannung der Photozelle abgefragt wird, muß einerseits hinreichend groß sein, um den zeitlichen Verlauf eines einzelnen Klumpens noch gut auflösen zu können, sie darf aber andererseits aus Speicherplatzgründen nicht zu groß gewählt werden.

Verwendet man eine mittlere Körnung des Sandes, einen Rohrinnendurchmesser von 6mm, einen Abstand von der Einschüttstelle von 1,33m und eine Samplefrequenz von 5000Hz, so sieht der zeitliche Verlauf der Spannung am Kanal 1 für eine Zehntel Sekunde typischerweise so aus:

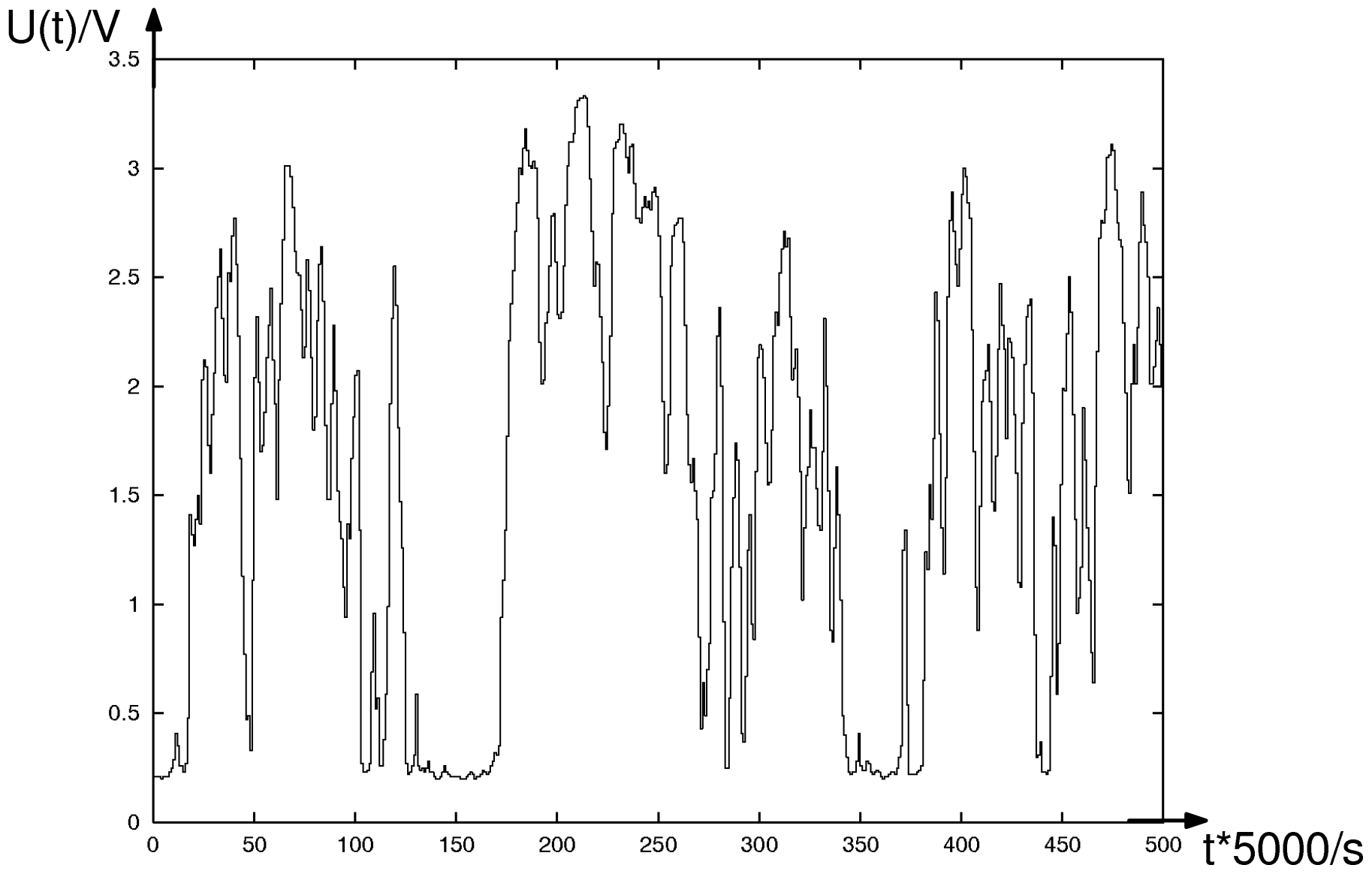

Bild 44 Spannung als Funktion der Zeit für 0.1s (0.1s = 500 Zeitschritte bei f = 5kHz)

Da das Spannungssignal sehr "zerhackt" ist, läßt sich schlecht entscheiden, was Klumpen ist und was nicht. Deshalb wird das Signal mit einem Tiefpaß gefiltert. Der Tiefpaß funktioniert dabei so, daß jeder Spannungswert auf den Mittelwert seiner 20 Nachbarwerte gesetzt wird (10 Nachbarn links, 10 Nachbarn rechts). Das so aus Bild 44 erhaltene Signal sieht wie folgt aus:

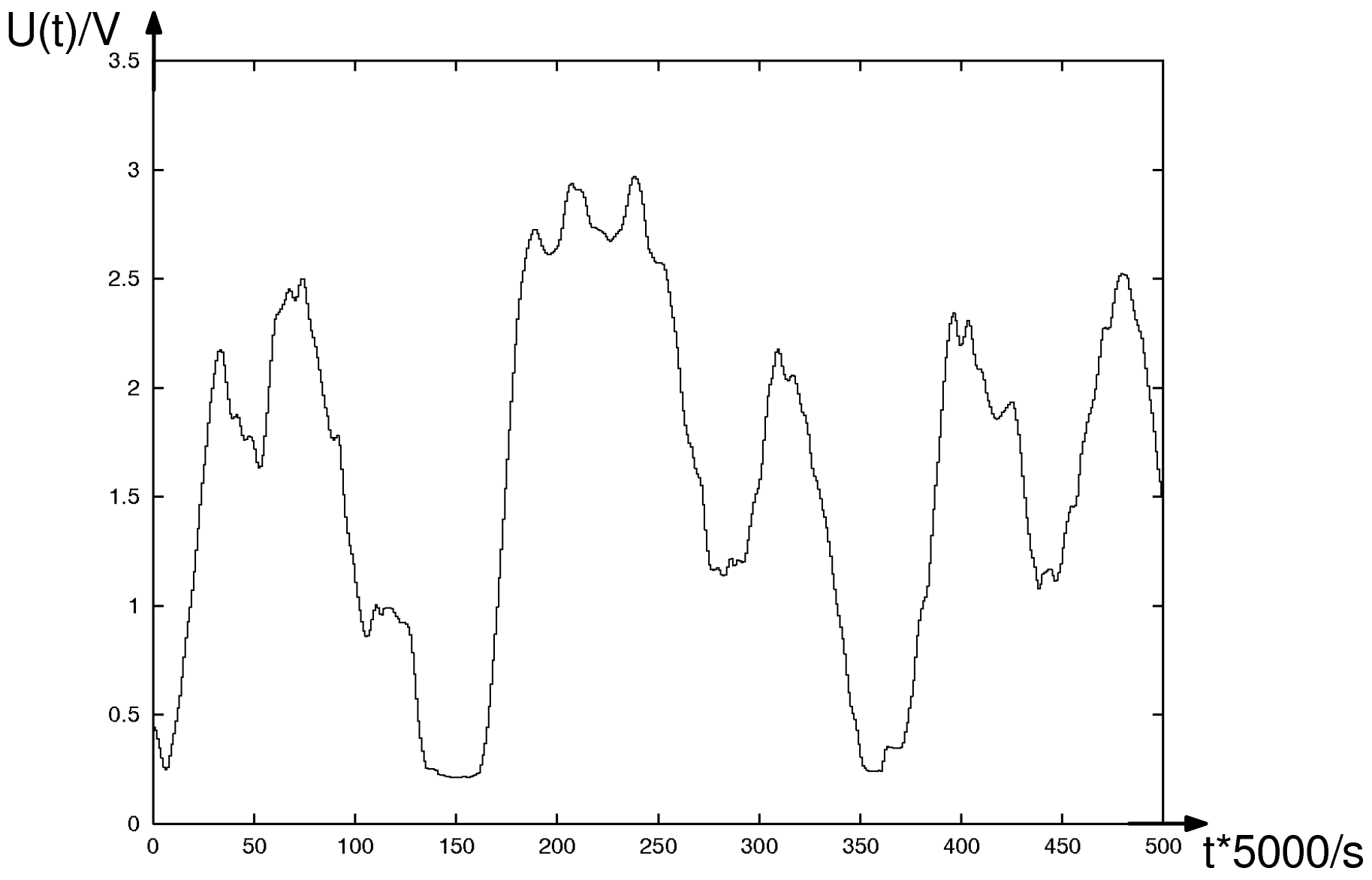

Bild 45 Zu Bild 44 gehörige gefilterte Spannung als Funktion der Zeit

Aus diesem relativ "weichen" Verlauf der Sanddichte muß nun die duale Information "Klumpen" oder "kein Klumpen" extrahiert werden. Dazu wird folgendes Verfahren benutzt. Ein Klumpen beginnt, wenn die Spannung über einen bestimmten Schwellwert (z.B. 1V) steigt und er endet, wenn die Spannung wieder unter diese Schwelle sinkt. Anschließend wird der Mittelwert aus Beginn- und Endzeit berechnet und die Differenz zweier aufeinanderfolgender Mittelwerte ist der gesuchte zeitliche Abstand $\tau$. (Siehe Bild 46 )

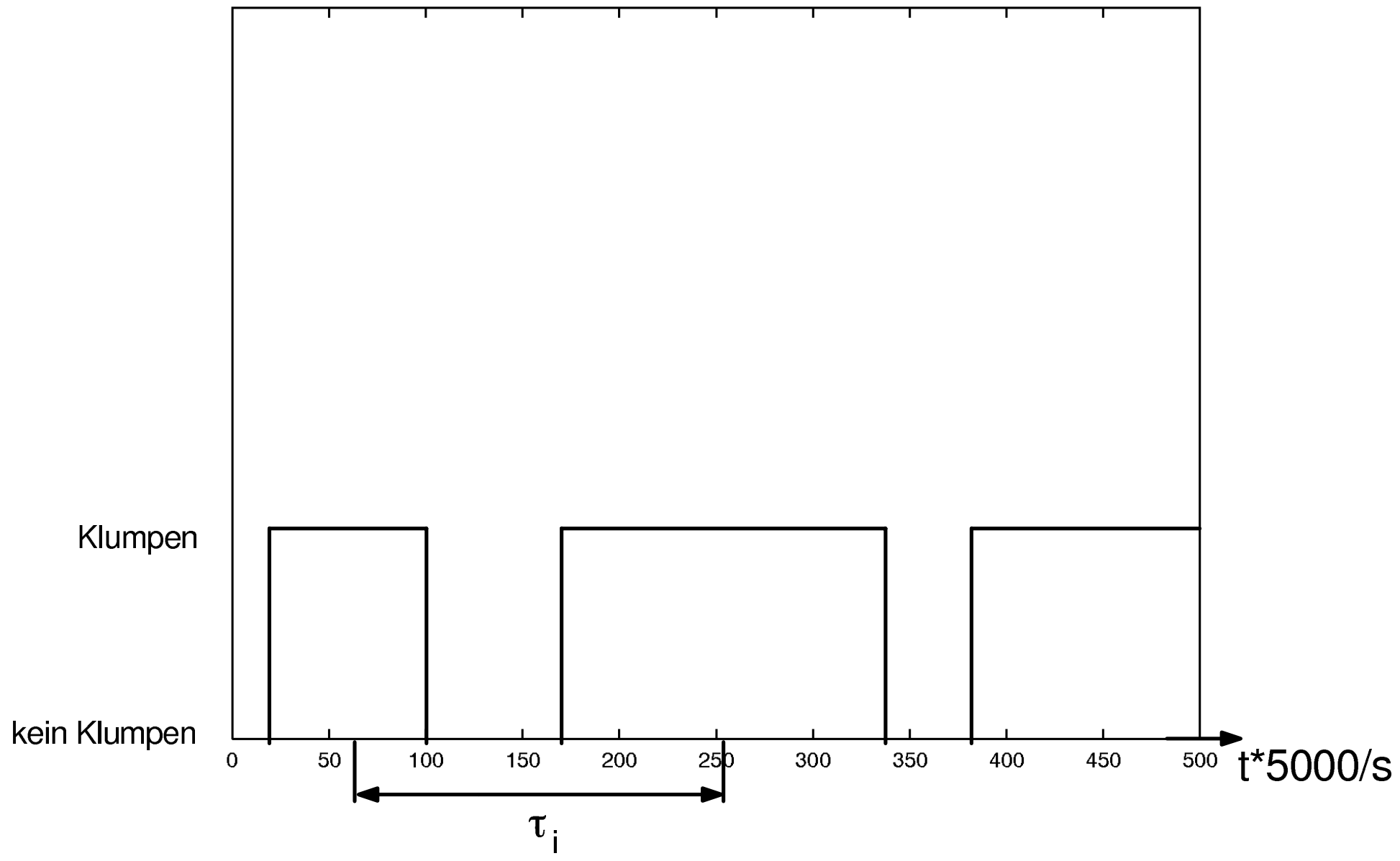

Bild 46 Zu Bild 45 gehörige duale Information mit Zeitabständen

An drei verschiedenen Rohrstellen wurde das Spannungssignal jeweils 60s lang aufgenommen. Anschließend wurden nach dem eben beschriebenen Verfahren die zeitlichen Abstände zweier aufeinanderfolgender Klumpen bestimmt und deren Häufigkeit ermittelt.

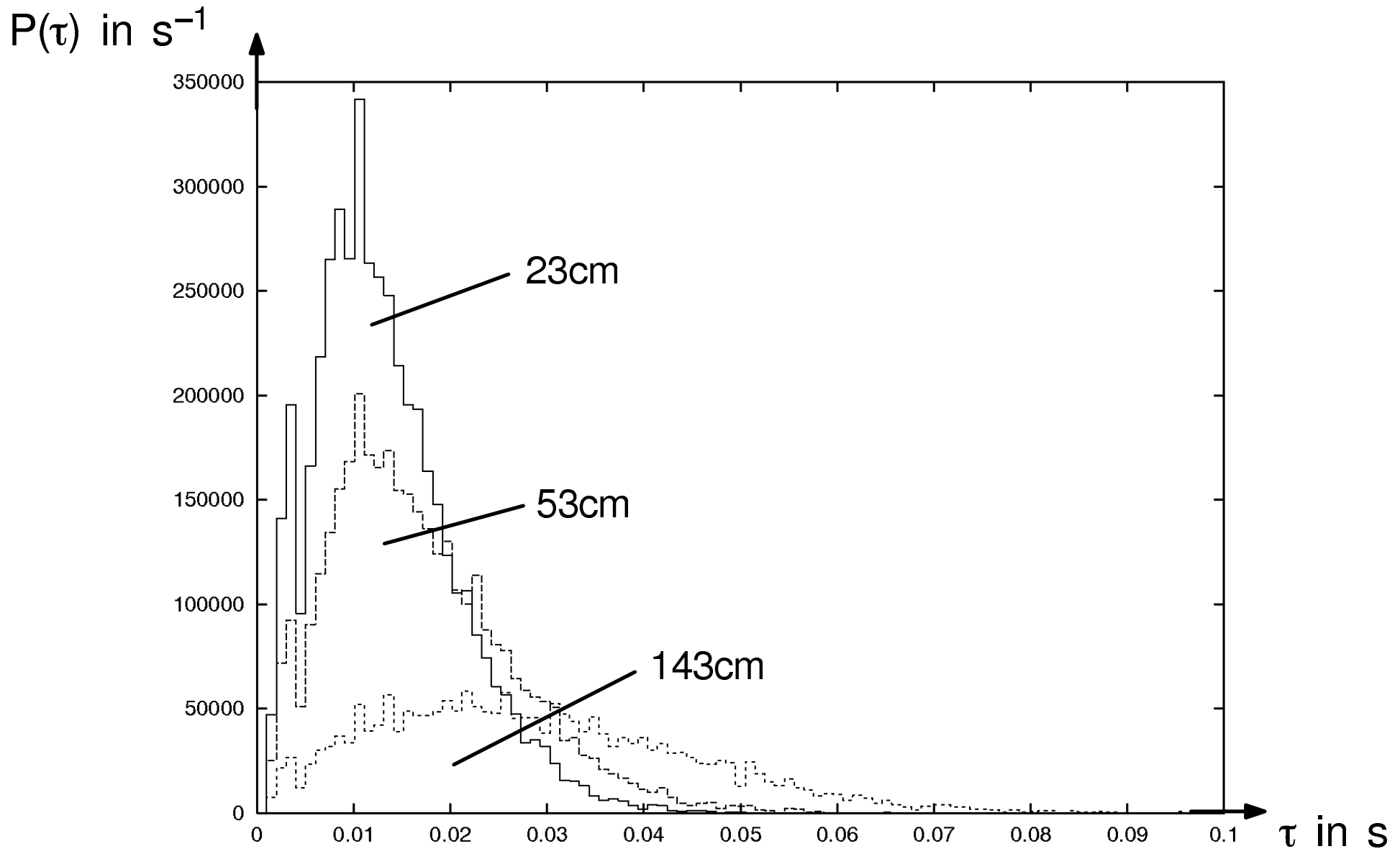

Bild 47 Häufigkeit der zeitlichen Abstände zweier aufeinanderfolgender Klumpen für 23cm, 53cm und 143cm Entfernung vom Trichter

Man erkennt, daß mit zunehmendem Abstand vom Trichter der mittlere Zeitabstand zwischen zwei aufeinanderfolgende Klumpen immer größer wird.

## 3.6 Messung der Anzahl Klumpen pro Zeit an verschiedenen Rohrstellen

Anhand des Signals der Photozelle können die Klumpen an einer bestimmten Rohrstelle gezählt werden. (Diese Anzahl entspricht dem Integral der Kurven aus Bild 47 .) Die Bilder 48 und 49 zeigen deutlich, daß die Zahl der Klumpen pro Minute mit zunehmender Entfernung vom Trichter abnimmt. Der erste Meßpunkt liegt noch im Bereich der Klumpenentstehung, deshalb schlägt der dort gemessene Wert aus der Reihe und besitzt eine sehr große Streuung. Dieses Ergebnis deckt sich mit den Erkenntnissen aus den Simulationsrechnungen. Dort liefen die einzelnen Klumpen mit der Zeit auch immer mehr zusammen, so daß die Anzahl der Klumpen abnimmt.

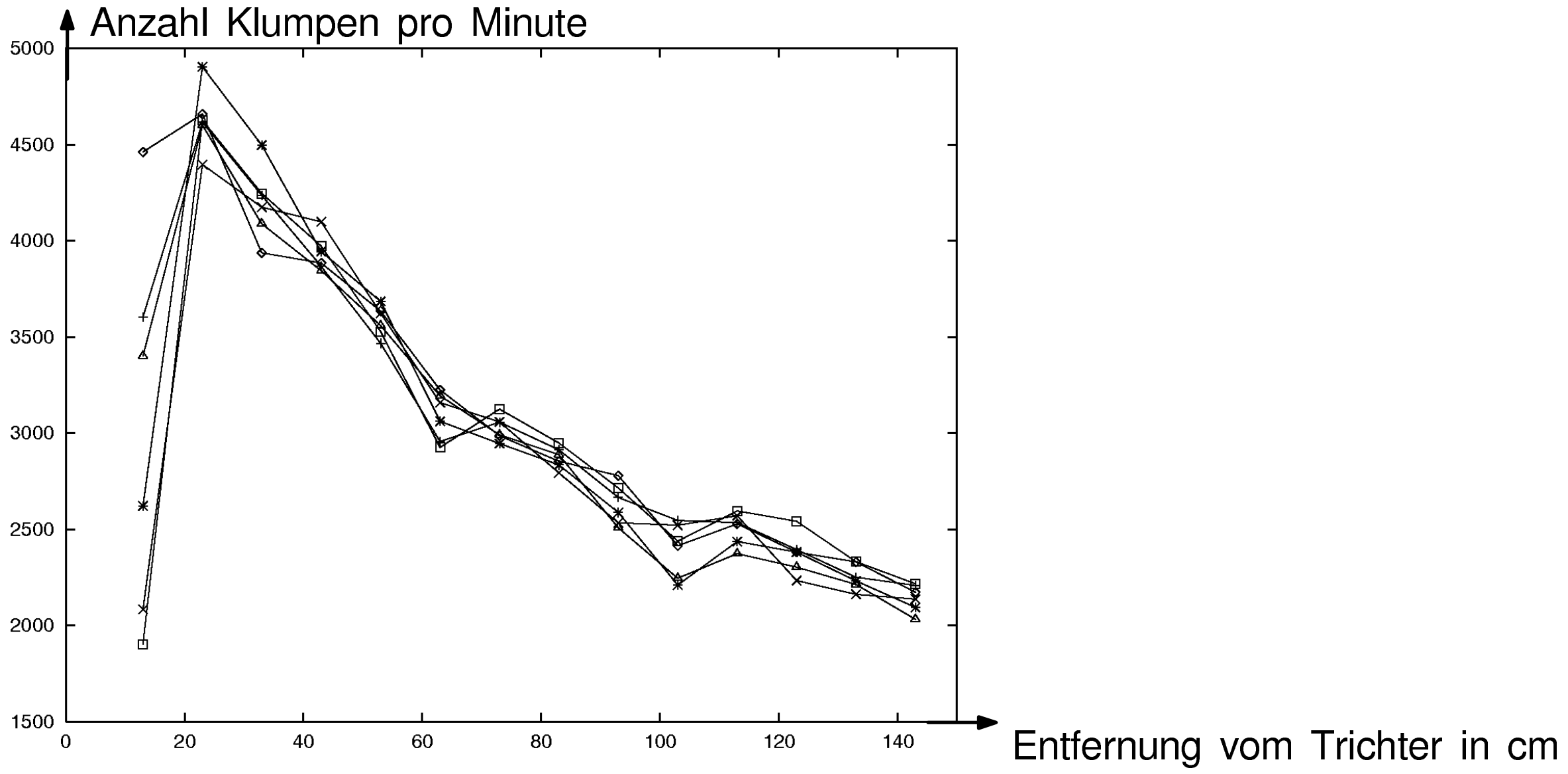

Bild 48 Anzahl Klumpen pro Minute in Abhängigkeit vom Abstand zur Einschüttstelle

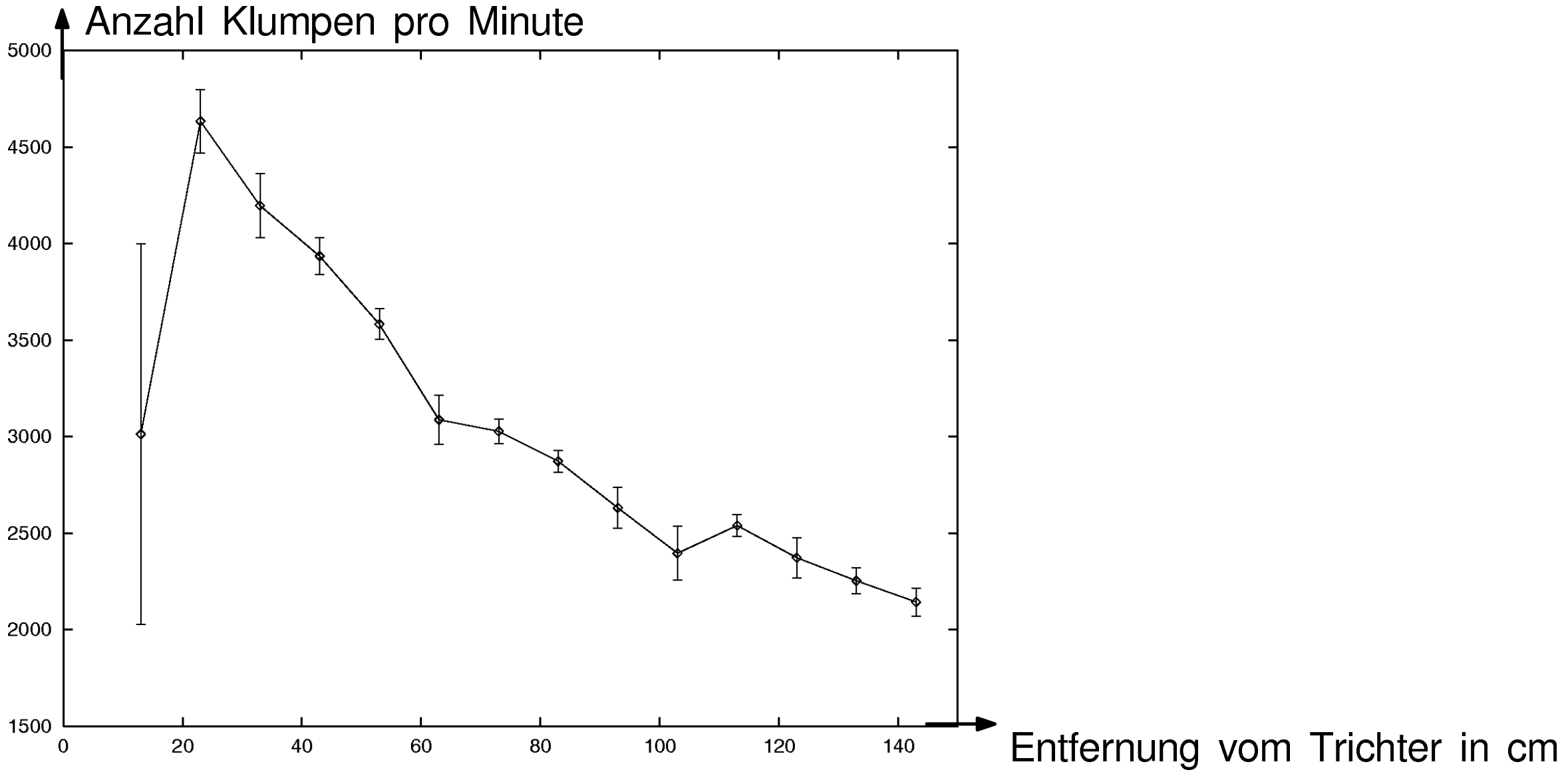

Bild 49 Mittelung von Bild 48 mit Standardabweichungen

# 4 Zusammenfassung und Ausblick

Nach der Erläuterung, was unter einem Granulat zu verstehen ist, wurden als erstes Langevin-Gleichungen zur wechselwirkungsfreien Beschreibung eines granularen Flusses in einem dünnen Rohr aufgestellt. Die zugehörige Fokker-Planck-Gleichung konnte für den homogenen Fall gelöst werden. Die Einbeziehung von inelastischen Stößen führte zu einer Boltzmann-Gleichung. Unter der Annahme des lokalen Gleichgewichts konnte zu einer hydrodynamischen Beschreibung übergegangen werden. Die erhaltenen hydrodynamischen Gleichungen und die homogene Lösung der Fokker-Planck-Gleichung veranlaßten zur Erweiterung der Langevin-Gleichungen auf den inhomogenen Fall. Für bestimmte Grenzfälle der Parameter konnten aus diesem Modell die hydrodynamischen Gleichungen von Kerner/Konhäuser abgeleitet werden, was den Zusammenhang zu Verkehrsflüssen herstellte. Anhand der hydrodynamischen Beschreibung war es möglich, lineare Stabilitätsuntersuchungen durchzuführen, in deren Ergebnis kritische Dichten für das Auftreten von Verklumpungserscheinungen berechnet werden konnten. Für einen sehr einfachen Fall konnte eine Bedingung angegeben werden, wann die Klumpen vorwärts oder rückwärts laufen.

In einem zweiten Kapitel wurden die Langevin-Gleichungen einmal für temperaturunabhängige Kraft und einmal für temperaturabhängige Kraft simuliert. Die Simulation von Langevin-Gleichungen bietet gegenüber einer Simulation von hydrodynamischen Gleichungen im wesentlichen folgende drei Vorteile:

– Man muß sich keine Gedanken um die numerische Stabilität machen.

– Es wird relativ wenig Rechenleistung benötigt.

– Es kann auch die Geschwindigkeitsverteilung der Teilchen bestimmt werden.

Anhand der Simulationen konnten die homogene Lösung und die kritischen Dichten verifiziert werden. Für den inhomogenen Fall wurde die Annahme des lokalen Gleichgewichts bestätigt. Außerdem war deutlich zu sehen, daß die Klumpen miteinander verschmelzen und die Zahl der Klumpen mit zunehmender Zeit abnimmt.

Im dritten Kapitel wurden schließlich Experimente mit Sand durchgeführt. Verschiedene Parameter des granularen Systems wurden abgeschätzt und dienten als Ausgangspunkt der numerischen Rechnungen des zweiten Kapitels. Mit einer Photozelle war es möglich, die Zahl der pro Zeiteinheit vorbeikommenden Klumpen an verschiedenen Rohrstellen zu messen. So konnte das Verschmelzen von Klumpen nachgewiesen werden.

Nach dem Nennen der Vorteile und Ergebnisse dieser Arbeit sollen auch die kritisch zu betrachtenden Punkte nicht verschwiegen werden. Die dargelegte Theorie hat im wesentlichen zwei Schwachstellen. Zum einen wird bei der verwendeten Form des Stoßintegrals der Impulserhaltungssatz verletzt, zum anderen kann eine reale Rohrwand immer nur dämp-fend auf die Geschwindigkeit eines stoßenden Teilchens wirken, während die Langevin-Kraft sowohl dämpfend als auch beschleunigend wirken kann.

Auf experimenteller Seite bleibt noch zu prüfen, welchen Einfluß die Luft auf die betrachteten Verklumpungserscheinungen hat. Darüber kann nur ein ähnliches Experiment im Vakuum Aufschluß geben.

# 5 Literatur

# 6 Sachwortregister

**V**

**W**

## Danksagung

Bedanken möchte ich mich bei allen Mitarbeitern des Lehrstuhls für ”Statistische Physik und Nichtlineare Dynamik” unter Leitung von Prof. Ebeling und des Lehrstuhls für ”Stochastische Prozesse” unter Leitung von Prof. Schimansky-Geier für die hilfreiche Zusammenarbeit und das gute Arbeitsklima.

Auch die zahlreichen Diskussionen mit meinem Betreuer Prof. Schimansky-Geier und Dr. Pöschel über allerlei theoretische und experimentelle Aspekte dieser Arbeit sollen hier nicht unerwähnt bleiben, da sie mich immer wieder motiviert haben und so wesentlich zum Gelingen meiner Diplomarbeit beigetragen haben.

Ein letzter Dank sei auch Heike Rheinländer, meinen Eltern und allen anderen meiner Bekannten ausgesprochen, die mich im privaten Bereich unterstützt haben.

Ich versichere, daß ich die vorliegende Arbeit selbständig und ohne die Verwendung unzulässiger Hilfsmittel angefertigt habe.

Ich bin damit einverstanden, daß diese Arbeit durch Bibliotheken der Öffentlichkeit zugänglich gemacht wird.

Berlin, am 13. 09. 1995